\documentclass[reprint,aps,amsmath,amssymb,prx, superscriptaddress]{revtex4-2}
\usepackage{graphicx}
\usepackage{dcolumn}
\usepackage{bm}
\usepackage{xcolor}
\usepackage[normalem]{ulem}
\usepackage[english]{babel}
\usepackage{float}
\usepackage{physics}
\usepackage[T1]{fontenc}
\usepackage[colorlinks]{hyperref}
\usepackage{tabularx}
\newcolumntype{Y}{>{\centering\arraybackslash}X}
\usepackage{accents}
\usepackage{mathtools}
\usepackage{amsmath}
\usepackage{amssymb}
\usepackage{soul}
\usepackage{tocloft}
\usepackage{titletoc}

\graphicspath{ {Figures/} }

\newcommand{\bk}[1]{ \hat{b}_{#1}^{\ } }
\newcommand{\bkd}[1]{ \hat{b}_{#1}^{\dagger} }

\newcommand{\rhoc}{ \hat{\rho}^c }
\newcommand{\rhou}{ \hat{\rho} }
\newcommand{\avg}[1]{ \langle #1 \rangle }
\newcommand{\avgc}[1]{ \langle #1 \rangle^c }
\newcommand{\Cc}[1]{ C^c_{#1} }

\newcommand{\varg}{ \text{\it{g}} }

\newcommand{\Ltot}{\mathcal{L}}
\newcommand{\Lqrc}{\mathcal{L}_{\rm \qrc}}
\newcommand{\Lsys}{\mathcal{L}_{\rm QS}}

\newcommand{\Lc}{\mathcal{L}_{c}}
\newcommand{\cmax}{\mathcal{C}_{\rm max}}

\newcommand{\CNL}{\mathcal{E}}
\newcommand{\CNLE}{\mathcal{E}_{\rm eff}}

\newcommand{\FD}{D_{\rm F}}
\newcommand{\amp}{\mathcal{A}}
\newcommand{\gammahet}{\gamma_{\rm H}}
\newcommand{\ntot}{R}

\newcommand{\frc}[1]{ {F}\!\left[ #1 \right] }
\newcommand{\frcl}[1]{ {F}_{\rm L}\!\left[ #1 \right] }
\newcommand{\frcnl}[1]{ {F}_{\rm NL}\!\left[ #1 \right] }

\newcommand{\qrc}{QNP}
\newcommand{\qrcs}{QNPs}

\newcommand\tbar[1]{\accentset{\rule{.4em}{.8pt}}{#1}}
\newcommand\vbar[1]{#1\hspace{0.1em}\rule[-0.25em]{0.15em}{0.15em}\rule[-0.25em]{0.15em}{1.25em}}
\newcommand\vvbar[1]{#1\hspace{0.1em}\rule[-1.0em]{0.15em}{0.15em}\rule[-1.0em]{0.15em}{2.5em}}

\newcommand\bigzero{\makebox(0,0){{\Large$\mathbf{0}$} } }

\definecolor{teal}{RGB}{32,178,170}
\definecolor{st1}{RGB}{31,120,180}
\definecolor{st2}{RGB}{255,127,14}
\definecolor{st3}{RGB}{25,87,179}
\definecolor{st4}{RGB}{255,191,0}
\definecolor{st5}{RGB}{32,178,170}
\definecolor{st6}{RGB}{176,83,203}
\definecolor{st7}{RGB}{32,178,170}
\definecolor{st8}{RGB}{176,83,203}
\newcommand{\ch}[1]{{\textcolor{black}{#1}}}

\makeatletter
\let\ORIbbl@fixname\bbl@fixname
\def\bbl@fixname#1{%
  \@ifundefined{languagealias@\expandafter\string#1}
    {\ORIbbl@fixname#1}
    {\edef\languagename{\@nameuse{languagealias@#1}}}%
}
\newcommand{\definelanguagealias}[2]{%
  \@namedef{languagealias@#1}{#2}%
}
\makeatother

\definelanguagealias{en}{english}
\definelanguagealias{EN}{english}

\begin{document}

\title[]{A neural processing approach to quantum state discrimination}

\makeatletter

\date{\today}

\author{Saeed A. Khan}
\affiliation{Department of Electrical and Computer Engineering, Princeton University, Princeton, NJ 08544}
\author{Fangjun Hu}
\affiliation{Department of Electrical and Computer Engineering, Princeton University, Princeton, NJ 08544}
\author{Gerasimos Angelatos}
\affiliation{Department of Electrical and Computer Engineering, Princeton University, Princeton, NJ 08544}
\author{Michael Hatridge}
\affiliation{Department of Physics and Astronomy, University of Pittsburgh, Pittsburgh}
\author{Hakan E. T\"{u}reci}
\affiliation{Department of Electrical and Computer Engineering, Princeton University, Princeton, NJ 08544}

\begin{abstract}
Although linear quantum amplification has proven essential to the processing of weak quantum signals, extracting higher-order quantum features such as correlations in principle demands nonlinear operations. However, nonlinear processing of quantum signals is often associated with non-idealities and excess noise, and absent a general framework to harness nonlinearity, such regimes are typically avoided. Here we present a framework to uncover general quantum signal processing principles of a broad class of bosonic quantum nonlinear processors (\qrcs{}), inspired by a remarkably analogous paradigm in nature: the processing of environmental stimuli by nonlinear, noisy neural ensembles, to enable perception. Using a quantum-coherent description of a \qrc{} monitoring a quantum signal source, we show that quantum nonlinearity can be harnessed to calculate higher-order features of an incident quantum signal, concentrating them into linearly-measurable observables, a transduction not possible using linear amplifiers. Secondly, \qrcs{} provide coherent nonlinear control over quantum fluctuations including their own added noise, enabling noise suppression in an observable without suppressing transduced information, a paradigm that bears striking similarities to optimal neural codings that allow perception even under highly stochastic neural dynamics. Unlike the neural case, we show that \qrc{}-engineered noise distributions can exhibit non-classical correlations, providing a new means to harness resources such as entanglement. Finally, we show that even simple \qrcs{} in realistic measurement chains can provide enhancements of signal-to-noise ratio for practical tasks such as quantum state discrimination. Our work provides pathways to utilize nonlinear quantum systems as general computation devices, and enables a new paradigm for nonlinear quantum information processing. 
\end{abstract}

\maketitle

\section{Introduction}

While engineering quantum systems hinges on isolating quantum components from their noisy classical environment, the observation of any such system necessitates extracting its emitted weak quantum signals back to the classical world. It is therefore essential to bolster quantum signals prior to their interaction with noisy classical modes and the readout noise they introduce, using signal processors that are themselves quantum-mechanical~\cite{caves_quantum_1982}. The most widely-used such processors are quantum amplifiers~\cite{yamamoto_flux-driven_2008, clerk_introduction_2010, roy_introduction_2016, aumentado_superconducting_2020}, which provide linear gain to an input signal and have been critical to quantum signal processing applications such as high-fidelity quantum state readout~\cite{bergeal_analog_2010, macklin_nearquantum-limited_2015} and generation of non-classical light~\cite{castellanos-beltran_amplification_2008, qiu_broadband_2023}. In spite of this success, the linearity of quantum amplifiers means they are only able to reveal a subset of the features of a complex quantum signal. A well-known limitation is the estimation of higher-order correlations. When only linear readout is available, the nonlinearity demanded by correlation calculations must be provided by classical post-processing: this process amplifies readout noise, leading to a signal-to-noise ratio that degrades exponentially with the order of the desired correlation function~\cite{da_silva_schemes_2010, eichler_experimental_2011, ryan_tomography_2015, boutin_effect_2017, rovny_nanoscale_2022}. For signals that exhibit non-classical, higher-order quantum correlations~\cite{boutin_effect_2017, schweigler_decay_2021}, linear processing restricts our ability to fully resolve the quantum domain, thereby limiting our capacity to control it.






Interestingly, even though a rich expanse of quantum devices can be realized beyond the confines of linearity~\cite{blais_quantum_2020}, the use of nonlinear quantum systems as general purpose quantum signal processors has been limited. Partly, this is because there are no established general principles of \textit{quantum} information processing with nonlinear quantum systems.  
The very fact that nonlinear systems are \textit{a priori} not bound by constraints of a linear dependence between input and output - which could hold the key to new ways of processing quantum signals - makes their operation much more complicated to analyze. More importantly, a complete description of the operation of any general quantum processor must be constrained by quantum mechanics, namely the uncertainty principle~\cite{heisenberg_uber_1927}. For linear quantum amplifiers, this translates to well-known limits on noise added during the amplification process~\cite{haus_quantum_1962, caves_quantum_1982, clerk_introduction_2010}. Analogous limits are much less explored for nonlinear amplifiers (with some exceptions~\cite{bondurant_quantum_1993, epstein_quantum_2021, yanagimoto_quantum_2023}), and nonlinear regimes are often even associated with excess noise~\cite{laflamme_quantum-limited_2011, boutin_effect_2017}. As a result, in spite of some measurement schemes effectively utilizing nonlinear systems~\cite{Vijay2009, Reed2010, sunada_photon-noise-tolerant_2024}, nonlinearity is often considered an inconvenience in quantum signal processing, being the source of gain saturation~\cite{roy_introduction_2016, boutin_effect_2017, liu_josephson_2017, liu_optimizing_2020} and operational instabilities~\cite{roy_introduction_2016}, which has led to significant efforts to mitigate nonlinear effects~\cite{frattini_3-wave_2017, sivak_kerr-free_2019}.

\begin{figure}[t]
    \centering
    \includegraphics[scale=1.0]{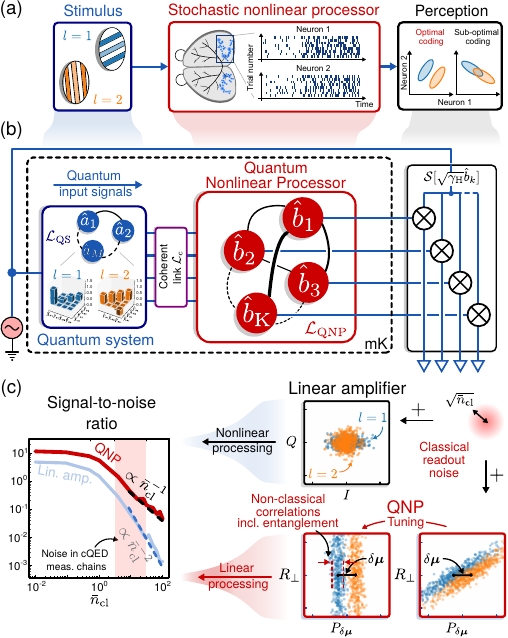}
    \caption{(a) Schematic representation of processing of visual signals by the brain. Visual stimuli lead to firing activity in neural ensembles in the visual cortex; the activity is highly stochastic even for identical stimuli, but can be strongly correlated for distinct neurons. Optimal neural codings encode visual information to avoid large noise in neuronal dynamics, whereas a sub-optimal encoding is more susceptible to noise. (b) Schematic representation of processing of quantum signals using a quantum nonlinear processor (\qrc{}). Quantum signals are generated by a quantum system (QS) governed by Liouvillian $\Lsys$, and are incident on the \qrc{} via a coherent link $\Lc$. We demonstrate the \qrc{} for the task of discriminating two states $\rhou_{\rm QS}^{(l)}$ of the QS using heterodyne readout of the \qrc{}, described by $\mathcal{S}[\sqrt{\gammahet }\hat{b}_k]$. (c) The \qrc{} enables such tasks using only linear processing of measured outputs, in contrast to linear quantum amplifiers, which require nonlinear post-processing. } 
    \label{fig:schematic}
\end{figure}

Remarkably, although the processing of weak stochastic signals using nonlinear devices is still not thoroughly explored in the quantum domain, this paradigm is fundamental to perhaps the most well-known information processor: the brain. It is understood that neurons, the signalling cells that constitute all nervous matter, function as part of large neural ensembles whose collective firing properties can be highly nonlinear~\cite{izhikevich_dynamical_2006} and exhibit great stochasticity~\cite{faisal_noise_2008, deco_stochastic_2009}. Experiments recording the activity of neural ensembles in the visual cortex of animals have provided robust evidence that neuronal response exhibits significant variability even under repeated presentations of identical stimuli (see Fig.~\ref{fig:schematic}(a)), with the total noise power even increasing with the size of the neural ensemble~\cite{rumyantsev_fundamental_2020}. Nevertheless, such noisy ensembles are able to successfully process distinct stimuli to enable perception~\cite{newsome_neuronal_1989, britten_analysis_1992}. Research over the past few decades has begun to establish an explanation~\cite{averbeck_neural_2006, urai_large-scale_2022, panzeri_structures_2022}. Nature appears to prefer encoding cortical signals in the response of collective neuron modes, as opposed to highly-variable single neuron dynamics (referred to as \textit{population} coding~\cite{abbott_effect_1999, panzeri_population_2010, montijn_population-level_2016, langdon_unifying_2023}). Furthermore, by exploiting the nonlinear response of neurons~\cite{shamir_nonlinear_2004}, the encoding of sensory information in collective neuron modes (or coding directions) that have the largest noise power is avoided~\cite{romo_correlated_2003, rumyantsev_fundamental_2020}. While neural noise still places limits on sensory signal discrimination~\cite{moreno-bote_information-limiting_2014, kanitscheider_origin_2015, bartolo_information-limiting_2020}, this coding principle allows the perception of weaker stimuli than would be resolvable if the largest noise mode was used for coding. This mechanism provides an example of how Nature is able to harness nonlinearity to facilitate signal processing in the presence of seemingly overwhelming stochasticity. 





Motivated by the example of stochastic neural processing, we therefore ask the question: can \textit{nonlinear}, fundamentally stochastic quantum systems provide an advantage in the detection and processing of weak quantum signals \textit{beyond the paradigm of linear amplification}? If so, can any general principles be established? Answering this question in its most general setting requires a description of the joint dynamics of the quantum nonlinear processor (QNP) and the signal generating quantum system (QS), often excluded in descriptions of linear quantum amplifiers~\cite{caves_quantum_1982} (with exceptions e.g. Ref.~\cite{silveri_theory_2016}). Here we develop the necessary framework to model a \qrc{} as an \textit{in situ} processor of signals generated by a quantum system (QS) it is coherently linked to~(see Fig.~\ref{fig:schematic}(b)); our approach thus accounts for the \qrc{} nonlinearity (with some restrictions), the QS-\qrc{} interaction (including the possibility of their entanglement), but most importantly for not only the quantum fluctuations of the \qrc{} itself, but also those due to the quantum nature of the input signals. Building on this framework, our analysis is able to identify two extremely general aspects of quantum information processing enabled by \qrcs{}, which we summarize here. First, by characterizing their now nonlinear input-output map, we show that \qrcs{} enable \textit{higher-order transduction}: the mapping of higher-order observables, such as the correlation of quantum signals, to lower-order observables such as quadrature means, making them accessible using much simpler linear readout protocols. Secondly, \qrcs{} are able to \textit{coherently manipulate} incident quantum fluctuations so as to direct amplified noise to observables that do not encode the transduced signal. This feature, which is particularly reminiscent of population coding in noisy neural ensembles, relies on both the \qrc{} nonlinearity and its deployment as a \textit{quantum} processor. 




To demonstrate the above results, we show \qrcs{} at work for a practical quantum information processing task inspired by Gaussian state tomography~\cite{sandbo_chang_generating_2018}: the binary discrimination of states for which the quadratures of constituent QS modes possess identical mean values, but differ only in their quantum fluctuation signatures~(Fig.~\ref{fig:schematic}(b)). These signatures may be evident in the quantum fluctuations of signals emanating from a single QS mode, or may only be revealed in non-classical noise correlations across distinct modes, such as for entangled states. When signals from a QS in such states are processed using standard linear amplifiers and heterodyne readout, the limitations of a linear input-output map ensure that obtained output distributions exhibit no difference in their mean values (Fig.~\ref{fig:schematic}(c), top). As a result, the distributions are impossible to distinguish perfectly using only a linear decision boundary. 

In contrast, we show that \qrcs{} are able to transduce the quantum fluctuation signatures of incident signals into displacements observable via heterodyne measurement, effectively functioning as quantum `cross-correlators' across multiple QS modes. As a result the obtained output distributions~(Fig.~\ref{fig:schematic}(c), bottom) exhibit a nonzero mean separation $\delta{\bm{\mu}}$, enabling a linear classifier to separate them perfectly. Most importantly, \qrcs{} enable all the nonlinear processing demanded by this discrimination task to be performed on pristine quantum signals, prior to their corruption by classical readout noise $\bar{n}_{\rm cl}$. Consequently, the signal-to-noise ratio for \qrc{}-enabled discrimination degrades only as $\bar{n}_{\rm cl}^{-1}$. In contrast, with standard linear amplifiers a nonlinear post-processing step must be performed on the final classical signals; this worsens readout noise, furnishing a signal-to-noise ratio that scales as $\bar{n}^{-2}_{\rm cl}$. We note that the concept of transduction using nonlinear systems has been explored previously~\cite{epstein_quantum_2021}, often via a lossless Hamiltonian description; here we consider a much more general setting, and account for practical constraints of losses and measurement. Secondly, here \qrcs{} are deployed as nonlinear processors of stochastic \textit{quantum} signals; this paradigm is therefore distinct from the more routinely explored use of quantum systems to compute nonlinear functions of deterministic \textit{classical} signals.

Finally, we show how \qrcs{} can control quantum fluctuations for information processing in ways unavailable to their linear counterparts. Linear amplifiers are restricted to perform the same transformation on an incident signal and its noise; for a fixed input signal, therefore, a decrease in the magnitude of the amplifier output noise is accompanied by a decrease in the magnitude of the output signal, and vice versa. Unconstrained by linearity, \qrcs{} enable a much more complex interplay between the transduced signal and the output noise. In particular, we show how tuning \qrc{} parameters alone, the output noise in the signal carrying quadrature $P_{\delta\bm{\mu}}$ (the unique quadrature parallel to $\delta\bm{\mu}$) can be decreased while the transduced signal magnitude $||\delta\bm{\mu}||$ remains fixed~(Fig.~\ref{fig:schematic}(c), bottom). Most strikingly, a tuning condition can be reached such that the signal carrying quadrature is parallel to the quadrature with minimal output noise, and thus orthogonal to all other quadrature combinations which possess larger noise, a scenario that appears very similar to optimal coding in neural circuits. Crucially, for \qrcs{} processing quantum input signals, the large noise modes include quantum fluctuations generated by the upstream QS, but also fluctuations amplified by the \qrc{} during processing itself, referred to as `added noise'. The minimum noise mode can even reach non-classical (i.e. sub-vacuum) levels, in the presence of squeezing or entanglement. Therefore the ability to control noise using \qrcs{} provides a practical means to harness non-classical correlations for quantum state discrimination.

A quantum-coherent description of a QS and \qrc{} in the same measurement chain, valid for multi-mode systems, and across a range of excitation conditions and interaction types, demands multiple theoretical techniques. Our key analytic results are built upon a nonlinear van Kampen expansion~\cite{van_kampen_stochastic_2007} of the Fokker-Planck equation associated with the quantum state of the complete measurement chain. At the expense of being restricted to weakly nonlinear \qrcs{}, this approach has the advantage of applying to arbitrarily-multimode bosonic quantum systems. As such, we expect the uncovered operating principles to hold for general \qrcs{} under heterodyne monitoring, especially relevant to cQED. These results are numerically verified and extended to stronger nonlinearities using a truncated cumulants approach that simulates measurement-conditioned dynamics of the measurement chain, with a complexity that is only quadratic in the number of total modes. Supplementing these results via full (stochastic) master equation simulations and exact results for select few-mode, multi-system measurement chains~\cite{SI}, we are able to provide a comprehensive description of quantum information processing using \qrcs{}.


The remainder of this paper is organized as follows. In Sec.~\ref{sec:qrc}, we introduce our model for processing quantum signals using a \qrc{}, provide a summary of the key results of our theoretical approach, and introduce the particular classification tasks we analyze. Sec.~\ref{sec:taski} then analyzes quantum state classification using a single-mode \qrc{}, based on a model that can be easily realized in cQED experiments. We highlight the crucial role of \qrc{} nonlinearity and discuss how \qrc{} parameters enable control over quantum fluctuations, including the ability to harness non-classical correlations. In Sec.~\ref{sec:comp} we provide a comparison of quantum state measurement using a \qrc{} against standard linear quantum amplifiers, finding improved robustness to added classical noise for the \qrc{}-based scheme. Sec.~\ref{sec:taskii} then considers quantum state classification tasks that require the non-local processing of quantum fluctuations, where a multi-mode \qrc{} becomes essential. Here, we demonstrate how output field entanglement can be engineered using a \qrc{} to produce sub-vacuum noise in a multi-mode quadrature. The paper concludes with a discussion of possible applications of \qrcs{} in the quantum information processing landscape.

\section{Quantum nonlinear processors for quantum signals}
\label{sec:qrc}

\subsection{Quantum measurement chain for processing quantum signals}

The complete measurement chain we consider is depicted in Fig.~\ref{fig:schematic}(b). The conditional evolution of the quantum state $\rhoc$ of the chain under continuous measurement is formally described by the stochastic master equation (SME)
\begin{align}
d\rhoc =  \left(\Lsys + \Lc + \Lqrc\right)\rhoc~dt +\sum_k\mathcal{S}[\sqrt{\gammahet}\hat{b}_k]\rhoc.
\label{eq:sme}
\end{align}
The superoperator $\Lsys$ describes a quantum system (QS) that is the source of quantum signals to be processed for a given quantum information processing task. We consider the form
\begin{align}
    \Lsys\hat{\rho} &= \! -i\big[\hat{\mathcal{H}}^{(l)}_{\rm QS},\hat{\rho}  \big]\! + \!\sum_m\kappa_m \mathcal{D}[\hat{a}_m]\hat{\rho},
    \label{eq:lsysgen}
\end{align}
where $\hat{\mathcal{H}}_{\rm QS}$ is a linear Hamiltonian governing the dynamics of $M$ bosonic modes $\hat{a}_m, m \in [1,M]$, and experiencing damping $\{\kappa_m\}$. The superscript $(l)$ indexes distinct states of the QS, to be distinguished in a classification task. 




This QS is coupled to the quantum nonlinear processor (\qrc{}) in the \textit{same measurement chain} (via a non-reciprocal coupling described by $\Lc$, see Appendix~\ref{app:measChain}), enabling the \qrc{} to monitor the QS as an \textit{in situ} measurement apparatus. The \qrc{} is governed by $\Lqrc$, 
\begin{align}
    \Lqrc\rhou = -i\Big[ \hat{\mathcal{H}}_{\rm \qrc} + \hat{\mathcal{N}}_{\rm \qrc} ,\rhou \Big],  
    \label{eq:Lqrc}
\end{align}
where $\hat{\mathcal{H}}_{\rm \qrc}$ is a general linear Hamiltonian governing the dynamics of $K$ bosonic modes $\hat{b}_k, k \in [1,K]$. The \qrc{} nonlinearity can then be furnished by any general bosonic nonlinear interaction; for simplicity we consider a Hamiltonian describing $K$ Kerr modes with common nonlinearity strength $\Lambda$, $\hat{\mathcal{N}}_{\rm \qrc{}} = -\sum_k \frac{\Lambda}{2} \bkd{k}\bkd{k}\bk{k}\bk{k}$.

The chosen measurement scheme is of the standard heterodyne type: it comprises continuous monitoring of the \qrc{} decay channels (with unit efficiency). The resulting measurement-conditioned evolution of the \textit{complete} measurement chain in accordance with quantum measurement theory~\cite{wiseman_quantum_2009} is governed by the stochastic heterodyne measurement superoperator $\mathcal{S}[\sqrt{\gammahet}\hat{b}_k]$, with a rate $\gammahet$ assumed equal for all modes (see Appendix~\ref{app:measChain}). This yields heterodyne records $\mathcal{I}_{k}(t),\mathcal{Q}_{k}(t)$ for each monitored \qrc{} mode~\cite{wiseman_quantum_2009}:
\begin{align}
    \mathcal{I}_{k}(t) &= \xi_{\mathcal{I}_k}\!(t) +\sqrt{\gamma_{\rm H}}\left[ \avg{\hat{X}_{k}} +  \xi_{\mathcal{I}_k}^{\rm qm}\!(t)\right] + \sqrt{\bar{n}_{\rm cl}}~\xi^{\rm cl}_{\mathcal{I}_k}\!(t), \nonumber \\
    \mathcal{Q}_{k}(t) &= \xi_{\mathcal{Q}_k}\!(t) + \sqrt{\gamma_{\rm H}}\left[\avg{\hat{P}_{k}} + \xi_{\mathcal{Q}_k}^{\rm qm}\!(t)\right] + \sqrt{\bar{n}_{\rm cl}}~\xi^{\rm cl}_{\mathcal{Q}_k}\!(t). \label{eq:IQkraw}
\end{align}
Heterodyne monitoring probes the canonically-conjugate \qrc{} quadratures $\hat{X}_k = \frac{1}{\sqrt{2}}(\bk{k}+\bkd{k})$, $\hat{P}_k = -\frac{i}{\sqrt{2}}(\bk{k}-\bkd{k})$, but a given measurement record also contains contributions from multiple noise sources $\xi$. The terms $\xi_{\mathcal{I}_k}\!,\xi_{\mathcal{Q}_k}\!$ describe vacuum noise that would be present even if no signal was emanating from the monitored quantum modes (i.e. $\gammahet\to 0$). The much more interesting terms are marked `${\rm qm}$': these describe noise contributions of a quantum origin, such as non-classical correlations due to squeezing or entanglement, or added noise by quantum dynamics, and are contingent on the measurement superoperator $\mathcal{S}$ (see Appendix~\ref{app:measChain}). In contrast, the terms `${\rm cl}$'  define classical readout noise in the measurement chain; these are not associated with a stochastic measurement superoperator and hence have no backaction on the quantum measurement chain. Equivalently, $\bar{n}_{\rm cl}$ quantifies noise added after the so-called \textit{Heisenberg-von Neumann cut}~\cite{heisenberg_physical_1930, wiseman_quantum_2009}. 

These stochastic records are thus often filtered to obtained heterodyne quadratures $(I_k,Q_k) \equiv \int_{t_0}^{t_0+\mathcal{T}} d\tau~\mathcal{K}(\tau)\times(\mathcal{I}_k(\tau),\mathcal{Q}_k(\tau))$, where $\mathcal{K}(\tau)$ is the filter function (we assume a boxcar filter $\mathcal{K}(\tau) = \frac{1}{\sqrt{2\mathcal{T}}}~\forall~\tau$) over a window of length $\mathcal{T}$ starting from an initial time $t_0$. The quadratures can be compactly represented via the vector $\bm{x} = (I_1,Q_1,\ldots,I_K,Q_K)^T \in \mathbb{R}^{2K}$. For any quantum information processing task, the data $\bm{x}$ is typically further processed to obtain a vector $\bm{y}$ of output features, a step we define generally via $\bm{y} = \frc{\bm{x}}$, which can include ensemble averaging over distinct shots, but also nonlinear processing to be clarified in due course. The complete measurement chain thus described is designed to measure QS properties via the obtained outputs $\bm{y}$; note that our description subsumes the standard paradigm of linear quantum amplification plus heterodyne measurement if $\hat{\mathcal{N}}_{\rm \qrc} \to 0$ and an appropriate choice of $\hat{\mathcal{H}}_{\rm \qrc{}}$ is made.





We can now specialize to the consideration of binary quantum state discrimination (although \qrcs{} can be deployed for more general processing tasks as well). Binary discrimination has a simple objective: to distinguish QS state $\rho_{\rm QS}^{(l)}$ from $\rho_{\rm QS}^{(p)}$, where $\rho_{\rm QS}^{(l)} \equiv {\rm tr}_{\rm \qrc{}}[\rhou^{(l)}]$, based on the corresponding outputs $\bm{y}^{(l)}$ and $\bm{y}^{(p)}$ from the measurement chain. To this end, we introduce a standard measure of the distinguishability of the two measured distributions of $\bm{y}^{(l)}$ and $\bm{y}^{(p)}$: Fisher's discriminant $\FD$, defined as
\begin{align}
\FD(\bm{y}) = \delta \bm{\mu}^T \cdot \mathbf{V}^{-1} \cdot \delta \bm{\mu},
\label{eq:fda}
\end{align}
where $\delta\bm{\mu} = \bm{\mu}^{(l)}-\bm{\mu}^{(p)}$ is the difference of means of the two measured distributions, while $\mathbf{V} = \frac{1}{2}\left(\mathbf{\Sigma}^{(l)}+\mathbf{\Sigma}^{(p)} \right)$ is a measure of their combined variance:
\begin{align}
    \bm{\mu}^{(l)} = \mathbb{E}\!\left[ \bm{y}^{(l)} \right],~\mathbf{\Sigma}^{(l)} = \mathbb{E}\!\left[\bm{y}^{(l)}\bm{y}^{(l)T} \right] - \mathbb{E}\!\left[\bm{y}^{(l)}\right]\mathbb{E}\!\left[\bm{y}^{(l)} \right]^T\!\!\!\!.
    \label{eq:Vl}
\end{align}
Note that $\FD$ has the intuitive form of a generalized signal-to-noise ratio (SNR). Furthermore, the fidelity $\mathcal{C}$ of classifying two Gaussian distributions with identical covariance matrices $\mathbf{\Sigma}^{(1)}=\mathbf{\Sigma}^{(2)}$, is simply $\mathcal{C} = \frac{1}{2}(1 + {\rm erf} \frac{\FD}{2\sqrt{2}})$, where ${\rm erf} z$ is the standard Gaussian error function; as $\FD\to\infty$, $\mathcal{C} \to 1$. For binary QS state discrimination, the $\FD$ of measured distributions is correlated with the fidelity of discriminating the two QS states that give rise to these distributions. 

Using such binary classification tasks, where the concepts of signal $\delta\bm{\mu}$ and noise $\mathbf{V}$ are precisely defined, we will demonstrate how \qrcs{} can process quantum signals \textit{in situ} for practical quantum information processing applications.

\subsection{Approximate input-output map for quantum nonlinear processors}
\label{subsec:nlsnr}

In the context of binary state discrimination, the role of \qrcs{} can be characterized by evaluating the central quantities $\delta{\bm{\mu}}$ and $\mathbf{V}$ identified in the previous section. However, the nonlinearity introduced by the \qrc{} - central to its operation - generally excludes exact theoretical treatments that would be valid if the measurement chain comprised only linear bosonic modes and interactions. Furthermore, allowing the QS and \qrc{} to comprise arbitrary numbers of modes renders exact numerical integration of Eq.~(\ref{eq:sme}) unfeasible for all but the lowest excitation numbers.

Our solution begins by introducing an alternate set of dynamical variables to describe the quantum state $\rhou$ of the entire nonlinear measurement chain: \textit{quantum cumulants}~(see Appendix~\ref{app:cumulants}). Formally, cumulants are an infinite set of dynamical variables parameterizing $\rhou$, indexed by an integer order $n_{\rm ord} \in \mathbb{Z}^+$. Crucially, however, we show that retaining cumulants only up to a certain \textit{finite} order $n_{\rm ord} \leq n_{\rm trunc}$ can very accurately describe the nonlinear measurement chains we consider, contingent on the nonlinearity strength. In this work, we show that $n_{\rm trunc} = 2$ is an excellent approximation provided certain well-understood and achievable constraints on the nonlinearity strength are met. In this case, defining $\hat{\bm{b}} \equiv (\hat{b}_1,\hat{b}_1^{\dagger},\ldots,\hat{b}_K,\hat{b}_K^{\dagger})^T$ and $\hat{\bm{a}}$ analogously for QS operators, the retained first-order cumulants are simply single-operator expectation values $\begin{psmallmatrix}
    \avg{\hat{\bm{a}}} \\
    \avg{\hat{\bm{b}}}
\end{psmallmatrix}$, while second-order cumulants are \textit{normal-ordered} covariances, $\mathbf{C} = \langle:\!\!\begin{psmallmatrix}
    \hat{\bm{a}} \\
    \hat{\bm{b}}
\end{psmallmatrix}\begin{psmallmatrix}
    \hat{\bm{a}}^T & \hat{\bm{b}}^T
\end{psmallmatrix} \!\!:\rangle - \avg{ \begin{psmallmatrix}
    \hat{\bm{a}} \\
    \hat{\bm{b}}
\end{psmallmatrix}}\avg{\begin{psmallmatrix}
    \hat{\bm{a}}^T & \hat{\bm{b}}^T
\end{psmallmatrix}}$. This constitutes a \textit{truncated cumulants} ansatz for $\rhou$: an efficient description of multimode nonlinear quantum dynamics, with the number of retained cumulants scaling only quadratically, and not exponentially, with the total number of modes $\ntot=K+M$ for $n_{\rm trunc}=2$.

We then develop an approximation that allows the truncated cumulants to be solved for analytically, formally employing a nonlinear van Kampen (NVK) expansion~\cite{van_kampen_stochastic_2007} in the \qrc{} nonlinearity (also referred to as a `system size' expansion, see full details in SI Sec.~\ref{app:FI}). Under the NVK approximation, first-order cumulants are written as $\begin{psmallmatrix}
    \avg{\hat{\bm{a}}} \\
    \avg{\hat{\bm{b}}}   
\end{psmallmatrix}= \tbar{\Lambda}^{-\frac{1}{2}}\begin{psmallmatrix}
    \avg{\hat{\tbar{\bm{a}}}} \\
    \avg{\hat{\tbar{\bm{b}}}}   
\end{psmallmatrix} + \begin{psmallmatrix}
    \avg{\bm{\delta}\hat{\bm{a}}} \\
    \avg{\bm{\delta}\hat{\bm{b}}}   
\end{psmallmatrix}$ where we have introduced the dimensionless nonlinearity $\tbar{\Lambda} = \frac{\Lambda}{\gamma}$, and $\gamma$, the \qrc{} mode decay rate, serves as a normalization factor. $\tbar{\Lambda} \to 0$ describes the classical (large occupation) limit, where $\avg{\hat{\tbar{\bm{a}}}}, \avg{\hat{\tbar{\bm{b}}}}$ become the dominant contributions to first-order cumulants; these define the expansion point and satisfy the equations of motion:
\begin{subequations}
    \begin{align}
    d_t \avg{\hat{\tbar{\bm{a}}}}&= \mathbf{L}^{(l)}_a \avg{\hat{\tbar{\bm{a}}}} + \tbar{\Lambda}^{\frac{1}{2}}\vec{\eta}^{(l)} \label{eq:teoma} \\
    d_t \avg{\hat{\tbar{\bm{b}}}} &= \mathbf{L}_b \avg{\hat{\tbar{\bm{b}}}} + \vec{N}_b(\avg{\hat{\tbar{\bm{b}}}}) - \vec{\eta}_b^{(l)} \label{eq:teomb}
\end{align}
\end{subequations}
These equations immediately provide useful insight. Being linear, the QS response is governed entirely by $\mathbf{L}_a$ and possible coherent drives $\vec{\eta}^{(l)}$, as dictated by $\hat{\mathcal{H}}_{\rm QS}^{(l)}$ (and dissipative terms). The non-reciprocal QS$\to$\qrc{} interaction defined by $\Lc$ ensures that the QS drives the \qrc{} via the coupling $\mathbf{\Gamma}$ and not vice-versa, leading to an effective QS state-dependent drive $\vec{\eta}_b^{(l)} \equiv \mathbf{\Gamma}\avg{\hat{\tbar{\bm{a}}}^{(l)}}$ on the \qrc{}. The \qrc{} dynamics, in contrast, contain both a linear contribution $\mathbf{L}_b$ as well as a nonlinear contribution $\vec{N}_b(\avg{\hat{\tbar{\bm{b}}}})$, the latter determined by the nonlinear Hamiltonian $\hat{\mathcal{N}}_{\rm \qrc}$. Thus Eq.~(\ref{eq:teomb}) is nonlinear in $\avg{\hat{\tbar{\bm{b}}}}$.


The full quantum state further requires specifying the deviation $\avg{\bm{\delta}\hat{\bm{b}}}$ (we can show that $\avg{\bm{\delta}\hat{\bm{a}}} \to 0$) and the second-order cumulants $\mathbf{C}$. Starting with the latter and introducing the block form $\mathbf{C} = \begin{psmallmatrix}
    \mathbf{C}_a & \mathbf{C}_{ab} \\
    \mathbf{C}_{ab}^T & \mathbf{C}_b
\end{psmallmatrix}$, we show that second-order cumulants satisfy the Lyapunov differential equation:
\begin{align}
    d_t \mathbf{C} =  
    \left[ \mathbf{C}
        \begin{pmatrix}
            \mathbf{L}_a & \mathbf{0} \\
            -\mathbf{\Gamma} & \tbar{\mathbf{J}}_b 
        \end{pmatrix}
    + \textit{m.t.} \right] +
    \begin{pmatrix}
    \mathbf{D}_a^{(l)} & \mathbf{0} \\ 
    \mathbf{0} &  \tbar{\mathbf{D}}_b
    \end{pmatrix},
    \label{eq:lyapunov}
\end{align}
where $m.t.$ is the (unconjugated) matrix transpose. Here $\tbar{\mathbf{J}}_b$ is the Jacobian matrix of the \qrc{}, $[\tbar{\mathbf{J}}_b]_{ij} = [\mathbf{L}_b]_{ij} + \frac{\partial [\vec{N}_b]_i}{\partial \avg{\hat{\tbar{\bm{b}}}}_j}$, with $[\cdot]_{ij}$ indicating tensor notation, and $(\tbar{\cdot})$ specifying evaluation at the expansion point $\avg{\hat{\tbar{\bm{b}}}}$. Then $\mathbf{D}_a$,  $\tbar{\mathbf{D}}_b$ are diffusion matrices for the QS and \qrc{} respectively; these describe incident fluctuations - quantum or classical - \textit{beyond} vacuum fluctuations, and must be nonvanishing to yield nontrivial $\mathbf{C}$.

Finally, we introduce the most important dynamical equation in the NVK approximation, for $\avg{\bm{\delta}\hat{\bm{b}}}$:
\begin{align}
    d_t \avg{\bm{\delta}\hat{\bm{b}}} = \tbar{\mathbf{J}}_b \avg{\bm{\delta}\hat{\bm{b}}} + \tbar{\Lambda}^{\frac{1}{2}}\tbar{\mathbf{H}}_b : \mathbf{C}_b^{(l)}.
\end{align}
Crucially, the change in first-order \qrc{} cumulants depends on the second-order cumulants, via the final term. This dependence is quantified by the \textit{Hessian tensor} $\tbar{\mathbf{H}}_b$ of the \qrc{}, defined as $[\tbar{\mathbf{H}}_b]_{ijk} = \frac{\partial [\vec{N}_b]_i}{\partial \avg{\hat{\tbar{\bm{b}}}}_j \partial \avg{\hat{\tbar{\bm{b}}}}_k}$. The Hessian operates on matrices - here on $\mathbf{C}_b$ - to return a vector, via the tensor double contraction $(:)$ over pairs of indices. The Hessian contribution is often neglected in standard linearization schemes; we will see its crucial role in quantum nonlinear processing. Note that if the \qrc{} is linear, $\vec{N}_b \to 0$, and the Hessian vanishes.


The NVK approximation, combined with input-output theory~\cite{gardiner_input_1985}, allows us to analytically obtain ${\bm{\mu}}^{(l)}$ and $\bm{\Sigma}^{(l)}$ as defined in Eq.~(\ref{eq:Vl}), and eventually $\FD$. The quadrature means after a filtering time $\mathcal{T}$ are given by 
\begin{align}
    {\bm{\mu}}^{(l)} = \sqrt{\frac{\gammahet\mathcal{T}}{2}}~\mathbf{U}_K\left( \tbar{\Lambda}^{-\frac{1}{2}}\avg{\hat{\tbar{\bm{b}}}^{(l)}} - \tbar{\Lambda}^{\frac{1}{2}}\tbar{\mathbf{J}}_b^{-1} \tbar{\mathbf{H}}_b :  \mathbf{C}_b^{(l)} \right),
    \label{eq:nvkmu}
\end{align}
where $\mathbf{U}_K$ is the $K$-mode quadrature change-of-basis matrix. The second term demonstrates transduction: the sensitivity of linearly-measurable \qrc{} quadratures to its higher-order (here, second-order) quantum cumulants. 

The covariance matrix in the long-filter limit ($\gamma \mathcal{T} \gg 1$, see SI Sec.~\ref{si:outputcov2}) is given by:
\begin{align}
    &\bm{\Sigma}^{(l)} \stackrel{\gamma \mathcal{T} \gg 1}{=} \sigma_{\rm vac}^2(\bar{n}_{\rm cl} + 1)\mathbf{I}_K \nonumber \\
    &+\sigma_{\rm vac}^2\frac{\gammahet}{\gamma}\mathbf{U}_K \tbar{\mathbf{J}}_b^{-1}\!\left[ \tbar{\mathbf{D}}_b + \mathbf{\Gamma}\mathbf{L}_a^{-1}\mathbf{D}_a^{(l)} (\mathbf{L}_a^{-1})^{T}\mathbf{\Gamma}^T \right]\!(\tbar{\mathbf{J}}_b^{-1})^{T}\mathbf{U}_K^T. 
    \label{eq:nvksigma}
\end{align}
The first term describes output vacuum fluctuations $\sigma_{\rm vac}^2 = \frac{1}{2}$ and classical readout noise $\propto \bar{n}_{\rm cl}$, the latter also in units of $\sigma_{\rm vac}^2$ ($\mathbf{I}_K \in \mathbb{R}^{2K}$ is the identity matrix). The second line then describes all other contributions from quantum noise. The term $\propto \tbar{\mathbf{D}}_b$ describes noise added by the \qrc{} itself. The term $\propto \mathbf{D}_a^{(l)}$, on the other hand, arises due to our quantum-coherent description of both the \qrc{} and QS: it describes noise originating from the QS, which arrives at the \qrc{} via the coupling $\mathbf{\Gamma}$, after undergoing QS evolution via $\mathbf{L}_a^{-1}$. Both noise terms are processed by the \qrc{}, as indicated by the appearance of the Jacobian $\tbar{\mathbf{J}}_b^{-1}$. 

The above expressions can be used to understand quantum state classification using general weakly-nonlinear bosonic quantum systems. First, we consider the limit of ideal linear quantum amplifiers, processing input drive signals ($\mathbf{D}_a \to 0$). In this case $\tbar{\mathbf{H}}_b \to 0$, $\tbar{\mathbf{J}}_b^{-1} \to \mathbf{L}_b^{-1}$, and $\tbar{\Lambda}^{-\frac{1}{2}}\avg{\hat{\tbar{\bm{b}}}} \to -\mathbf{L}_b^{-1}(\mathbf{L}_a^{(l)})^{-1}\vec{\eta}^{(l)}$; hence both the mean and covariance are determined entirely by the matrix $\mathbf{L}_b$, as must be the case for a linear system. The resulting expressions can be used to obtain the standard quantum limits~\cite{caves_quantum_1982, clerk_introduction_2010} on amplification~(see SI Sec.~\ref{app:noise}). 

\textit{Practical} quantum amplifiers, in contrast, can exhibit nonlinear behaviour for sufficiently strong input signals. Then, the leading contribution $\avg{\hat{\tbar{\bm{b}}}}$ to $\bm{\mu}$ is determined by the nonlinear Eq.~(\ref{eq:teomb}), while $\mathbf{\Sigma}$ is determined by $\tbar{\mathbf{J}}_b$, an indicator of the difference in response of nonlinear quantum systems to signal and noise. However, if the aim is still to process input drives $\vec{\eta}^{(l)}$, the large drives typically needed to reach such regimes lead to high signal-to-noise ratios where this difference is often ignored, with more attention instead paid to mitigating the nonlinear response of $\avg{\hat{\tbar{\bm{b}}}}$ (to limit gain compression, for example).


However, in other quantum information processing tasks, noise takes center stage. A simple case arises when discriminating quantum states $l$ and $p$ such that $\avg{\hat{{\tbar{\bm{a}}}}^{(l)}} = \avg{\hat{\tbar{\bm{a}}}^{(p)}}$, so that the leading term $\propto \avg{\hat{\tbar{\bm{b}}}}$ in $\bm{\mu}$ makes no contribution to $\delta\bm{\mu}$. Then, quantum state classification becomes the task of processing quantum fluctuations encoded in $\mathbf{D}_a^{(l)}$, and consequently in $\mathbf{C}_b^{(l)}$. The \qrc{} nonlinearity enables such processing \textit{in situ}, via the Hessian tensor. Furthermore, while the covariance is still determined by $\tbar{\mathbf{J}}_b^{-1}$ alone, $\delta\bm{\mu}$ is now determined both by $\tbar{\mathbf{J}}_b^{-1}$ and by the Hessian tensor $\tbar{\mathbf{H}}_b$. This enables a complex interplay of signal and noise only possible using nonlinear quantum systems. We demonstrate the implications of this result using various examples in this paper.  

While these expressions are derived using the NVK approximation, we show that the quantum information processing principles they reveal persist more generally. Importantly, the truncated cumulants ansatz holds past the NVK regime; we use this to develop a computational approach, the Stochastic Truncated Equations of Motion (STEOMs), that can be used to simulate measurement-conditioned dynamics of the nonlinear multimode measurement chain beyond the NVK approximation~(see SI Sec.~\ref{si:singleNodeQRC}). The STEOMs allow us to account for classification under finite sampling as in real experiments, numerically verify NVK results, and analyze performance in strongly-nonlinear \qrc{} regimes. Finally, we provide select comparisons using exact (S)ME integration to qualitatively verify our results without any assumptions on nonlinearity strength~(see SI Sec.~\ref{si:verify}).

\subsection{Quantum state discrimination tasks}
\label{subsec:taskdef}

We begin by defining the QS of Eq.~(\ref{eq:lsysgen}) whose states we wish to classify. Precisely, we consider the QS to be a general $M$-mode \textit{linear} quantum system under coherent driving at frequencies $\{\omega_{dm}\}$; written in the interaction picture and assuming resonant driving, it is described by the Hamiltonian
\begin{align}
    \hat{\mathcal{H}}_{\rm QS}^{(l)} &= \! \Big( \frac{1}{2}\!\sum_{m}\! G_m^{(l)}e^{-i\phi_m^{(l)}}\hat{a}_m^{2} + \!\! \sum_{n\neq m} \!\! G_{nm}^{(l)}e^{-i\phi_{nm}^{(l)}}\hat{a}_n\hat{a}_m + h.c. \Big) \nonumber \\
    &~~~~+\!\sum_{m}\!\eta_m^{(l)}(-i\hat{a}_m+i\hat{a}_m^{\dagger}).
    \label{eq:hsysamp}
\end{align}
The integer $l$ then indexes the states we wish to classify, generated by distinct choices of parameters also indexed by $l$. Such a Hamiltonian can be realized in the cQED architecture using tunable parametric drives~\cite{abdo_full_2013, abdo_nondegenerate_2013, sliwa_reconfigurable_2015}.

\begin{figure}[t!]
    \centering
    \includegraphics[scale=1.0]{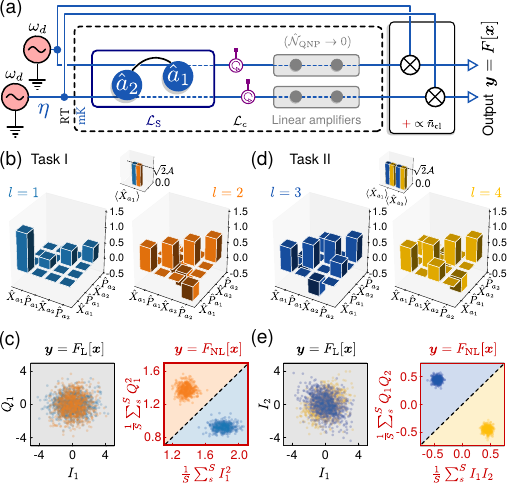}
    \caption{(a) Measurement chain defined by Eq.~(\ref{eq:sme}) for heterodyne monitoring of the QS (shown for $M=2$ modes) using \textit{linear} quantum amplifiers. (b) Task I: distinguishing a single-mode squeezed state from a two-mode squeezed state, using readout of mode $\hat{a}_1$ only. (c) Distributions of measured features for Task~I under linear (left) and nonlinear (right) processing (centered by subtracting off mean values) (d) Task~II: distinguishing two-mode squeezed states with orthogonal squeezed quadratures, using joint readout of modes $\hat{a}_1$ and $\hat{a}_2$. (e) Same as (c), for Task~II. }
    \label{fig:classifyAmpOnly}
\end{figure}

We are ultimately interested in coupling the QS directly to the \qrc{} - a nonlinear quantum device - for quantum signal processing. However, to provide a benchmark for later comparisons, we begin by introducing the classification tasks and how they may be performed in a simpler, more usual context: heterodyne readout of the QS using \textit{linear} quantum amplifiers. Recall that this simplified measurement chain, depicted in Fig.~\ref{fig:classifyAmpOnly}(a) for $M=2$, is still described by the same general form of SME, Eq.~(\ref{eq:sme}), provided $\hat{\mathcal{N}}_{\rm \qrc} \to 0$ and $\hat{\mathcal{H}}_{\rm \qrc{}} \to \hat{\mathcal{H}}_{\rm PP}$, where the latter Hamiltonian specifically describes the phase-preserving (PP) style of linear amplifier (for full model, see Appendix~\ref{app:linamp}).


We can finally define the binary quantum state discrimination tasks we wish to analyze: our work considers the discrimination of QS states which possess the unique feature that any monitored modes (i.e. modes that are coupled to the downstream processor) have identical steady-state quadrature expectation values for both states. More precisely, for two distinct states indexed by $l$ and $p$, $\avg{\hat{a}_{m}^{(l)}}=\avg{\hat{a}_{m}^{(p)}} = \amp~\forall~m~{\rm s.t.~}\Gamma_m~\neq~0$, where $\amp$ defines a common amplitude of monitored modes. Such states can then only be distinguished on the basis of their quantum fluctuation statistics, \textit{if} these are distinct.

For Task~I of this type, we consider distinguishing a single-mode squeezed state of mode $\hat{a}_1$ ($l=1$) from a two-mode squeezed state ($l=2$), by monitoring only mode $\hat{a}_1$ ($\Gamma_1 \neq 0, \Gamma_2 = 0$). The QS parameters used to realize these states are summarized in Table~\ref{tab:tasks}. The top plot in Fig.~\ref{fig:classifyAmpOnly}(b) shows the steady-state quadrature expectation values of the monitored mode $\hat{a}_1$, which are equal by construction. In contrast, the depicted full QS quadrature covariance matrices, related to the second-order QS cumulants $\mathbf{C}_a^{(l)}$ via a simple change-of-basis (see SI), clearly indicate the differences in quantum fluctuations of the two QS states. The discrimination task requires extracting these differences from measurement records ${\bm{x}}$, typically following some processing to obtain readout features $\bm{y}=\frc{{\bm{x}}}$. If we restrict this processing to only linear operations, the obtained distributions of features ${\bm{y}} = \frcl{{\bm{x}}} = ({I}_1,{Q}_1)$ (simulated using the STEOMs) are shown in the first panel of Fig.~\ref{fig:classifyAmpOnly}(c) for single-shot ($S=1$) readout of different realizations of each QS state. Clearly, the mean values of both measured feature distributions overlap; note that this mean would be unchanged by averaging over repeated shots ($S>1$). Hence $||\delta{\bm{\mu}}||\to 0$ and therefore $\FD \to 0$, leading to the conclusion that the two states cannot be distinguished in the space of $({I}_{1},{Q}_{1})$. 




\begin{table}[t!]
    \caption{QS parameters defining the two classification tasks considered, in units of common QS mode loss rate $\kappa$. Note that we set $\Delta^{(l)}_m = \phi^{(l)}_m = 0~\forall~m,~l$.}
    \begin{minipage}{.5\linewidth}
      \centering
    \begin{tabular}{m{0.07\columnwidth}<{\centering}|m{0.12\columnwidth}<{\centering}|m{0.12\columnwidth}<{\centering}|m{0.11\columnwidth}<{\centering}|m{0.19\columnwidth}<{\centering}|m{0.1\columnwidth}<{\centering}}
    \multicolumn{6}{c}{Task~I} \\
    $l$ & $G_{1}^{(l)}$ & $G_{12}^{(l)}$ & $\phi_{12}^{(l)}$ & $\eta_1^{(l)}$ & $\eta_2^{(l)}$  \\
    \hline
    \hline
    {\color{st1}$1$} & {\color{st1}$0.3$} & {\color{st1}$0.0$} & $0.0$ & {\color{st1}$0.20$}$\amp$ & $0.0$  \\
    {\color{st2}$2$} & {\color{st2}$0.0$} & {\color{st2}$0.3$} & $0.0$ & {\color{st2}$0.32$}$\amp$ & $0.0$  \\
    \end{tabular}
    \end{minipage}%
    \vrule height 0.038\textheight depth 0.038\textheight
    \begin{minipage}{.5\linewidth}
      \centering
    \begin{tabular}{m{0.07\columnwidth}<{\centering}|m{0.12\columnwidth}<{\centering}|m{0.12\columnwidth}<{\centering}|m{0.11\columnwidth}<{\centering}|m{0.16\columnwidth}<{\centering}|m{0.16\columnwidth}<{\centering}}
    \multicolumn{6}{c}{Task~II} \\
    $l$& $G_{1}^{(l)}$ & $G_{12}^{(l)}$ & $\phi_{12}^{(l)}$ & $\eta_1^{(l)}$ & $\eta_2^{(l)}$  \\
    \hline
    \hline
    {\color{st3}$3$} & $0.0$ & $0.3$ & {\color{st3}$-\frac{\pi}{2}$} & {\color{st3}$0.8$}$\amp$ & {\color{st3}$0.8$}$\amp$ \\
    {\color{st4}$4$} & $0.0$ & $0.3$ & {\color{st4}$+\frac{\pi}{2}$} & {\color{st4}$0.2$}$\amp$ & {\color{st4}$0.2$}$\amp$ \\
    \end{tabular}
    \end{minipage} 
    \label{tab:tasks}
\end{table}


However, the distributions of measured quadratures \textit{are} visibly distinct, just not in their mean values, but instead in their second-order moments or covariances. To estimate such second-order moments (routinely required for example for Gaussian state tomography~\cite{smithey_measurement_1993, sandbo_chang_generating_2018, kotler_direct_2021}), the standard approach is to obtain $S$ shots and estimate the variance of the measurements over the dataset yielding readout features ${\bm{y}} = \frcnl{{\bm{x}}} = \frac{1}{S}\sum_s^S({I}^2_1,{Q}^2_1)$ (shot $s$ dependence of quadratures $I_k,Q_k$ is implied).  For Task~I, we plot distributions of these nonlinear readout features in the second panel of Fig.~\ref{fig:classifyAmpOnly}(c). The mean values of these features are now estimators of the monitored mode covariances; as these are distinct, the centers of the distributions no longer coincide, rendering them linearly separable. Without this nonlinear processing step, it is impossible to perfectly distinguish the QS states in Task~I under QS readout using only linear amplifiers.


For Task~II we consider a more complex classification task, which requires both nonlinear and \textit{nonlocal} information processing. Specifically, we wish to distinguish a pair of two-mode squeezed states $l \in (3,4)$ that experience an identical amount of joint squeezing, but whose two-mode squeezed quadratures are mutually orthogonal. Such states can be generated using two-mode squeezing interactions of equal strength but opposite phase (see Table~\ref{tab:tasks}). The full QS covariance matrices (see Fig.~\ref{fig:classifyAmpOnly}(d)) then differ only in the \textit{sign of cross-correlations} between the two QS modes (off-diagonal blocks); all other covariance metrics are identical. 

Measuring such cross-correlations necessitates monitoring both QS modes ($\Gamma_1, \Gamma_2 \neq 0$). We again first show measured features under linear processing alone, now for two distinct modes, ${\bm{y}} = \frcl{{\bm{x}}} = ({I}_1,{I}_2)$, in the left panel of  Fig.~\ref{fig:classifyAmpOnly}(e). Clearly, the distributions have anisotropic profiles that differ in the axis of minimal fluctuations, indicative of two-mode squeezing of distinct joint quadratures. However, the centers of measured distributions overlap as before, due to their equal mean values. Also as before, nonlinear processing to estimate second-order moments provides the solution, but with an important caveat: all \textit{local} second-order moments are insensitive to the sign of cross-correlations of the QS modes. Instead, the estimation of \textit{nonlocal} second-order moments, ${\bm{y}} = \frcnl{{\bm{x}}} = \frac{1}{S}\sum_s^S({I}_1{I}_2,{Q}_1{Q}_2)$, is necessary to obtain linearly separable distributions for Task~II.

Therefore, both our considered classification tasks - instances of the broad family of tasks that require measuring correlations between observables to distinguish quantum states - demand nonlinear processing of heterodyne records at room temperature. However, such processing can be particularly sensitive to high-temperature noise $\propto \bar{n}_{\rm cl}$ in the measurement chain. A key result of our work is that using \qrcs{}{} to process signals from the QS in the same cryogenic environment where they are generated can circumvent the need for nonlinear processing at room temperature. This renders classification schemes incorporating \qrcs{}{} much more robust to excess noise in the measurement chain (see Sec.~\ref{sec:comp}).



\section{Quantum state discrimination using a single-mode \qrc}
\label{sec:taski}



\begin{figure*}[t]
    \centering
    \includegraphics[scale=1.0]{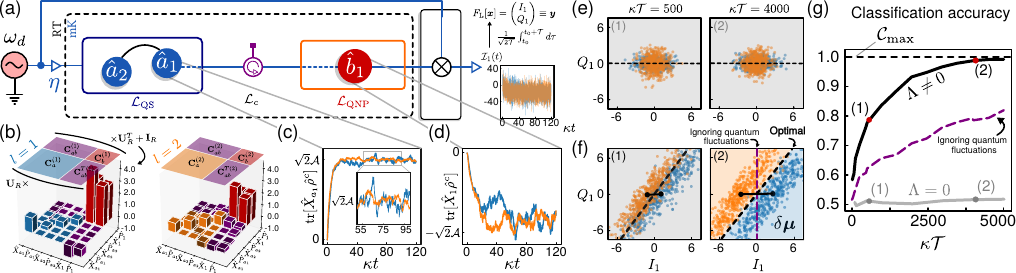}
    \caption{(a) Measurement chain for Task~I: using a single-mode \qrc{} to distinguish a single-mode squeezed state $\rhou_{\rm QS}^{(1)}$ from a two-mode squeezed state $\rhou_{\rm QS}^{(2)}$. (b) Covariance matrices of the joint quantum state of the measurement chain when the QS is in state $\rhou_{\rm QS}^{(1)}$ (left) and $\rhou_{\rm QS}^{(2)}$ (right); $\mathbf{U}_{\ntot}, \mathbf{I}_{\ntot}$ are the $R$-mode change-of-basis and identity matrix respectively. (c) STEOMs simulations of quantum trajectories and measurement records: for each QS state, we show the expectation value of $\hat{X}$ conditioned on an individual quantum trajectory $\rhoc$ for (i) mode $\hat{a}_1$ and (ii) mode $\hat{b}_1$, while the resulting heterodyne record $\mathcal{I}_1(t)$ is shown in (iii). (e) Distribution of readout features ${\bm{y}}$ under linear processing $\frcl{\cdot}$ for a \qrc{} with $\Lambda=0$ and for filtering times (1) $\kappa\mathcal{T} = 500$, and (2) $\kappa\mathcal{T} = 4000$. (f) Same as (e) but now for a \qrc{} with $\Lambda \neq 0$. (g) Classification accuracy as a function of $\kappa\mathcal{T}$. Black curve is for the \qrc{} with $\Lambda \neq 0$ and optimal decision boundary in (f), purple curve is obtained if quantum fluctuations are ignored (purple decision boundary in (f)), while the gray curve corresponds to the \qrc{} with $\Lambda = 0$ in (e).   }
    \label{fig:classifyAmpStates}
\end{figure*}


We now demonstrate the paradigm of \textit{in situ} nonlinear processing of quantum signals enabled by the \qrc{} in the context of quantum state discrimination. In particular, we will show that such processing would not be possible using standard linear amplifiers in the same measurement configuration. Our analysis begins with the simpler Task~I; Task~II is analyzed in Sec.~\ref{sec:taskii}. 

Recalling that only a single QS mode is read out in Task~I, we consider a $K=1$ \qrc{} to monitor this QS mode, which proves sufficient. Our chosen \qrc{} is defined by a single Kerr-nonlinear mode with frequency $\omega_1$ and nonlinearity $\Lambda$ (for details, see Appendix~\ref{app:measChain}). The resulting quantum measurement chain is then depicted in Fig.~\ref{fig:classifyAmpStates}(a). When coupled, the quantum state of the \qrc{} is determined by the state of the QS, as desired: this dependence is shown in Fig.~\ref{fig:classifyAmpStates}(b) via the covariance matrices of the complete measurement chain for the two QS states to be disntinguished. In the NVK approximation, the dependence is quantified by the linear Lyapunov system, Eq.~(\ref{eq:lyapunov}), although the exact relationship can be more complex. From Eq.~(\ref{eq:lyapunov}), it is also clear that this inter-dependence itself does not require the \qrc{} to be nonlinear; it only requires a nonzero coupling $\mathbf{\Gamma}$.

Crucially, we require this dependence to be transduced to \qrc{} readout features to enable successful QS state discrimination by \textit{linear} readout of the \qrc{} alone. It is here that the role of \qrc{} nonlinearity becomes clear. To demonstrate this, we analyze \qrc{} readout distributions for features obtained under linear processing only, ${\bm{y}} = \frcl{{\bm{x}}} = (I_1,Q_1)$, as a function of nonlinearity. Our STEOMs framework enables simulating individual quantum trajectories of the QS and \qrc{} (crucially accounting for the latter's nonlinearity), providing the resulting heterodyne measurement records defined by Eq.~(\ref{eq:IQkraw}) that are used to construct ${\bm{y}}$; typical examples are shown in Fig.~\ref{fig:classifyAmpStates}(c), (d). By repeating over several measurement chain initializations for each QS state, here $l=1,2$, we obtain distributions of measured features shown in Fig.~\ref{fig:classifyAmpStates}(e), (f). Then Fisher's discriminant $\FD$ computed for the two distributions determines the fidelity of classifying the QS states.

Recall that $\FD$ depends on the mean separation $\delta{\bm{\mu}}$ of the distributions. The NVK approximation provides a very useful form of $\delta{\bm{\mu}}$, Eqs.~(\ref{eq:nvkmu}), which specialized to the single-mode Kerr \qrc{} takes the form:
\begin{align}
    \delta{\bm{\mu}} = \sqrt{\frac{\gammahet \mathcal{T}}{2}}\mathbf{U}_K \left[  \sqrt{\frac{\Lambda}{\gamma}}~ \tbar{\mathbf{J}}_b^{-1}\tbar{\mathbf{H}}_b : \left( \mathbf{C}_b^{(l)}-\mathbf{C}_b^{(p)} \right) \right],
    \label{eq:nvkmukerr}
\end{align}
where $\gamma = \gammahet + \Gamma_1$ is the total \qrc{} damping rate. The nonlinear dependence is associated with the Hessian $\tbar{\mathbf{H}}_b$, which vanishes for linear systems, so that $||\delta{\bm{\mu}}|| = 0$. This is exactly what is observed for readout with linear amplifiers in Sec.~\ref{subsec:taskdef}, as well as with vanishing \qrc{} nonlinearity ($\Lambda = 0$), shown in Fig.~\ref{fig:classifyAmpStates}(e). In this case, increasing the filtering time $\mathcal{T}$ also has no effect on $||\delta{\bm{\mu}}||$, and consequently on $\FD$. 



The measured distributions change qualitatively when the \qrc{} is nonlinear ($\Lambda \neq 0$), as shown in Fig.~\ref{fig:classifyAmpStates}(f). Now $||\delta{\bm{\mu}}||\neq 0$, and the mean separation increases with $\mathcal{T}$ as predicted by Eq.~(\ref{eq:nvkmukerr}). The \qrc{} is able to transduce information encoded in \qrc{} fluctuations $\mathbf{C}_b^{(l)}$, which in turn depend on the QS state, to \qrc{} quadrature mean values, an operation governed by the Hessian $\tbar{\mathbf{H}}_b$. Now even with only linear processing $\frcl{\cdot}$ of measured quadratures, $\FD \neq 0$: a suitable \qrc{} is thus able to circumvent the need for nonlinear post-processing of measurement records, and the resulting classification accuracy shown in Fig.~\ref{fig:classifyAmpStates}(f) approaches unity. In contrast, for vanishing nonlinearity, perfect classification is impossible using only $\frcl{\cdot}$.


This simple example highlights the extremely general principle of transduction using \qrcs{}. However, making successful use of this principle requires asking some more complex questions, each of which we address in this paper. First, one may ask whether \textit{any} \qrc{} works for a given quantum state discrimination task. Unsurprisingly, this is not the case; instead, \qrc{} optimization is important, and features a landscape far richer than that of linear quantum amplifiers. The reason for this is simple: the very nonlinearity that enables a nonzero `signal' $||\delta{\bm{\mu}}||$ also determines the fluctuations $\mathbf{C}_b^{(l)}$ that appear in the `noise', characterized by the measured covariance matrix $\mathbf{\Sigma}^{(l)}$. This leads to a complex interplay between signal and noise for classification using \qrcs{}, where one cannot be optimized without considering the other. Further complicating this interplay is the noise added by the \qrc{} itself, which must also be accounted for in the optimization. While the limits of added noise have been well-established for linear quantum amplifiers~\cite{caves_quantum_1982, clerk_introduction_2010}, much less is known about added noise in the context of nonlinear quantum systems, and even less so for driven-dissipative cases relevant in practical settings. 

Our framework characterizes this nontrivial relationship between signal and noise in \qrcs{}. Even more importantly, it also accounts for noise added by \qrcs{} during processing. The latter crucially includes the possibility of the noise being correlated, even non-classically, either due to correlations in the incident quantum signals arriving from the QS, or even due to correlations generated by the \qrc{} dynamics itself. Crucially, we find that the nonlinear nature of \qrcs{} implies that the signal and noise contributions exhibit different relative functional dependencies, a freedom not available to linear quantum systems. This distinction provides the ability to manipulate the readout noise for \qrcs{} so as to limit its impact on the signal, by tuning parameters of the \qrc{} alone. This capability proves crucial to the optimization of a \qrc{} for learning, as is discussed in the next subsection.

We emphasize that a precise understanding of the noise physics and optimal operating regimes of \qrcs{} is not merely an academic question, but one that must be addressed to understand the ultimate limits of information processing with such devices. For example, for classification with minimal resources (e.g. filtering time $\mathcal{T}$), information processing must be carried out in regimes where the signal $||\delta{\bm{\mu}}||$ is of comparable magnitude to the standard deviation of measured distributions. Such a case is depicted in Fig.~\ref{fig:classifyAmpStates}(f). From the observed non-isotropic distributions, it is clear that \qrc{} readout can exhibit correlated noise statistics. If the correlated nature of fluctuations was ignored (in other words, if the added noise was assumed to be uncorrelated and isotropic), the decision boundary would simply be the bisector of the mean separation $||\delta{\bm{\mu}}||$, depicted by the dashed purple line. Unless $\mathcal{T}$ is very large so that $||\delta{\bm{\mu}}||$ dominates over noise (a significant increase in measurement resources), the resulting classification accuracy (purple) is substantially lower than the optimal. Therefore the correlation properties of noise in \qrc{} readout - including of noise added by the \qrc{} itself - must be accounted for, as we do here and analyze in detail next.

\begin{figure*}[t]
    \centering
    \includegraphics[scale=1.0]{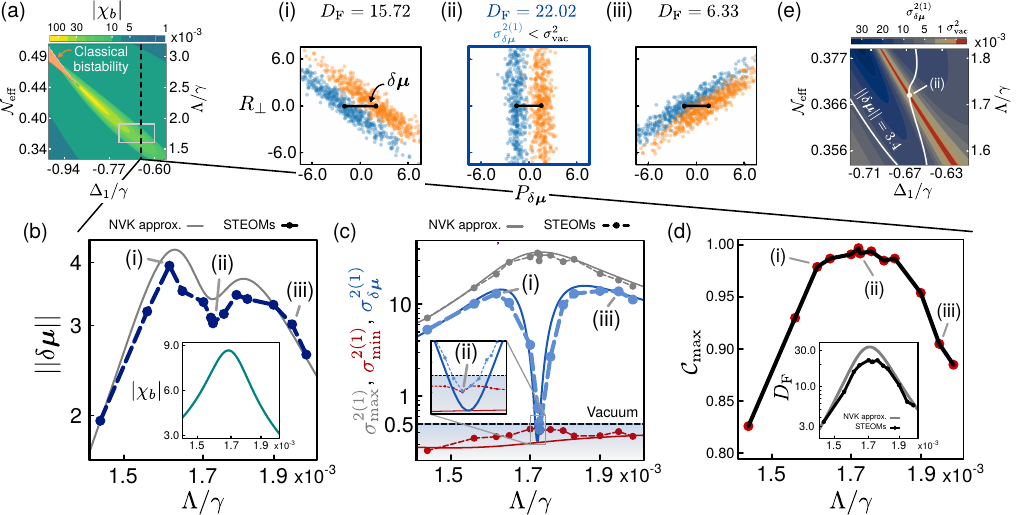}
    \caption{(a) Susceptibility $|\chi_b|$ as a function of nonlinearity ${\Lambda}/\gamma$ and detuning $\Delta_1/\gamma$, which exhibits an increase as the classical Kerr oscillator bistability is approached. (b) $|\chi_b|$ and $||\delta{\bm{\mu}}||$, (c) noise properties, and (d) $\cmax$ as a function of nonlinearity ${\Lambda}/\gamma$, for the cross-section of (a) at fixed detuning ${\Delta}_1/{\gamma} = -0.67$. Inset in (d) indicates Fisher's linear discriminant $\FD$. Results are shown using both the NVK approximation and numerical STEOMs (apart from $\cmax$). (i)-(iii) show measured \qrc{} quadrature distributions under only linear processing $\frcl{\cdot}$, simulated using the STEOMs at the indicated nonlinearity values, depicting the ability to manipulate quantum fluctuations. (ii) maximises $\FD$ within the NVK approximation by minimizing projected noise. (e) $\sigma_{\delta{\bm{\mu}}}^{2(1)}$ as a function of nonlinearity ${\Lambda}/\gamma$ and detuning $\Delta_1/\gamma$ in the NVK approximation. The white curve shows operating conditions with constant $||\delta{\bm{\mu}}||$, but with clearly varying projected noise values.  }
    \label{fig:ampClassifyLambda}
\end{figure*}


\subsection{Coherent control of quantum fluctuations using a \qrc{}}
\label{subsec:squeezing}

We have seen that the \qrc{} nonlinearity is essential to learn QS states using linear readout. However, not all \qrcs{} will be equally effective at learning: performance metrics such as $\cmax$ for a fixed $\mathcal{T}$ are strongly dependent on \qrc{} operating parameters. In this section, we address how \qrcs{} can be optimized for learning. Our analysis centers around understanding how Fisher's discriminant $\FD$ for \qrc-enabled classification can be maximised, to which end analytic expressions obtained using the NVK approximation are particularly powerful. In particular, while our analysis is presented for Task~I using a $K=1$ mode \qrc{}, the NVK approximation allows us to uncover general principles for \qrc{} operation that apply to more complex tasks with larger \qrcs{}, as we show in Sec.~\ref{sec:taskii} with Task~II.

A nonzero Fisher's discriminant $\FD$ requires a non-vanishing $\delta{\bm{\mu}}$. Within the NVK approximation, Eq.~(\ref{eq:nvkmukerr}), it is clear that the magnitude $||\delta{\bm{\mu}}||$ must be determined by the inverse of the Jacobian $\tbar{\mathbf{J}}_b^{-1}$ describing the \qrc{}. This is unsurprising, as the Jacobian typically determines the linearized response of a nonlinear system, with the dimensionless susceptibility $|\chi_b|$ given by the largest eigenvalue of $\tbar{\mathbf{J}}_b^{-1}$. As discussed in Appendix~\ref{app:FI}, for a $K=1$ mode Kerr-type \qrc{} this takes the form
\begin{align}
    |\chi_b| = {\rm max}\{|{\rm eig}~\tbar{\mathbf{J}}_b^{-1}|\} = \frac{\gamma}{ \frac{\gamma}{2} - \sqrt{(2\tbar{n}_1)^2- (\Delta_1+4\tbar{n}_1)^2  } },
\end{align}
where $\tbar{n}_1 = |\avg{\tbar{b}_1}|^2$, obtained by solving Eq.~(\ref{eq:teomb}). Importantly, it can be shown that $|\chi_b|$ depends on the \qrc{} nonlinearity only via the dimensionless effective parameter $\CNLE$,
\begin{align}
    \CNLE = \sqrt{\frac{\Lambda}{\gamma}}\cdot \frac{\Gamma_1}{\gamma}\avg{\hat{\tbar{a}}_1^{(l)}} = \sqrt{\frac{\Lambda}{\gamma}}\cdot\amp\frac{\Gamma_1}{\gamma} 
    \label{eq:cnle}
\end{align}
which is independent of $l$ due to our choice of classification tasks. The above form of $|\chi_b|$ is ubiquitous in studies of the linearized response of a variety of Kerr-based systems, from parametric amplifiers~\cite{laflamme_quantum-limited_2011} to frequency comb generators~\cite{khan_frequency_2018}. The susceptibility can be made large by suitable parameter choices, including $\CNLE$; typically this is determined by the strength of a separately applied pump tone. Our use of Kerr-based \qrcs{} exhibits a slight difference: $\CNLE$ is instead set by the amplitude $\amp$ of signals incident from the QS upstream. The QS state to be measured hence serves to `pump' the very \qrc{} being used to measure it (although a distinct \qrc{} pump tone could equivalently be applied). 

This difference notwithstanding, suitable choices of $(\CNLE,\Delta_1)$ can similarly cause $|\chi_b|$ to become large, increasing the magnitude of \qrc{} response to any input, including to quantum fluctuations from the QS. In Fig.~\ref{fig:ampClassifyLambda}(a), we plot $|\chi_b|$ as a function of $\Delta_1/\gamma$ and $\CNLE$ (also $\Lambda/\gamma$ for $\amp\cdot\frac{\Gamma_1}{\gamma}=\frac{80}{9}$). The orange region, where $|\chi_b|$ diverges, marks the well-known classical bistability of the single Kerr oscillator~\cite{dykman_fluctuating_2012}, here brought about by the mean amplitude $\amp$ of incident QS signals. Operating near this bistability - and more generally, any instability of the Jacobian $\tbar{\mathbf{J}}_b$ - will increase $|\chi_b|$. 

We therefore consider \qrcs{} near the bistability, more precisely for the fixed detuning $\Delta_1/\gamma = -0.67$ and for varying $\CNLE$ across the vertical dashed line in Fig.~\ref{fig:ampClassifyLambda}(a). Here the susceptibility $|\chi_b|$ is plotted in the inset of Fig.~\ref{fig:ampClassifyLambda}(b), which exhibits a Lorentzian-like profile with a single maximum. Note that the important quantity for Fisher's discriminant is instead the magnitude of measured mean separation $||\delta{\bm{\mu}}||$; this is also shown in Fig.~\ref{fig:ampClassifyLambda}(b), using both the NVK approximation and STEOMs integration. Interestingly, while $||\delta{\bm{\mu}}||$ does increase in conjunction with $|\chi_b|$, it also displays a double-peak structure that is manifestly distinct. Thus, while $|\chi_b|$ is clearly important, it does not completely define the physics of QS state learning using a \qrc{}. Here, the difference can be attributed to the fact that $||\delta{\bm{\mu}}||$ is not simply the linearized response to an input field, but the response to quantum fluctuations encoded in $\mathbf{C}_{b}^{(l)}$, which drive the \qrc{} via the Hessian tensor $\tbar{\mathbf{H}}_b$ (which is absent in standard linearization approaches). 

The other factor determining $\FD$ is the noise in measured quadratures, quantified by the covariance matrices $\mathbf{\Sigma}^{(l)}$. To aid our analysis of noise properties, we first introduce the minimum and maximum noise eigenvalues and corresponding eigenvectors of the $2K$-by-$2K$ covariance matrix $\mathbf{\Sigma}^{(l)}$ as $\{\sigma_{\rm min}^{2(l)},{\bm{v}}_{\rm min}^{(l)}\}$ and $\{\sigma_{\rm max}^{2(l)},{\bm{v}}_{\rm max}^{(l)}\}$ respectively. These eigenpairs denote combinations of measured \qrc{} quadratures with minimal and maximal noise, and for $K=1$, completely define $\mathbf{\Sigma}^{(l)}$. We next note that binary classification in an arbitrary-dimensional measured space can be cast into a two-dimensional subspace for visualization. We introduce a vector ${\bm{v}}_{\parallel} = \frac{1}{||\delta{\bm{\mu}}||}\delta{\bm{\mu}}$ parallel to $\delta{\bm{\mu}}$, which is unique upto normalization, and a vector $\bm{v}_{\perp}$ orthogonal to ${\bm{v}}_{\parallel}$, namely ${\bm{v}}_{\parallel}^T{\bm{v}}_{\perp} = 0$, of which there are $2K-1$ choices. These vectors allow us to a define a measured \qrc{} quadrature $P_{\delta{\bm{\mu}}}$ that is parallel to $\delta{\bm{\mu}}$, and $R_{\perp}$ as one of $2K-1$ measured quadratures orthogonal to $\delta{\bm{\mu}}$,
\begin{align}
    P_{\delta{\bm{\mu}}} = {\bm{v}}_{\parallel}^T{\bm{y}},~R_{\perp} = {\bm{v}}_{\perp}^T{\bm{y}}.
\end{align}
The quadrature $P_{\delta{\bm{\mu}}}$ has the property that $\mathbb{E}[P_{\delta{\bm{\mu}}}^{(l)}-P_{\delta{\bm{\mu}}}^{(p)}] = ||\delta{\bm{\mu}}||$, as may be readily verified. For isotropic noise distributions where only the mean separation determines distinguishability, $P_{\delta{\bm{\mu}}}$ then defines the only feature that need be measured for classification. Of course, noise in our situation is far from isotropic. We therefore also introduce the noise projected along $\delta{\bm{\mu}}$, 
\begin{align}
    \sigma^{2(l)}_{\delta{\bm{\mu}}} = {\bm{v}}_{\parallel}^{T}\cdot\mathbf{\Sigma}^{(l)}\cdot{\bm{v}}_{\parallel}.
    \label{eq:sigmadeltamu}
\end{align}
Note that the noise in the $P_{\delta{\bm{\mu}}}$ quadrature is exactly $\sigma^{2(l)}_{\delta{\bm{\mu}}}$. 


In Fig.~\ref{fig:ampClassifyLambda}(c), we now plot this quantity, together with the minimum and maximum noise eigenvalues, for $l=1$ and for the same parameters as Fig.~\ref{fig:ampClassifyLambda}(b). Finally, $\cmax$ and $\FD$ are plotted over the same parameter range in Fig.~\ref{fig:ampClassifyLambda}(d). Clearly, $\sigma_{\rm max}^{2(1)}$ increases following $|\chi_b|$, describing the amplification of fluctuations by the \qrc. In contrast, $\sigma_{\rm min}^{2(1)}$, which should represent squeezing, does not undergo a corresponding dip; this is because the Kerr-based \qrc{} does not generate ideal squeezed states. Regardless, we still have $\sigma_{\rm min}^{2(1)} < \sigma_{\rm vac}^2$, so the measured distributions exhibit squeezing below vacuum. The projected noise $\sigma_{\delta{\bm{\mu}}}^{2(1)}$ varies in between these maximum and minimum noise eigenvalues, with some important features that we now describe.

To illustrate the interplay of mean separation and noise, we consider three \qrcs{} defined by nonlinearity values marked (i)-(iii) in Fig.~\ref{fig:ampClassifyLambda}. As seen in Fig.~\ref{fig:ampClassifyLambda}(b), measured output distributions from \qrc{} (i) have the largest mean separation. However, the projected noise plotted in Fig.~\ref{fig:ampClassifyLambda}(c) is far from minimal. From the resulting measured distributions depicted in Fig.~\ref{fig:ampClassifyLambda}(i), it is visually clear that noise in the direction of $\delta{\bm{\mu}}$ is not minimized.

\qrc{} (ii) ends up being the most interesting operating point. While the mean separation of outputs from \qrc{} (ii) is lower than (i), the projected noise is minimized, $\sigma_{\delta{\bm{\mu}}}^{2(1)} = \sigma_{\rm min}^{2(1)}$, as seen in Fig.~\ref{fig:ampClassifyLambda}(ii). From the measured distributions, we see that $\delta{\bm{\mu}}$ is now aligned with the direction of minimum noise, defined by the noise eigenvector $\bm{v}^{(1)}_{\rm min}$. The resulting Fisher's discriminant $\FD$ and $\cmax$ are in fact larger than for \qrc{} (i), even though $||\delta{\bm{\mu}}||$ is slightly smaller.

Finally, \qrc{} (iii) is specifically chosen so that its output $||\delta{\bm{\mu}}||$ is equal to that for \qrc{} (ii); however, the projected noise is much larger. From Fig.~\ref{fig:ampClassifyLambda}(iii), it is visually clear that the measured distributions from \qrc{} (iii) are the least distinguishable of the three \qrcs{} considered, leading to the smallest value of $\FD$ and hence $\cmax$.



This analysis has several important conclusions. First, it emphasizes the necessity of accounting for quantum fluctuations for optimal classification. This is the only factor that distinguishes \qrcs{} (ii) and (iii). Secondly, we are able to clarify the role of \qrc-mediated squeezing of quantum fluctuations in classification. As seen from Fig.~\ref{fig:ampClassifyLambda}(d), the measured outputs from all three \qrcs{} exhibit distributions with squeezing below vacuum ($\sigma_{\rm min}^{2(1)} < \sigma_{\rm vac}^2$). However, this squeezing does not optimally aid classification unless the specific measured \qrc{} quadrature defined by $P_{\delta{\bm{\mu}}}$ is the one undergoing squeezing, here \qrc{} (ii). 

Finally, and most remarkably, even a simple, single-mode Kerr \qrc{} provides the ability to manipulate quantum fluctuations separately from mean values, so that the aforementioned optimal squeezing scenario can actually be engineered using suitable parameter choices. This capability is emphasized by \qrcs{} (ii) and (iii), which exhibit the same `signal' $||\delta{\bm{\mu}}||$ but very different projected noise properties, a scenario not possible in, for example, linear amplifiers. The remarkable freedom in adjusting the projected noise independently of the mean separation is demonstrated via the surface plot of $\sigma_{\rm min}^{2(1)}$ in Fig.~\ref{fig:ampClassifyLambda}(e); here the white curve, which defines operating points with fixed $||\delta{\bm{\mu}}||$, traverses through regions of both very large projected noise, as well as minimal projected noise that is sub-vacuum. Clearly, optimal performance requires operating in regimes where the projected noise can be lowered (not necessarily always exactly minimized, as $\FD$ also depends on $||\delta{\bm{\mu}}||$).




Before proceeding, we make two further observations. First, it is clear that $||\delta{\bm{\mu}}||$ (and hence $\FD$) exhibits a complex dependence on the \qrc{} nonlinearity, in contrast to the monotonic dependence alluded to by Eq.~(\ref{eq:nvkmukerr}). This is because simply varying $\Lambda$ also modifies $\CNLE$ via Eq.~(\ref{eq:cnle}) if $\amp$ is fixed, thus changing the \qrc{} operating conditions in a nontrivial fashion. To explore the performance improvement with $\Lambda$ hinted at by Eq.~(\ref{eq:nvkmukerr}) a more careful analysis is needed; this is presented in Appendix~\ref{app:ampnl}. Secondly, note that all quantities calculated using the NVK approximation agree well with their counterparts obtained using exact integration of the STEOMs, especially in qualitative terms. This highlights the utility and validity of the NVK approximation in analyzing \qrcs{} and, more generally, networks of coupled nonlinear quantum systems.

\section{\qrc{} advantage for quantum signals}
\label{sec:comp}

Thus far we have demonstrated that \qrcs{} enable processing of quantum signals in a way that is distinct from linear quantum amplifiers. However, it is not yet clear that this difference will lead to a practical advantage for quantum information processing. We now address this key question.

\begin{figure*}[t]
    \centering
    \includegraphics[scale=1.0]{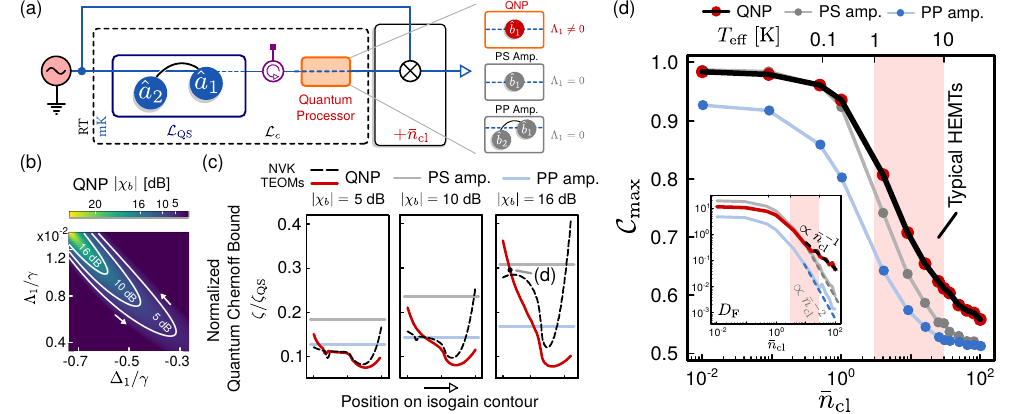}
    \caption{(a) Measurement chain for quantum state discrimination using different quantum processors: phase-sensitive (PS) or phase-preserving (PP) quantum amplifiers, or \qrcs{}. (b) \qrc{} susceptibility $|\chi_b|$ in dB, as a function of operating parameters. Contours of fixed $|\chi_b|$ are shown in white. (c) Normalized QCB $\zeta/\zeta_{\rm QS}$ for \qrc{} parameters along the three `isogain' contours in (b), using both the NVK approximation and using TEOMs. Horizontal lines show normalized QCB values at the same $|\chi_b|$ for linear PS and PP amplifiers. (d) $\cmax$ for \qrc{} readout (black), PS amplifier readout (gray), and PP amplifier readout (blue) as a function of excess classical noise power in the measurement chain, all for $|\chi_b|=$16 dB indicated in (c). Top axis calibrates the classical readout noise in terms of an effective temperature $T_{\rm eff}$. Red shaded region marks the noise added by typical HEMT amplifiers in modern cQED measurement chains. Inset: Fisher's linear discriminant $\FD$, a measure of the \textit{signal-to-noise} ratio for quantum state discrimination, for the \qrc{} and both linear amplifiers as a function of $\bar{n}_{\rm cl}$. }
    \label{fig:ampComp}
\end{figure*}

To do so, we will compare readout using the \qrc{} against standard linear quantum amplifiers for the task of quantum state discrimination. Note that both these schemes have already been discussed separately in Sec.~\ref{sec:taski} and Sec.~\ref{subsec:taskdef} respectively, albeit the latter only for phase-preserving amplification. Now, we take advantage of the sufficiently general and comprehensive description of quantum measurement chains enabled by our framework to consider the configuration in Fig.~\ref{fig:ampComp}, where a general quantum processor follows the QS to be read out. The processor can be the \qrc{}, or one of a phase-preserving (PP) or phase-sensitive (PS) linear quantum amplifier (see Appendix~\ref{app:linamp} for exact models). In this way, the individual couplings, losses, and readout rates of modes constituting the quantum processor can be held fixed from one model to the next, ensuring a direct comparison. Secondly, we now also include the effect of excess classical noise $\propto \bar{n}_{\rm cl}$, which necessitates the use of quantum processors in the first place. We can then identify what practical advantages (if any) are enabled by the \qrc{}.


We briefly note that if one considers only binary quantum state discrimination tasks, provably optimal measurement protocols are known, namely Helstrom measurements~\cite{wiseman_quantum_2009}. While a sufficiently broad definition of a \qrc{} would also incorporate such schemes, such measurements are specific to the particular states to be distinguished, and are not always straightforward to perform as they typically require non-Gaussian operations or strong constraints on coherence. We envision the \qrcs{} considered in this paper to be akin to linear quantum amplifiers in this respect: they may not necessarily be the optimal choice for a single task, but can provide general quantum processing for quantum signals for much more general tasks. Nevertheless, we will still compare the \qrc{} and linear amplifiers against bounds set by optimal discrimination schemes, to be discussed shortly.

Even for processors in the same measurement chain, one last degree of freedom remains: the operating parameters. For linear PS and PP amplifiers, it is straightforward to identify equivalent operating regimes: we can simply choose them to operate at the same gain, or equivalently the same susceptibility $|\chi_b|$. The PS amplifier has only one additional degree of freedom, namely the phase of the quadrature to be amplified. Unlike these amplifiers, the \qrc{} does not provide only linear amplification, and so its operation is not determined entirely by $|\chi_b|$, as shown earlier. Nevertheless, we choose to operate the \qrc{} at the same fixed $|\chi_b|$ as the linear amplifiers to ensure comparable operating points. These operating points are defined by `isogain' contours of $|\chi_b|$ in $\Lambda$-$\Delta_1$ space; considered examples are shown in Fig.~\ref{fig:ampComp}(b). 

\subsection{Quantum Chernoff Bound and optimal discrimination}

In assessing the relative performance of the various processors at quantum state discrimination, we first note that the upper bound on the discrimination accuracy must depend on the fundamental distinguishability of the two reduced QS quantum states $\rhou_{\rm QS}^{(1)},\rhou_{\rm QS}^{(2)}$, independent of the specific processing or measurement scheme. One measure of this distinguishability is the Quantum Chernoff Bound (QCB) $\zeta$. The QCB bounds the discrimination accuracy according to $1 - \mathcal{C}_{\rm max} \sim \exp ( -N \zeta )$ where access to $N$ identical copies of each state is assumed, all of which may be measured~\cite{audenaert_discriminating_2007, calsamiglia_quantum_2008, chernoff_measure_1952, pirandola_computable_2008, nielsen_revisiting_2022}. Clearly, the larger the value of the QCB $\zeta$, the more easily distinguishable the two quantum states are for fixed $N$. While strictly speaking a bound that holds in the asymptotic limit of many copies $N$, we use the QCB here due to the ease of its calculation for Gaussian states. We then define $\zeta_{\rm QS}$ as the QCB for the QS states $\rhou_{\rm QS}^{(1)},\rhou_{\rm QS}^{(2)}$. Similarly, when such states are processed using a downstream \qrc{}, the QCB $\zeta$ can also be computed to quantify the distinguishability of the \qrc{} states $\rhou_{\rm \qrc{}}^{(1)},\rhou_{\rm \qrc{}}^{(2)}$, where $\rhou_{\rm \qrc{}}^{(l)} = {\rm tr}_{\rm QS}\!\left[\rhou^{(l)}\right]$. We will consider Task~I in what follows, so that the QCB is computed for the single mode to be read out (see Appendix~\ref{app:qcb} for full details). 

We plot the ratio of these two QCB values, $\zeta/\zeta_{\rm QS}$, for three different $|\chi_b|$ values for the different quantum processors in Fig.~\ref{fig:ampComp}(c). For the PS amplifier the optimal QCB is chosen by varying the PS amplification phase. For the \qrc{}, the QCB is plotted along the isogain contours in Fig.~\ref{fig:ampComp}(b), using both TEOMs and within the NVK approximation. Immediately, we note that $\zeta/\zeta_{\rm QS} < 1$: the use of a quantum processor will generally introduce loss channels that are independent of the states to be distinguished and therefore hinder classification. We also note that for the present task, PS amplification enables a larger QCB than PP amplification: even though the PS amplifier can only amplify information in a single quadrature of the quantum input signal, it does so without adding any noise. Interestingly, we see that the \qrc{} can approach and at certain operating points even exceed the QCB obtainable using linear quantum amplifiers in the same measurement chain. We emphasize that this was not \textit{a priori} guaranteed in the presence of the added noise of the \qrc{}; hence this observation is promising for the use of \qrcs{} for quantum state discrimination. 


\subsection{\qrc{} robustness to readout noise}

Given the reduction in the QCB when using a downstream quantum processor, we reiterate why such processors are necessary in the first place. Excess classical noise $\bar{n}_{\rm cl}$ in the measurement chain can swamp weak quantum signals, such as those emanating from the QS directly, as they are extracted to the classical observer. The QCB is a discrimination bound that assumes optimal measurements, and therefore does not account for the limitations on readout imposed by $\bar{n}_{\rm cl}$. In practical measurements subject to readout noise, quantum processors such as linear amplifiers then provide the pre-amplification necessary to overcome this noise and enable visibility of weak quantum signals. 

To quantify the impact of this practical constraint, we instead compute the classification accuracy for the different quantum processors under heterodyne readout and in the presence of $\bar{n}_{\rm cl}$. We consider the operating point labelled (d) in Fig.~\ref{fig:ampComp}(c), where $\zeta/\zeta_{\rm QS}$ agrees well between the NVK and TEOMs. We also deliberately consider an operating point where the PS amplifier enables a larger QCB $\zeta$ than the \qrc{}; as we will show, the particular \qrc{} advantage we wish to highlight prevails in spite of this. Finally, we define readout features obtained using linear ($\frcl{\cdot}$) and nonlinear ($\frcnl{\cdot}$) post-processing of measurement records over multiple shots, ${\bm{y}} = \frac{1}{S}\sum_s^S(I_1,Q_1,I_1^2,Q_1^2,I_1Q_1)$. This enables a fair comparison of the \qrc{} against linear amplifiers, as the latter can only perform this discrimination task when nonlinear post-processing is allowed. 

We plot the obtained $\mathcal{C}_{\rm max}$ using the different quantum processors as a function of $\bar{n}_{\rm cl}$ in Fig.~\ref{fig:ampComp}(d). For low enough $\bar{n}_{\rm cl}$, both \qrc{} and PS readout exhibit similar performance while the PP amplifier is worse, consistent with the obtained QCB values. With increasing $\bar{n}_{\rm cl}$, the performance of all readout schemes degrades as expected. However \qrc{} readout is clearly the most robust, exhibiting the smallest reduction. The red shaded region marks typical $\bar{n}_{\rm cl}$ values for cQED measurement chains, attributed primarily to noisy HEMT amplifiers (for calibration and mapping to equivalent noise temperatures as shown, see SI Sec.~\ref{app:noise}). \qrc{} readout therefore provides enhanced robustness against experimentally relevant levels of excess classical noise.

The observed advantage stems from a simple principle: \qrcs{} enable nonlinear processing of quantum signals from the QS in the same quantum environment where they are generated, and - crucially - \textit{prior} to corruption by excess readout noise $\bar{n}_{\rm cl}$. In contrast, linear quantum amplifiers provide only \textit{linear} gain to quantum signals, so that nonlinear processing of the extracted noisy signals is still required. By reducing or even eliminating the need for nonlinear post-processing, \qrcs{} therefore purvey a fundamentally different scaling of the signal-to-noise ratio with readout noise $\bar{n}_{\rm cl}$. As shown in the inset of Fig.~\ref{fig:ampComp}(d), Fisher's discriminant (which defines the signal-to-noise ratio for binary discrimination) degrades only linearly with $\bar{n}_{\rm cl}$ for the \qrc{}. This is markedly different to linear amplifiers, in which case $\FD$ degrades \textit{quadratically} with $\bar{n}_{\rm cl}$, as the post-processing required to compute second-order nonlinear features necessary for this task must also amplify the added readout noise. Most importantly, our results emphasize that this \qrc{} processing advantage does not demand large-scale or highly-coherent quantum systems as \qrcs{}: few- and even single-mode nonlinear quantum systems suffice, deployed in lossy quantum measurement chains with minimal additional components. 

We conclude with two practical observations. First, while other choices of readout features $\bm{y}$ (e.g. for different temporal filters $\mathcal{K}(\tau)$) could influence the quantitative performance of either type of processor, the \textit{qualitative} difference in $\FD$ scaling with $\bar{n}_{\rm cl}$, which relies on the \textit{in situ} nonlinear processing paradigm enabled by \qrcs{}, will remain. 

Secondly, in the interest of a fair comparison, here we have considered \qrcs{} as a direct replacement for linear quantum amplifiers, demanding \qrcs{} to offer both transduction \textit{and} amplification to overcome readout noise. In this configuration, we are thus not able to take full advantage of the \qrc{}'s ability to engineer minimal projected noise $\sigma^2_{\delta\bm{\mu}} = \sigma^2_{\rm min}$, as this minimal noise can be swamped by the following readout noise $\bar{n}_{\rm cl} \gg \sigma^2_{\rm min}$. Using the \qrc{} to control quantum fluctuations is still important, but to instead operate in a regime where $\sigma^2_{\delta\bm{\mu}} \gtrsim \bar{n}_{\rm cl}$ (to the extent possible for  a given $|\chi_b|$). Alternatively, \qrcs{} can be deployed in conjunction with linear amplifiers. Then, a \qrc{} would first be used to engineer a transduced signal quadrature $P_{\delta\bm{\mu}}$ with minimal projected noise $\sigma^2_{\rm min}$. Next, a phase-sensitive amplifier would provide noise-free gain $\sqrt{\mathcal{G}_{\rm PS}}$ to this quadrature, yielding $||\delta\bm{\mu}||^2 = \mathcal{G}_{\rm PS}(\mathbb{E}[P^{(l)}_{\delta\bm{\mu}}-P^{(p)}_{\delta\bm{\mu}}])^2$ and $\sigma^2_{\delta\bm{\mu}} = \mathcal{G}_{\rm PS}\sigma^2_{\rm min} + (\bar{n}_{\rm cl}+1)\sigma_{\rm vac}^2$. Provided $\mathcal{G}_{\rm PS}$ is large enough to satisfy $\mathcal{G}_{\rm PS}\sigma^2_{\rm min} \gg \bar{n}_{\rm cl}$, the output $\FD$ becomes $\FD \simeq (\mathbb{E}[P^{(l)}_{\delta\bm{\mu}}-P^{(p)}_{\delta\bm{\mu}}])^2/\sigma^2_{\rm min}$, taking advantage of both transduction using \qrcs{} (which determines the `signal'), and their ability to engineer non-classical noise distributions (which determines the `noise').

\section{Multi-mode \qrcs{}}
\label{sec:taskii}

In this final section, we analyze the use of a \qrc{} to perform Task~II: distinguishing a pair of two mode squeezed states. Here, we see that the need for multi-mode \qrcs{} naturally arises, which then allows us to identify the role of entanglement in quantum information processing tasks using \qrcs{}.

\subsection{Role of coupling}

As shown in Sec.~\ref{subsec:taskdef}, successfully performing Task~II under direct QS readout requires monitoring of both QS modes, and computation of cross-correlations. For QS modes with disparate frequencies $|\nu_1 - \nu_2| \gg \kappa$ (as required to realize non-degenerate squeezed states), quantum signals from the QS modes at $\omega_{dn} \sim \nu_n$ are also at widely separated frequencies. A single \qrc{} mode with similar bandwidth $\gamma_1+\Gamma_1 \simeq \kappa_1+\Gamma_1$, will thus not exhibit a strong response to two signals with such a wide spectral separation. Note that increasing the \qrc{} bandwidth relative to the QS requires $\gamma_1 \gg \Gamma_1$; however, this undercouples the QS to the \qrc{}, reducing the influence of input QS signals and hence the \qrc{} response (see Eq.~(\ref{eq:cnle})). 

\begin{figure}[t]
    \centering
    \includegraphics[scale=1.0]{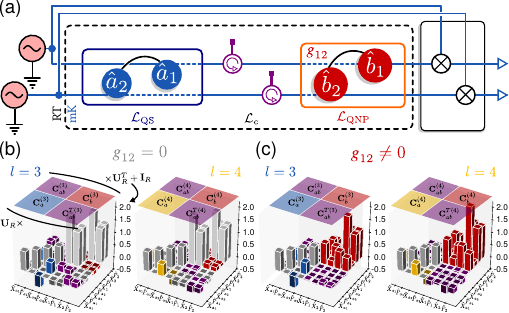}
    \caption{(a) Measurement chain using a two-mode \qrc{} for Task~II: distinguishing two-mode squeezed states $\rhou_{\rm QS}^{(3)},\rhou_{\rm QS}^{(4)}$ of a QS. (b) Covariance matrices of the joint quantum state of the chain when the QS is in state $\rhou_{\rm QS}^{(3)}$ (left) and $\rhou_{\rm QS}^{(4)}$ (right). Only matrix elements that are distinct between the two cases are shown in color. (c) Same as (b) but now for $\varg_{12}\neq 0$.}
    \label{fig:twoNodeQRC}
\end{figure}

The processing of quantum signals from multiple non-degenerate quantum modes is therefore a situation where the need for a multi-mode \qrc{} naturally emerges. We consider the measurement chain shown in Fig.~\ref{fig:twoNodeQRC}(a) incorporating a $K=2$ mode \qrc{}, and where each QS mode is monitored by a \qrc{} mode via the non-reciprocal interaction defined by $\mathbf{\Gamma}$. A tunable (e.g. parametric) coupling $\varg_{12}$ between the \qrc{} modes allows us to clearly identify the role played by the multi-mode nature of the \qrc{}.


The full measurement chain covariance matrices are shown in Fig.~\ref{fig:twoNodeQRC}(b) for the two QS states $l=3,4$ constituting Task~II, first for $\varg_{12} = 0$. Only matrix elements that are different between the two states are shown in color, to aid visibility. In particular, when $\varg_{12} = 0$, the local \qrc{} mode covariances are independent of $l$; this is because the two QS states also have identical local covariances. Instead, even if the \qrc{} modes are uncoupled, non-trivial \textit{correlations} that vary with the QS state are established between the \qrc{} modes. While this may appear surprising at first glance, it is merely a consequence of the fact that the two uncoupled \qrc{} modes are nevertheless driven by a correlated quantum input signal originating from the QS, which means the two \qrc{} modes' states must also be correlated. However, it is only when the coupling is turned on, Fig.~\ref{fig:twoNodeQRC}(c), that these cross-correlations can be transduced to differences in the local \qrc{} mode covariances. We will see how, in contrast to Task~I, this means that now both the nonlinearity and coupling of \qrc{} modes is required for success at Task~II.



\begin{figure*}[t]
    \centering
    \includegraphics[scale=1.0]{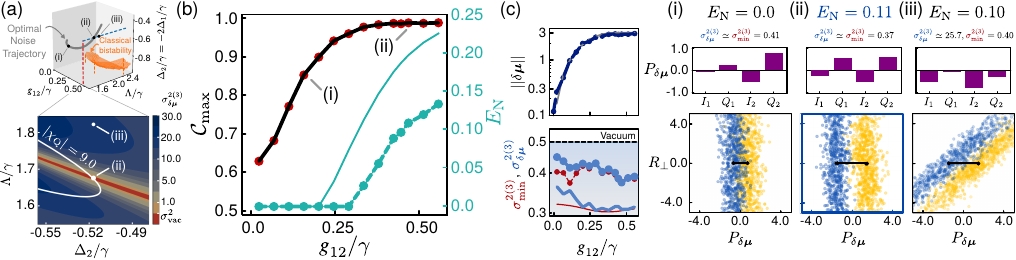}
    \caption{Performance metrics for Task~II using a $K=2$ mode \qrc{}. (a) Top panel: \qrc{} phase diagram in $(\varg_{12},\Lambda,\Delta_2)/\gamma$ parameter space, where $\Delta_2 = -2\Delta_1$. Orange region marks the bistability region of Kerr dimer. Gray curve marks the optimal noise trajectory of parameters for which $|\chi_b|=9.0$ \textit{and} projected noise is minimized. Lower panel: Cross-section at fixed coupling, showing the contour of fixed $|\chi_b|=9.0$, and surface plot of projected noise. (b) $\cmax$ (black, left-hand axis) and measured output logarithmic negativity $E_{\rm N}$ (green, right-hand axis) characterizing entanglement between \qrc{} mode outputs, as a function of \qrc{} mode coupling ${\varg}_{12}/\gamma$. (c) $||\delta{\bm{\mu}}||$ and $\sigma^{2(3)}_{\delta{\bm{\mu}}}$ as a function of \qrc{} mode coupling along the optimal noise trajectory, using both the NVK approximation and STEOMs. (i)-(iii) show measured distributions in $(P_{\delta{\bm{\mu}}},R_{\perp})$ subspace for \qrcs{} with parameters labelled (i)-(iii) in (a). Top bar plots show the respective quadrature composition of $P_{\delta{\bm{\mu}}}$. }
    \label{fig:ampClassifyG}
\end{figure*}


Unfortunately, with increasing \qrc{} size, the number of tunable parameters and the complexity of their interplay grows, making it \textit{a priori} difficult to isolate the importance of a particular parameter, such as the coupling $\varg_{12}$. Here, our analysis of \qrc{} optimization for Task~I in Sec.~\ref{subsec:squeezing} proves its generality, and hence its value. For a given $\amp$, two principles were found: the importance of $|\chi_b|$ (the largest eigenvalue of the \qrc{} Jacobian) in enhancing the response of the \qrc{} to any input, and the control of quantum fluctuations to minimize projected noise $\sigma^{2(l)}_{\delta{\bm{\mu}}}$. Our strategy to isolate the role of coupling is to therefore fix other parameters such that $|\chi_b|$ can be held constant and $\sigma^{2(l)}_{\delta{\bm{\mu}}}$ is simultaneously minimized.

The two-mode \qrc{} model we consider is a realization of the Kerr or Bose-Hubbard dimer that has been analyzed in prior work, but only under coherent driving~\cite{sarchi_coherent_2008}, not with both modes driven by quantum signals from a QS upstream as illustrated schematically in Fig.~\ref{fig:ampClassifyG}(a). By analyzing the Jacobian of this two-mode \qrc{}, we identify regions of parameter space where the susceptibility $|\chi_b|$ grows and diverges, namely near the classical bistability of the Kerr dimer, marked as the orange region in the phase diagram of Fig.~\ref{fig:ampClassifyG}(a). Then, using the NVK approximation, we identify a trajectory through this parameter space near the classical bistability - labelled the \textit{optimal noise trajectory} (gray curve) - where $|\chi_b| = 9.0$ at every coupling strength, and simultaneously the projected noise is minimized, $\sigma^{2(3)}_{\delta{\bm{\mu}}} \simeq \sigma^{2(3)}_{\rm min}$. An example of how this choice constraints parameters is shown for point (ii) on this trajectory in the lower panel of Fig.~\ref{fig:ampClassifyG}(a).

We again consider single-shot readout features obtained under linear processing only, now from \textit{both} \qrc{} modes, ${\bm{y}} = \frcl{{\bm{x}}} = (I_1,Q_1,I_2,Q_2)$. Performing Task~II for parameters along this trajectory, we obtain $\cmax$ as a function of coupling in Fig.~\ref{fig:ampClassifyG}(c), which follows a smooth curve. We also plot the corresponding mean separation $||\delta{\bm{\mu}}||$ and projected noise in Fig.~\ref{fig:ampClassifyG}(c). For $\varg_{12} \to 0$, we see that $||\delta{\bm{\mu}}||\to 0$, even though $\mathbf{C}_b^{(3)}\neq \mathbf{C}_b^{(4)}$ as seen in Fig.~\ref{fig:twoNodeQRC}(b) and the nonlinearity is nonzero. This is because the Hessian tensor in Eq.~(\ref{eq:nvkmukerr}) is still local, as the nonlinearity is on-site, so that $\delta{\bm{\mu}}$ is only sensitive to local \qrc{} mode covariances which are independent of $l$ when $\varg_{12} = 0$. This observation is the result of a much simpler fact: since each \qrc{} mode is coupled to only one QS mode, and the difference between QS states $l=3,4$ is present in correlations of \textit{different} QS modes, it is necessary for the two \qrc{} modes to `communicate' (hence, be coupled) to be able to distinguish the inputs they receive. If this communication is not performed \textit{in situ} using $\varg_{12}$, it will have to be performed in post-processing, by computing correlations of measured \qrc{} quadratures for distinct \qrc{} modes explicitly, which constitutes a nonlinear processing step. Turning on $\varg_{12}$ allows these correlations to be computed via the \qrc{} dynamics, allowing the difference in QS states to translate to a nonzero $||\delta{\bm{\mu}}||$ of measured \qrc{} quadratures, enabling $\cmax \to 1$.

\subsection{Engineering entanglement for classification}

We are now well-placed to explore the role of entanglement in the processing of quantum signals using a \qrc{}. Importantly, we are interested in output field entanglement, which requires non-classical correlations between \textit{measured} \qrc{} quadratures. For the two-mode \qrc{}, the 4-by-4 measured covariance matrix can be expressed in the block form
\begin{align}
    \bm{\Sigma} = 
    \begin{pmatrix}
        \bm{\Sigma}_{11} & \bm{\Sigma}_{12} \\
        \bm{\Sigma}_{12}^T & \bm{\Sigma}_{22}
    \end{pmatrix}
    \label{eq:sigma2}
\end{align}
where $\mathbf{\Sigma}_{jk}$ is the contribution from covariances between \qrc{} modes $j$ and $k$ (see Appendix~\ref{app:io}), and we have suppressed the superscript $(l)$ indicating the corresponding QS state. The metric we then use to quantify the degree of entanglement in \qrc{} outputs is the \textit{logarithmic negativity} $E_{\rm N}$~\cite{vidal_computable_2002}, a standard entanglement monotone~\cite{plenio_logarithmic_2005}. In terms of Eq.~(\ref{eq:sigma2}), $E_{\rm N}$ can be defined as
\begin{align}
    E_{\rm N} &= {\rm max}\{0,-\ln 2\nu^-\},~\nu^- = \sqrt{ \frac{d-\sqrt{d^2-4\det\bm{\Sigma}}}{2} }, \nonumber \\
    d &= \det\bm{\Sigma}_{11}+\det\bm{\Sigma}_{22}-2\det\bm{\Sigma}_{12}
\end{align}
In  Fig.~\ref{fig:ampClassifyG}(b), we then plot $E_{\rm N}$ as a function of coupling $\varg_{12}$ along the optimal noise trajectory, computed using the NVK approximation as well as using simulated covariance matrices via the STEOMs. At first glance, the role of entanglement appears straightforward: increasing coupling coincides with an increase in the degree of entanglement between measured \qrc{} quadratures, and leads to improved classification performance.

However, this observation does not indicate whether entanglement is \textit{always} useful, or if it provides an advantage that goes beyond non-classical correlations of only single-mode variables. The context of binary quantum state discrimination allows us to probe these questions directly. We once again consider the reduced quadrature description introduced in Sec.~\ref{subsec:squeezing}. In Fig.~\ref{fig:ampClassifyG}(d), we then plot measured \qrc{} distributions in the projected space spanned by $(P_{\delta{\bm{\mu}}},R_{\perp})$ for three different operating parameters (i)-(iii), labelled in Fig.~\ref{fig:ampClassifyG}(a). 

By construction, the separation of distribution means for all \qrcs{} (i)-(iii) is entirely confined to the $P_{\delta{\bm{\mu}}}$ quadrature. \qrcs{} (i) and (ii) fall along the optimal noise trajectory and are thus engineered to minimize noise in the $P_{\delta{\bm{\mu}}}$ quadrature, as is clearly visible in the distributions. Fig.~\ref{fig:ampClassifyG}(d). In the top panel of Fig.~\ref{fig:ampClassifyG}(d), we also show the content of the $P_{\delta{\bm{\mu}}}$ quadrature. For (i), $P_{\delta{\bm{\mu}}}$ is predominantly comprised of quadratures of only a single \qrc{} mode (here $k=2$). Hence, the sub-vacuum noise in $P_{\delta{\bm{\mu}}}$ is the result of quantum correlations between outputs of only a single \qrc{} mode, indicative of single-mode squeezing. In this weak coupling regime, entanglement is not necessary to obtain sub-vacuum noise; particular, the measured \qrc{} outputs do not exhibit entanglement as $E_{\rm N}=0.0$.

However, Task~II benefits from increasing coupling between the \qrc{} modes, as observed previously; this not only increases the mean separation $||\delta{\bm{\mu}}||$, but also generates output entanglement, as seen for \qrc{} (ii) where $E_{\rm N} = 0.11$. Now $P_{\delta{\bm{\mu}}}$ is comprised of quadratures of \textit{distinct} \qrc{} modes, and thus is a non-local quadrature. Sub-vacuum noise in $P_{\delta{\bm{\mu}}}$ therefore must arise due to non-classical correlations amongst these non-local quantum modes, namely entanglement. Here, entanglement of measured quadratures directly improves classification performance by ensuring that non-local quadratures that carry useful information (encoded in $||\delta{\bm{\mu}}||$) are also non-classically correlated, such that their fluctuations are reduced below the vacuum limit.

Perhaps most interestingly, just the presence of output field entanglement is not always guaranteed to be beneficial for classification. A simple counterexample is \qrc{} (iii), which demonstrates a nonzero logarithmic negativity $E_{\rm N} = 0.10$ like \qrc{} (ii), but is not on the optimal noise trajectory as seen from the lower panel of Fig.~\ref{fig:ampClassifyG}(a). $P_{\delta{\bm{\mu}}}$ is still a non-local quadrature, but its noise is clearly not minimal, and is in fact above vacuum. The presence of output field entanglement means that there is some set of non-local quadratures that exhibit non-classical correlations that lead to sub-vacuum noise; however, these quadratures are not always guaranteed to carry useful information encoded in $||\delta{\bm{\mu}}||$, and hence such non-classical correlations may not always be useful for the task at hand. 

Crucially, the use of nonlinear \qrcs{} provides the ability to control quantum fluctuations, such that the situation in (ii) can be engineered, and non-classical correlations can be manipulated for computational benefit. For binary classification, this entails finding operating points where the unique quadrature with maximal signal $P_{\delta{\bm{\mu}}}$ can simultaneously exhibit minimal, non-classical (e.g. sub-vacuum) fluctuations. The ability to engineer this unique quadrature also opens up avenues to only amplify and measure said quadrature using noiseless phase-sensitive amplification, instead of phase-preserving amplification of all $2K$ quadratures for a $K$-mode \qrc{}, which necessarily adds noise.

\section{Conclusions and Outlook}

Nonlinearity is an essential component of signal processing, playing a key role from digital processing to neural circuits. However, the role of general, fundamentally stochastic quantum nonlinear devices in processing quantum signals - beneficial or otherwise - is much less explored. In this paper, we have addressed this limitation by identifying key general principles of quantum information processing enabled by a broad class of nonlinear bosonic quantum systems, which we refer to as \qrcs{}.

Our two main results, which hold beyond the quantum state discrimination tasks we have chosen to demonstrate them, and which make explicit use of \qrcs{} being both quantum \textit{and} nonlinear, can be simply stated. We show that \qrcs{} can be efficient information transducers (cf. Eq.~(\ref{eq:nvkmu})): by processing quantum signals \textit{in situ}, \qrcs{} can render nonlinear properties of quantum signals such as correlation functions accessible via linear readout schemes like heterodyne monitoring. Secondly, by harnessing a nonlinear input-output map, \qrcs{} can coherently manipulate quantum fluctuations in a manner unavailable to linear amplifiers that are linearly coupled to a quantum signal source: without suppressing the transduced signal magnitude, \qrcs{} can modify its output noise properties. This requires \qrcs{} to control not only quantum fluctuations emanating from the quantum system the \qrc{} is monitoring, but also those resulting from the \qrc{} operation itself, referred to as `added noise' for linear amplifiers (cf. Eq.~(\ref{eq:nvksigma})). Remarkably, both these capabilities can be accessed using small-scale \qrcs{}, making them relevant for implementations in current experiments. In fact, by analyzing realistic measurement chains, we show that even single-mode \qrcs{} can provide robustness against classical readout noise. 

On the other hand, the key theoretical tools these results are built on - the analytic NVK approximation of nonlinear measurement chains and the numerical STEOMs framework to efficiently simulate their conditional dynamics, both verified against exact master equation methods - provide the means to analyze quantum information processing using very general, arbitrarily-multimode nonlinear bosonic quantum systems. As such they can be used to study the processing of signals from many-body quantum systems and non-classical correlations or entanglement across several modes, as we study in Task~II. Furthermore, they can enable the exploration of more general paradigms of quantum information processing. One example pointed out in the conclusion of Sec.~\ref{sec:comp} is the use of measurement chains employing both \qrcs{} and phase-\textit{sensitive} amplifiers. Related paradigms enabled by our framework include the possibility of entangling the \qrc{} with the QS, via for example non-reciprocal entangling operations~\cite{wang_quantum_2023, orr_high-purity_2023}. 


Taking an even broader view, our work has direct applications to general information processing and computation paradigms such as quantum machine learning and quantum sensing. Nonlinearity is considered essential to the expressive capacity of physical neural networks including quantum systems~\cite{markovic_reservoir_2019, angelatos_reservoir_2021, govia_quantum_2021, bravo_quantum_2022, mujal_time-series_2023, hu_overcoming_2024, senanian_microwave_2024}. However, several successful bosonic quantum machine learning platforms are still linear~\cite{nokkala_gaussian_2021, nokkala_high-performance_2021, garcia-beni_scalable_2023, dudas_quantum_2023}, instead enabling nonlinear processing by careful use of nonlinear input encoding schemes. Our work provides both tools and possible directions to explore the utility of multi-mode quantum nonlinear devices for learning applications; distinct from other approaches, our framework also describes learning on \textit{quantum} inputs. In particular, it is expected that genuinely quantum advantages in such applications must make use of decidedly quantum properties, such as squeezing and entanglement. \qrcs{} harness nonlinearity to control quantum fluctuations such that non-classical correlations appear in desired observables only. Furthermore, we elucidate general principles that can enable \qrcs{} to operate in regimes to enable such processing. The resulting quantum mechanism to harness quantum correlations for the enhancement of classification accuracy could prove useful in extracting quantum advantages for quantum machine learning and quantum sensing.

Our work also invites exploration of more complex \qrcs{} beyond the weakly-nonlinear regime. Firstly, the scaling advantage with $\bar{n}_{\rm cl}$ can be expected to be even more significant with increasing complexity of the required nonlinear computation, such as the calculation of higher-order or many-body correlations~\cite{da_silva_schemes_2010}. Such computations will demand the analysis of \qrcs{} with higher-order nonlinearities, or operation in increasingly non-Gaussian regimes. Secondly, for the specific case of quantum state discrimination, we have shown that \qrcs{} can approach and even exceed the optimal discrimination bound, quantified here by the Quantum Chernoff bound (QCB), achievable using linear quantum processors. However, a study of the maximum attainable QCB limits using \qrcs{} requires going beyond the NVK approximation, and could quantify the ultimate constraints on quantum information processing using nonlinear systems. Both these directions are natural extensions that we leave for future work.

Finally, we return full circle to our original motivation: by extracting more information from the quantum domain, we hope \qrcs{} can ultimately improve our ability to control quantum systems. By enabling the efficient simulation of measurement-conditioned dynamics of measurement chains including \qrcs{}, our STEOMs framework provides the necessary first step in the study of quantum feedback and control using \qrcs{}. Such control is necessary for important quantum information processing tasks such as error correction~\cite{ahn_continuous_2002,gregoratti_quantum_2004, cramer_repeated_2016}, either via continuous monitoring and feedback~\cite{doherty_feedback_1999, minev_catch_2019} or autonomous protocols via coherent quantum feedback~\cite{lloyd_coherent_2000}.


\begin{acknowledgements}
We would like to thank Dan Gauthier, Luke Govia, Peter McMahon, Ioan Pop, Graham Rowlands, Guilhem Ribeill, Tatsuhiro Onodera, Logan Wright, Ryan Kaufman, Boris Mesits, Shyam Shankar, Florian Marquardt, Danijela Markovi\'c, Ryotatsu Yanagimoto, and Sridhar Prabhu for useful discussions. This work is supported by AFOSR under Grant No. FA9550-20-1-0177, AFOSR-MURI under Grant No. FA9550-22-1-0203, and the Army Research Office under Grant No. W911NF18-1-0144. Simulations in this paper were performed using the Princeton Research Computing resources at Princeton University, which is a consortium of groups led by the Princeton Institute for Computational Science and Engineering (PICSciE) and Office of Information Technology's Research Computing.
\end{acknowledgements}

\begin{center}
    \rule{30mm}{1pt}
\end{center}

\appendix

\section{Measurement chain description and table of symbols}
\label{app:measChain}

In this Appendix section we expand on details of the measurement chain considered described by Eq.~(\ref{eq:sme}) that were omitted for brevity in the main text.

First, we define the interaction between the QS and \qrc{} described by $\Lc$. We engineer this interaction to be non-reciprocal by balancing a coherent and a dissipative hopping interaction with an appropriately-chosen phase~\cite{metelmann_nonreciprocal_2015},
\begin{align}
    \mathcal{L}_c\hat{\rho} = -i\!\left[\frac{i}{2}\sum_m\Gamma_{m}\hat{a}_m^{\dagger}\hat{b}_m + h.c. ,\hat{\rho}\right]\!+\! \sum_m\Gamma_{m}\mathcal{D}[\hat{a}_m + \hat{b}_m]\hat{\rho}.
    \label{eq:couplingCirc}
\end{align}
Some features of this interaction are of note. Firstly, although the interaction has a dissipative component, it defines a coherent \textit{link}, such that it allows for a non-separable joint quantum state of the QS and \qrc{}. Secondly, for simplicity we require that one mode of the QS couples to at most one mode of the \qrc{}, with strength $\Gamma_m$ (although the coupling can also vanish if $\Gamma_m = 0$ for a given $\hat{a}_m$). Finally, we note that Eq.~(\ref{eq:couplingCirc}) describes a standard circulator~\cite{metelmann_nonreciprocal_2015}, and is therefore realized as a matter of course in cQED experiments.

For practical implementations in the cQED architecture, we consider for the \qrc{} a model of $K$ coupled Kerr nonlinear modes $\hat{b}_k, k\in[1,K]$, furnished by capacitively-shunted Josephson junctions~\cite{squidHandBook}. The nonlinear modes have frequencies $\{\omega_k\}$ and nonlinearity strengths $\{\Lambda_k\}$, with linear coupling $\{\varg_{jk}\}$ between modes $i$ and $j$. Then, in an appropriate interaction picture at frequencies $\{\omega_{dk}\}$ (close to $\{\omega_k\}$ respectively), the linear \qrc{} Hamiltonian takes the form
\begin{align}
    \hat{\mathcal{H}}_{\rm \qrc{}} = -\!\sum_k\! \Delta_k \bkd{k}\bk{k} + \sum_{jk}\varg_{jk}(\bk{j}\bkd{k} + \bkd{j}\bk{k}), 
\end{align}
with detunings $\Delta_k = \omega_{dk}-\omega_k$, while the nonlinear component of the \qrc{} Hamiltonian is given by
\begin{align}
    \hat{\mathcal{N}}_{\rm \qrc} = - \sum_k \frac{\Lambda_k}{2}\bkd{k}\bkd{k}\bk{k}\bk{k}.
\end{align}
For simplicity, we assume $\Lambda_k \equiv \Lambda~\forall~k$, although we emphasize this is not necessary for \qrcs{}.


\newcolumntype{Y}{>{\centering\arraybackslash}X}
\begin{table}[h]
    \centering
    \caption{Definition of measurement chain parameters. Superscripts $(l)$ index parameters defining different quantum states for classification tasks.}
    \begin{tabularx}{246pt}{lY}
    \hline
    \hline 
    \multicolumn{2}{l}{ \textbf{Quantum System (QS)} }\\
    \hline
    $\eta_m^{(l)}$ & Drive strength on $\hat{a}_m$  \\
    $\omega_{dm}$ & Drive frequency on $\hat{a}_m$ \\
    $G_m^{(l)}$   & Single-mode squeezing strength on $\hat{a}_m$ \\
    $\phi_m^{(l)}$ & Single-mode squeezing phase \\
    $G_{nm}^{(l)}$ & Two-mode squeezing strength \\
    $\phi_{nm}^{(l)}$ & Two-mode squeezing phase \\
    $\kappa_m$ & Unmonitored loss for $\hat{a}_m$ \\
    $\kappa$ & Total loss $\kappa_m+\Gamma_m$, equal for all $m$ \\
    \hline
    \multicolumn{2}{l}{ \textbf{Non-reciprocal coupling} } \\
    \hline
    $\Gamma_m$ & Coupling strength between modes $\hat{a}_m$ and $\hat{b}_m$ \\
    \hline
    \multicolumn{2}{l}{ \textbf{Quantum Nonlinear Processor (\qrc)} } \\
    \hline
    $\Lambda$ & Kerr nonlinearity strength, equal for all $k$ \\
    $\Delta_k$ & Detuning $\omega_{dk}-\omega_k$ \\
    $\varg_{jk}$ & Linear coupling between modes $\hat{b}_j$ and $\hat{b}_k$ \\
    $\gammahet$ & Monitoring rate for $\hat{b}_k$, equal for all $k$ \\
    $\gamma$ & Total loss $\gammahet+\Gamma_k$, equal for all $k$ \\
    \hline
    \multicolumn{2}{l}{ \textbf{Output layer} } \\
    \hline
    $\mathcal{I}_k(t),\mathcal{Q}_k(t)$ & Raw heterodyne records for \qrc{} mode $\hat{b}_k$ \\
    $I_k,Q_k$ & Filtered records for \qrc{} mode $\hat{b}_k$ \\
    $\mathcal{T}$ & Filtering window length \\
    \hline
    \hline
    \end{tabularx}
    \label{tab:symbols}
\end{table}


Finally, the conditional dynamics of $\rhoc$ under heterodyne monitoring are governed by the stochastic measurement superoperator $\mathcal{S}[\sqrt{\gammahet}\hat{b}_k]$, given by:
\begin{align}
&\mathcal{S}[\sqrt{\gammahet}\hat{b}_k]\rhoc = \sqrt{\frac{\gammahet}{2}}\!\left( \hat{b}_k \rhoc + \rhoc \hat{b}_k^{\dagger} - \avgc{\hat{b}_k+\hat{b}_k^{\dagger}} \right)\! dW_{\mathcal{I}_k}\!(t)~ \nonumber \\
&~~~~~~+\sqrt{\frac{\gammahet}{2}}\!\left( -i\hat{b}_k \rhoc + i\rhoc \hat{b}_k^{\dagger} - \avgc{-i\hat{b}_k+i\hat{b}_k^{\dagger}} \right) \!dW_{\mathcal{Q}_k}\!(t) \nonumber \\
&~~~~~~+\gammahet \mathcal{D}[\hat{b}_k]\rhou, 
\label{eq:Smeas}
\end{align}
where $\avgc{\hat{o}}$ indicates the conditional expectation value of an arbitrary operator $\hat{o}$ with respect to the measurement-conditioned quantum state, $\avgc{\hat{o}} = {\rm tr}\{\rhoc\hat{o}\}$. Here $dW_{\mathcal{I}_k}, dW_{\mathcal{Q}_k}$ are independent Wiener increments describing measurement noise, satisfying $\avg{dW_{\mathcal{I}_k}}=\avg{dW_{\mathcal{Q}_k}}=0$, and $\avg{dW_{\mathcal{I}_k}dW_{\mathcal{Q}_{k'}}} = 0, \avg{dW_{\mathcal{I}_k}dW_{\mathcal{I}_{k'}}} = \avg{dW_{\mathcal{Q}_k}dW_{\mathcal{Q}_{k'}}} = \delta_{k,k'}dt$. The Wiener increments are therefore related to the white noise terms introduced in Eq.~(\ref{eq:IQkraw}) via $\xi_{\mathcal{I}_k} \equiv \frac{dW_{\mathcal{I}_k}}{dt}, \xi_{\mathcal{Q}_k} \equiv \frac{dW_{\mathcal{Q}_k}}{dt}$. The classical readout noise terms in Eq.~(\ref{eq:IQkraw}) are also taken to obey white noise statistics.

On the other hand, the quantum noise contributions $\xi^{\rm qm}_{\mathcal{I}_k}, \xi^{\rm qm}_{\mathcal{Q}_k}$ introduced in Eq.~(\ref{eq:IQkraw}) generally do \textit{not} obey white noise statistics. Quantum trajectories $\rhoc$ are conditioned on the heterodyne measurement record via the stochastic measurement superoperator $\mathcal{S}$. Quantum noise contributions depend on these quantum trajectories; more concretely, for example, $ \xi_{\mathcal{I}_k}^{\rm qm}\!(t) = \avgc{\hat{X}_{k}} - \avg{\hat{X}_{k}}$, and thus describes the stochastic deviation of an observable conditioned on a specific quantum trajectory from the ensemble average. Hence the quantum noise contributions inherit nontrivial correlation and noise properties from quantum trajectories. Lastly, monitoring of \qrc{} modes opens them up to linear damping at the rate $\gammahet$, described by the standard dissipative superoperator $\mathcal{D}[\hat{o}] = \hat{o}\rhou\hat{o}^{\dagger} - \frac{1}{2}\{ \hat{o}^{\dagger}\hat{o},\rhou\}$.


Definitions of all parameters characterizing the general measurement chain are summarized in Table~\ref{tab:symbols}. Specific parameter values used to generate the various figures in the main text are summarized in Table~\ref{tab:paramvals} in the SI.




\section{Truncated cumulants approach}
\label{app:cumulants}

For an arbitrary $\rhou$ describing a quantum measurement chain with $\ntot = M+K$ modes, we can define its associated \textit{characteristic function} $\chi(\vec{w},\vec{w}^*)$~\cite{carmichael_statistical_2002},
\begin{align}
    \chi(\vec{w},\vec{w}^*) = \mathrm{tr} \left\{ \! \rhou  \prod_{j=1}^{\ntot} e^{iw_j^{\ast} \hat{o}_j^{\dagger}} \! \prod_{l=1}^{\ntot} e^{iw_{l} \hat{o}_l}\! \right\}\!  
\end{align}
where $\hat{o}_j \in \{\hat{a}_m,\hat{b}_k\}$ describes any mode of the complete quantum measurement chain, and $\vec{w} = (w_1,\ldots,w_{\ntot})$ are auxiliary variables. Then, normal-ordered cumulants of the density matrix are formally defined via the log of the characteristic function (sometimes called the \textit{generating function}),
\begin{align}
    C_{o_1^{\dagger p_1} \cdots o_{\ntot}^{\dagger p_{\ntot} } o_1^{q_1} \cdots o_{\ntot}^{q_{\ntot}}} \equiv \left.  \frac{\partial^{n_{\rm ord}} \ln \chi(\vec{w},\vec{w}^*)  }{\prod_j\! \partial (i w_j^{\ast})^{p_j} \! \prod_l\! \partial (i w_l)^{q_l} }  \right|_{ \vec{w}=\vec{w}^{\ast}=0}
\end{align}
where $n_{\rm ord} = p_1 + \ldots + p_J + q_1 + \ldots + q_J$ defines the order of cumulants. Notwithstanding the complex formal definition, cumulants can be transparently related to more familiar operator expectation values: for a general operator $\hat{o}_j$, first-order cumulants $C_{o_j} \equiv \avg{\hat{o}_j}$ are simply expectation values, while second-order cumulants $C_{o_jo_l} = \avg{\hat{o}_j\hat{o}_l} - \avg{\hat{o}_j}\avg{\hat{o}_l}$ are their covariances. Expressions for higher-order cumulants become increasingly more involved, but can be systematically obtained, as discussed in the SI~\cite{SI}.

\subsection{Numerical scheme: Stochastic Truncated Equations of Motion}

The crucial advantage of cumulants as descriptors of a quantum state is that \textit{specific} multimode quantum states admit particularly efficient representations when expressed in terms of cumulants. In particular, a quantum system in a product of coherent states is described entirely by its nonzero first-order cumulants; all cumulants with $n_{\rm ord} > 1$ vanish~(see SI~\cite{SI} for derivations). Multimode quantum states that are defined entirely by their first \textit{and} second-order cumulants admit Gaussian phase-space representations, and are thus labelled Gaussian states. States with nonzero cumulants of third or higher-order are thus by definition non-Gaussian states~\cite{boutin_effect_2017, ra_non-gaussian_2020}. 

\begin{figure}[t]
    \centering
    \includegraphics[scale=1.0]{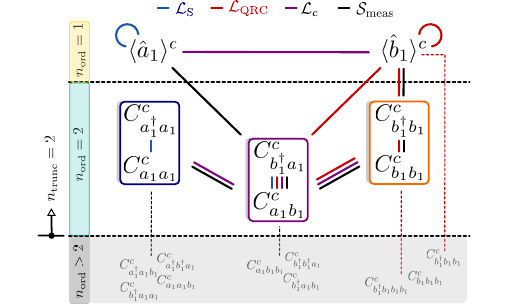}
    \caption{Schematic representation of stochastic truncated equations of motion (STEOMs), shown for simplicity for the $M=K=1$ case. Black, red, purple, and blue lines indicate dependencies between cumulants due to measurement, \qrc{} nonlinearity, system-\qrc{} coupling, and system dynamics respectively. Dashed lines indicate dependencies on cumulants of order higher than $n_{\rm trunc} = 2$ in this case, which can arise due to the nonlinearity generating non-Gaussian states.} 
    \label{fig:cumulantSchematic}
\end{figure}

Our numerical approach leverages this economy of representation by using normal-ordered cumulants as a set of dynamical variables for quantum modes of the measurement chain. For example, for specific systems such as coherently-driven linear bosonic systems initialized in a Gaussian state, cumulants of order $n_{\rm ord} \leq 2$ are sufficient, as such systems can be shown to persist in Gaussian states. If the \qrcs{} considered in this paper were also linear, our model would satisfy this requirement. However, our interest is precisely in the role of the \qrc{} nonlinearity. In this case, the cumulants describing the measurement chain and their dependencies are shown schematically in Fig.~\ref{fig:cumulantSchematic} for $M=K=1$. In particular, the nonlinearity generates states with nonzero cumulants of $n_{\rm ord}> 2$. In the most general case, the dynamical equations for these cumulants couple to all orders, forming an infinite hierarchy of equations that does not close.

To obtain a tractable numerical method, we consider an ansatz wherein the quantum state of the complete measurement is described entirely by cumulants up to a finite order $n_{\rm ord} \leq n_{\rm trunc}$; all cumulants of order $n_{\rm ord} > n_{\rm trunc}$ are thus set to zero, truncating the hierarchy and yielding a closed set of equations for the retained nonzero cumulants. In this paper, we choose $n_{\rm trunc} = 2$ for a quantum measurement chain defined entirely by its first and second-order cumulants, although the truncation can similarly be carried out at higher order. The resulting Stochastic Truncated Equations Of Motion (STEOMs) form the basis of our numerical approach in this paper.

We emphasize that the truncated cumulants ansatz has some important differences when compared to standard linearization approaches. As seen in Fig.~\ref{fig:cumulantSchematic}, note that both the nonlinearity and measurement terms couple second-order cumulants to first-order cumulants: this fact proves critical for the quantum state classification tasks we consider in this paper. 

The utility of the STEOMs is naturally determined by the validity of the truncated cumulants ansatz. Since we are specifically interested in nonlinear quantum systems, which can generate higher-order cumulants in dynamics, one must ask when such an ansatz may hold. In the SI~\cite{SI}, we carry out a detailed analysis of the regimes of validity of this framework, together with benchmarking against exact solutions and full SME integration. We find that the STEOMs provide a very good approximation of full SME integration provided the strength of nonlinearity of \qrc{} modes is weak relative to their loss rates; good agreement remains even up to $\Lambda_k/\gamma \sim 0.1$, which is around an order of magnitude larger than the strongest nonlinearity we consider in this paper.


\subsection{Semi-analytic scheme: Nonlinear van Kampen expansion based of the Fokker-Planck equation}

In addition to enabling the (S)TEOMs as a practical numerical method for multimode measurement chains, the truncated cumulants approach is also central to our main analytic tool: a description of quantum dynamics that is perturbative in the nonlinearity of the measurement chain. This analysis is enabled by the close connection between normal-ordered cumulants and the positive-${P}$ representation via the characteristic function. More precisely, the positive-${P}$ representation is simply the Fourier transform of the characteristic function~\cite{carmichael_statistical_2002}, 
\begin{align}
    \mathcal{P}(\vec{\mathcal{O}},\vec{\mathcal{O}}^{\dagger}) = \frac{1}{\pi^{2J} }\int\prod_{j=1}^{\ntot} d^2 w_j d^2 w_j^{*}~e^{-iw_j^* \mathcal{O}_j^{\dagger}}e^{-iw_j \mathcal{O}_j}\chi(\vec{w},\vec{w}^*)
\end{align}
The dynamics of the positive-${P}$ distribution follow a Fokker-Planck equation that is equivalent to Eq.~(\ref{eq:sme}), while also employing normal-ordered cumulants as its natural dynamical variable set. We use this connection to first obtain an approximate Fokker-Planck equation for dynamics of the measurement chain in powers of the \qrc{} nonlinearity. This directly allows us to obtain semi-analytic solutions for the TEOMs, Eqs.~(\ref{eq:nvkmu}),~(\ref{eq:nvksigma}), from which classification metrics such as the Fisher's discriminant can be readily evaluated. Our analysis and its results, which are used throughout the main text, are detailed in the SI, Sec.~\ref{app:lin} and~\ref{app:FI}.

\section{Scaling of \qrc{} performance with nonlinearity for fixed $\CNLE$}
\label{app:ampnl}

In Sec.~\ref{subsec:squeezing}, we explored how \qrc{} parameters can be optimized to improve classification performance. However, a plain reading of the expression for the mean separation $\delta{\bm{\mu}}$, Eq.~(\ref{eq:nvkmukerr}), would appear to suggest that an even simpler strategy may be to simply increase the \qrc{} nonlinearity strength $\Lambda$, which monotonically scales $\delta{\bm{\mu}}$. However, in practice this dependence is much more complex: for a classification task defined by a given $\amp$, varying $\Lambda$ also varies $\CNLE$ via Eq.~(\ref{eq:cnle}). This changes the operating conditions of the \qrc{} and directly influences the mean separation and noise properties.


\begin{figure}[t]
    \centering
    \includegraphics[scale=1.0]{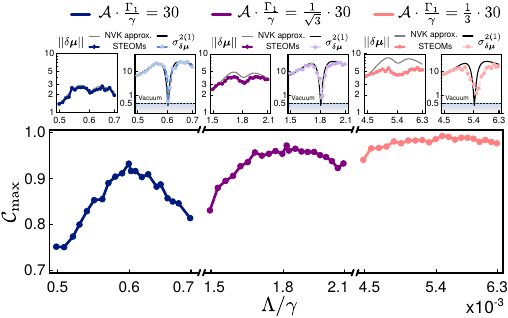}
    \caption{$\cmax$ as a function of \qrc{} nonlinearity ${\Lambda}$ for instances of Task~I with different amplitudes $\amp$ as indicated, calculated using numerical integration of the STEOMs. For each task instance, the top panel shows $||\delta{\bm{\mu}}||$ and $\sigma^{2(1)}_{\delta{\bm{\mu}}}$, calculated using the NVK approximation (solid curves), and STEOMs (dots). }
    \label{fig:ampClassifyDrive}
\end{figure}


Therefore, to observe and understand the scaling of $||\delta{\bm{\mu}}||$ with nonlinearity, we must keep $\CNLE$ fixed while varying the nonlinearity. Practically, we do so by using \qrcs{} of varying nonlinearity to perform separate instances of Task~I, characterized by different values of $\amp\cdot\frac{\Gamma_1}{\gamma}$. For each instance, the achieved $\cmax$ as a function of $\Lambda$ is plotted in Fig.~\ref{fig:ampClassifyDrive}. 

We immediately note that for decreasing $\amp\cdot\frac{\Gamma_1}{\gamma}$, the required \qrc{} nonlinearity to reach the optimal $\cmax$ increases, as required by the form of $\CNLE$. The crosses indicate three specific \qrcs{}, one for each instance of the considered task, with nonlinearity $\Lambda$ such that $\CNLE$ is the same. Comparing these \qrcs{}, we now clearly see that the optimal $\cmax$ increases with increasing $\Lambda$. In the top panel of Fig.~\ref{fig:ampClassifyDrive}, we plot the mean separation amplitude $||\delta{\bm{\mu}}||$ and the projected noise $\sigma^{2(l)}_{\delta{\bm{\mu}}}$ for $l=1$ for each considered instance of Task~I, under the NVK approximation and using integration of STEOMs. To lowest order in nonlinearity as captured by the NVK approximation, the projected noise properties remain the same for each instance. In contrast, the mean separation increases with nonlinearity, indicative of the sought-after scaling. Unchanged noise properties with increasing mean separation imply an increase in $\FD$ and hence classification accuracy. 

Note that with increasing nonlinearity $\Lambda$, the agreement between the NVK approximation and exact integration of the STEOMs is reduced. In particular, the exact mean separation is lower than the NVK approximation, while the projected noise is larger. Both effects serve to reduce $\FD$ and hence limit the improvement in $\cmax$ to below that predicted by the NVK approximation. These observations are an example of saturation that is higher-order in the \qrc{} nonlinearity and hence not captured by the NVK approximation, and are commonplace in strongly-driven nonlinear quantum systems~\cite{boutin_effect_2017}. 




\section{Models of standard linear quantum amplifiers}
\label{app:linamp}

In this section we provide the models for both phase-preserving and phase-sensitive linear quantum amplifiers deployed for readout in cQED, and which we use as benchmarks in Sec.~\ref{sec:comp} of the main text. Note that our description of the measurement chain, Eq.~(\ref{eq:sme}), is general enough to include this standard paradigm, by neglecting the nonlinear contribution to the Hamiltonian, $\hat{\mathcal{N}}_{\rm \qrc{}} \to 0$.

In particular, by taking $\hat{\mathcal{H}}_{\rm \qrc{}} \to \hat{\mathcal{H}}_{\rm PP}$ where
\begin{align}
    \hat{\mathcal{H}}_{\rm PP} =  -\sum_k \Delta_{k} \hat{b}_k^{\dagger}\hat{b}_k  + G_{\rm PP}(-i\hat{b}_1\hat{b}_2 + h.c.),
\end{align}
we are able to describe QS readout using linear phase-preserving quantum amplifiers.

For phase-sensitive (PS) amplifiers on the other hand, we can set $\hat{\mathcal{H}}_{\rm \qrc{}} \to \hat{\mathcal{H}}_{\rm PS}$, where
\begin{align}
    \hat{\mathcal{H}}_{\rm PS} = -\sum_k \Delta_{k} \hat{b}_k^{\dagger}\hat{b}_k  + G_{\rm PS}(-i\hat{b}_1^2 + h.c.).
\end{align}

\section{Quantum Chernoff Bound: additional details}
\label{app:qcb}

In this Appendix section we provide some additional details of the calculation of the Quantum Chernoff bound (QCB) used in Sec.~\ref{sec:comp} of the main text to determine the optimal discrimination bounds for quantum state discrimination tasks.

To define the QCB as used in the main text, we first introduce the quantity~\cite{audenaert_discriminating_2007, calsamiglia_quantum_2008}:
\begin{align}
    Q(\rhou^{(l)},\rhou^{(p)}) = - \min_{s \in [0,1]} \log {\rm tr} \left[ (\rhou^{(l)})^s(\rhou^{(p)})^{1-s} \right]
\end{align}
We compute the QCB using only the quantum states of the modes that are intended to be monitored. For Task~I, this is simply mode $\hat{a}_1$ for the QS. We can then define the QCB for the QS $\zeta_{\rm QS}$ as:
\begin{align}
    \zeta_{\rm QS} \equiv Q\left( {\rm tr}_{\hat{a}_2} \!\!\left[\rhou^{(1)}_{\rm QS}\right], {\rm tr}_{\hat{a}_2}\!\!\left[\rhou^{(2)}_{\rm QS}\right] \right)
    \label{appeq:qcbqs}
\end{align}
where ${\rm tr}_{\hat{o}}[\rhou]$ is used to trace out the sector corresponding to mode $\hat{o}$ from the density matrix $\rhou$. For the general quantum processor, only mode $\hat{b}_1$ is monitored, so the QCB $\zeta$ analogously becomes:
\begin{align}
    \zeta \equiv Q\left( {\rm tr}_{\hat{b}_2}\!\!\left[ \rhou^{(1)}_{\rm \qrc{}} \right], {\rm tr}_{\hat{b}_2}\!\!\left[\rhou^{(2)}_{\rm \qrc{}}\right] \right)
    \label{appeq:qcbqnp}
\end{align}
Note that the \qrc{} and the PS amplifier are both single-mode devices for Task~I, so the trace operation has no effect in those cases.

The expressions in Eqs.~(\ref{appeq:qcbqs}),~(\ref{appeq:qcbqnp}) can be computed straightforwardly if the quantum states are Gaussian. These simplified expressions are derived in Ref.~\cite{calsamiglia_quantum_2008}; we employ these results in plotting the QCB in the main text, Fig.~\ref{fig:ampComp}.

\section{Non-classical noise due to the \qrc{} alone: classifying thermal states}
\label{app:thermal}

In the main text, the considered Tasks~I and~II both require classification of states that exhibit some degree of quantum correlations, namely at least one state exhibits some degree of single or two-mode squeezing. As a result, the observed quantum (i.e. sub-vacuum) noise observed at the \qrc{} output will in general contain a contribution attributable to the quantum correlations of QS signals, and cannot be attributed to the \qrc{} alone.

In this Appendix section, we show that non-classical correlations can be generated solely by appropriate \qrc{} models, and they can be manipulated for use in classification in the same way as has been demonstrated for both Task~I and~II. To this end, we first introduce the modified Liouvillian for the QS:
\begin{align}
    \Lsys\rhou &= -i[\hat{\mathcal{H}}^{(l)}_{\rm QS},\rhou] + \nonumber \\ 
    &~~~~~~\sum_m \kappa_m (n^{{\rm th}(l)}_m+1)\mathcal{D}[\hat{a}_m]\rhou 
    + \kappa_m n^{{\rm th}(l)}_m \mathcal{D}[\hat{a}^{\dagger}_m]\rhou,
\end{align}
where in contrast to Eq.~(\ref{eq:lsysgen}) each QS mode is now coupled to a finite temperature bath. For simplicity, we further restrict ourselves to a single-mode QS being monitored by a single-mode \qrc{}, $M=K=1$; all QS parameters for mode $m=2$ are set to zero. Secondly, we turn off single-mode squeezing of the QS mode, $G_1^{(l)} = 0~\forall~l$, so that the QS states to be distinguished have no non-trivial quantum correlations. Instead, the states to be distinguished are characterized entirely by different thermal bath temperatures and hence occupation numbers, $n^{{\rm th}(l)}_m$. While not straightforward to generate in practice, this choice is ideal to test whether non-classical correlations can be generated by the \qrc{} alone, and whether they can still be manipulated by the \qrc. All parameters characterizing Task~III are summarized in Table~\ref{tab:task3}.


\begin{table}[t]
    \caption{QS parameters ($M=1$ only) defining the third classification task considered, in units of QS mode total loss rate $\kappa$.}
      \centering
    \begin{tabular}{m{0.035\columnwidth}<{\centering}|m{0.06\columnwidth}<{\centering}|m{0.080\columnwidth}<{\centering}|m{0.095\columnwidth}<{\centering}}
    \multicolumn{4}{c}{Task~III} \\
    $l$ & $G_{1}^{(l)}$ & $n^{{\rm th}(l)}_1$ & $\eta_1^{(l)}$  \\
    \hline
    \hline
    {\color{st5}$5$} & {\color{st5}$0.0$} & {\color{st5}$0.1$} & {\color{st5}$0.50$}$\amp$   \\
    {\color{st6}$6$} & {\color{st6}$0.0$} & {\color{st6}$0.8$} & {\color{st6}$0.50$}$\amp$  \\
    \end{tabular}
    \label{tab:task3}
\end{table}



\begin{figure}[t]
    \centering
    \includegraphics[scale=1.0]{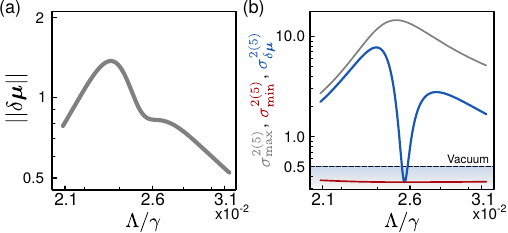}
    \caption{\qrc{} performance metrics for Task~III, classification of thermal states, within the NVK approximation. (a) Mean separation $||\delta{\bm{\mu}}||$ and (b) noise properties as a function of nonlinearity ${\Lambda}/\gamma$. The reduction of projected noise to below vacuum levels where input signals are classical (originating from a thermal state) indicates that the \qrc{} alone can give rise to non-classical correlations in the measured outputs. } 
    \label{fig:thermalClassify}
\end{figure}


It suffices to consider only the NVK approximation and to then plot both the the mean separation of measured \qrc{} quadratures $||\delta{\bm{\mu}}||$ and the projected noise in Fig.~\ref{fig:thermalClassify}, as a function of \qrc{} nonlinearity. We note that the qualitative features are as observed in Fig.~\ref{fig:ampClassifyLambda}, and therefore reiterate the main results in the paper: the ability to compute correlations \textit{in situ}, and the ability to control noise in measured quadratures differently from their mean value. Crucially, projected noise below vacuum in Fig.~\ref{fig:thermalClassify}(b) can now solely be attributed to the action of the \qrc{}, as the QS states are purely thermal and do not demonstrate such non-classical correlations; in fact, they exhibit fluctuations above the vacuum value due to the coupling to a finite temperature bath. This emphasizes that the \qrc{} can provide the ability to not just manipulate existing quantum correlations, but also provide useful quantum correlations itself.

We emphasize that the \qrc{} model we consider here is based on coherently-driven Kerr oscillators, which by themselves are not ideal squeezers. However, our general approach applies to much more general \qrc{} models, where more non-classical correlations could be generated.

\bibliography{refs}

\begin{widetext}
    \startcontents[Supplementary Information]
\printcontents[Supplementary Information]{l}{1}{\section*{Supplementary Information}\setcounter{tocdepth}{2}}

\setcounter{page}{1}
\setcounter{section}{0}

\makeatletter
\let\toc@pre\relax
\let\toc@post\relax
\makeatother

\newpage

\section{Table of simulation parameters}
\label{app:pars}

Simulation parameters that are used to generate plots in the various figures in the main text are summarized in Table~\ref{tab:paramvals}.


\begin{table}[h]
    \centering
    \caption{Specific parameters for classification tasks considered in this paper. Brackets [$\cdot$,$\cdot$] indicate range of values, while slashes (/) indicate discrete choices. }
    \begin{tabularx}{450pt}{Y|Y|Y|Y|Y|Y}
    \hline
    \hline 
    & \multicolumn{4}{c|}{\textbf{Task~I}} & \textbf{Task~II} \\
    Parameter & Fig.~3 & Fig.~4 & Fig.~5(d) & Fig.~9 & Fig.~7 \\
    \hline
    $\amp$ & 10 & 80 & 10 & 10/17.3/30 & 80  \\
    $\kappa_1$ & 0.5 & 0.5 & 0.5 & 0.5 & 0.5 \\
    $\kappa_2$ & 1.0 & 1.0 & 1.0 & 1.0 & 0.5 \\
    $\kappa$   & 1.0 & 1.0 & 1.0 & 1.0 & 1.0 \\
    \hline
    $\Gamma_1$ & 0.5 & 0.5 & 0.5 & 0.5 & 0.5 \\
    $\Gamma_2$ & 0.0 & 0.0 & 0.0 & 0.0 & 0.5 \\
    \hline
    $\Lambda_1~(\times 10^{-3})$  & 5.5 & [6.5,9.0] & 7.4 & $(\frac{30}{\amp})^2$[0.5,0.6] & [6.6,7.7] \\
    $\Lambda_2~(\times 10^{-3})$ & -- & -- & -- & -- & [6.6,7.7] \\
    $\Delta_1$ & -0.67 & -3.0 & -0.83 & -0.67 & [0.8,1.5] \\
    $\Delta_2$ & -- & -- & -- & -- & [-1.6,-3.0] \\
    $\varg_{12}$ & -- & -- & -- & -- & [0.1,2.5] \\
    $\gammahet$ & 0.5 & 4.0 & 0.5 & 0.5 & 4.0 \\
    $\gamma$ & 1.0 & 4.5 & 1.0 & 1.0 & 4.5 \\
    \hline
    \hline
    \end{tabularx}
    \label{tab:paramvals}
\end{table}


\newpage

\section{Quantum cumulants}
\label{si:cumulants}

The use of quantum cumulants as an efficient representation of certain \qrc{} states is essential to our work. In this section, we provide a simplified introduction to quantum cumulants as a description of quantum states.

\subsection{Definition in terms of the generating function}

The starting point for our discussion of cumulants is via the characteristic function $\chi(z,z^*)$, defined in terms of the density operator $\rhou$ for an arbitrary quantum state as
\begin{align}
    \chi(z,z^*) = {\rm tr}\left\{ \rhou e^{i z^* \hat{b}^{\dagger}} e^{i z \hat{b}} \right\}.
    \label{appeq:chi}
\end{align}
The characteristic function as defined in Eq.~(\ref{appeq:chi}) has the property that it defines all \textit{normal-ordered} operator averages with respect to the quantum state $\rhou$; simply evaluating the derivative of the characteristic function~\cite{carmichael_statistical_2002} at $z=z^* = 0$ yields
\begin{equation}
  \left. \frac{\partial^{p + q}}{\partial (iz^*)^p \partial (iz)^q} (\mathrm{tr} \{ \rho e^{i z^* \hat{b}^{\dagger}} e^{i z \hat{b}}\} )  \right|_{z = z^* = 0} = \langle \hat{b}^{\dagger p} \hat{b}^q \rangle.
\end{equation}
It is then possible to define a distribution function for calculating these normal-ordered moments,
\begin{align}
    P (\beta, \beta^{\dagger}) &=  \frac{1}{\pi^2}  \int d^2z~\chi (z, z^{\ast}) e^{- i z^{\ast} \beta^{\ast}} e^{- i z \beta} \label{eq:chi2P} 
\end{align}
for which we use the suggestive notation $P(\beta,\beta^{\dagger})$, as it is easily shown to be none other than the Glauber-Sudarshan $P$ distribution~\cite{glauber_coherent_1963,sudarshan_equivalence_1963,carmichael_statistical_2002}:
\begin{align}
    \rhou = \int d^2\beta~P(\beta,\beta^{\dagger})|\beta\rangle\langle\beta|
\end{align}



Having recalled the connection of the characteristic function to the quantum state defined by $\rhou$ and its normal-ordered moments, we now define a quantum generating function $G : \mathbb{C} \times \mathbb{C} \rightarrow
\mathbb{R}$,
\begin{equation}
  G (z, z^{\ast}) = \ln \mathrm{tr} \{ \chi(z,z^*) \} = \ln \mathrm{tr} \{\hat{\rho} e^{i z^{\ast} \hat{b}^{\dagger}} e^{i z \hat{b}}\}. \label{eq:GeneratingFunction}
\end{equation}
The \textit{quantum cumulants} are defined through Eq.~(\ref{eq:GeneratingFunction}) by the series expansion of the generating function $G (z, z^{\ast})$ which yields all quantum cumulants $C_{a^{\dagger p} a^q}$ for any $p,q \in \mathbb{N}^2$
\begin{equation}
  \ln \mathrm{tr} \{\hat{\rho} e^{i z^{\ast} \hat{b}^{\dagger}} e^{i z \hat{b}}\} = \sum_{n = 1}^{\infty}\!\!
  \sum_{\scriptsize{\begin{array}{c}
    p, q \in \mathbb{N} \\
    p\!+\!q \!=\! n
  \end{array}}} \frac{1}{p!q!} C_{b^{\dagger p} b^q} (i z^{\ast})^p (i z)^q,
\end{equation}
or equivalently, 
\begin{equation}
    C_{b^{\dagger p} b^q} \equiv \left. \frac{\partial^{p+q} G}{\partial (i z^{\ast})^p \partial (i z)^q} (z, z^{\ast})\right|_{z=z^{\ast}=0} 
\end{equation}

So far, cumulants may simply appear to be a mathematical construct. However, it can be shown that normal-ordered cumulants and normal-ordered moments can be mapped to each other one-to-one. To see this, we can think of $G (z,z^{\ast})$ as the composite of the logarithm function $\mathrm{log} (\chi) = \ln \chi$ and $\chi(z,z^*) = {\rm tr}\left\{ \rhou e^{i z^* \hat{b}^{\dagger}} e^{i z \hat{b}} \right\}$, and employ multi-variable version of Fa{\`a} di Bruno's formula for the $n$-th order partial derivative where $n=p+q$,
\begin{align}
    & \frac{\partial^{p+q} G}{\partial (i z^{\ast})^p \partial (i z)^q} (z, z^{\ast}) =~ \sum_{\pi} \left( \mathrm{log}^{(| \pi |)} (\chi (z, z^{\ast})) \prod_{B \in \pi} \frac{\partial^{p' + q'} \chi}{\partial (i z^{\ast})^{p'} \partial (i z)^{q'}} (z, z^{\ast}) \right),
\end{align}
where $\pi$ defines the set of possible partitions of the ordered set $\{ iz^{\ast}, \cdots, iz^{\ast}, iz, \cdots iz \}$ with $iz^{\ast}$ appearing $p$ times and $iz$ appearing $q$ times. $B$ indicates elements in this set of partitions, and $p'$ and $q'$ respectively counts the number of $iz^{\ast}$ and $iz$ in each partition element $B$. Finally, $f^{(|\pi|)}(\cdot)$ denotes the $|\pi|$-order derivative of the function $f(\cdot)$, where $|\pi|$ denotes the cardinality of the set $\pi$. By noting that $\chi (0, 0) = \mathrm{tr} (\rhou) = 1$, and then using $\mathrm{log}^{(| \pi |)} (\chi (0, 0)) = (\ln \chi)^{(| \pi |)} |_{\chi = 1} = (- 1)^{| \pi | - 1} (| \pi | - 1) !$ we obtain
\begin{align}
    C_{b^{\dagger p} b^q} & = \left. \frac{\partial^{p+q} G}{\partial (i z^{\ast})^p \partial (i z)^q} (z, z^{\ast})\right|_{z=z^{\ast}=0} \nonumber \\
    & = \sum_{\pi} \left( (- 1)^{| \pi | - 1} (| \pi | - 1) ! \prod_{B \in \pi} \langle \hat{b}^{\dagger p'} \hat{b}^{q'} \rangle \! \right). \label{eq:moment2cumulant}
\end{align}
As such, cumulants can be thought of as a reparameterization of the normal-ordered moments that typically appear in observables measured in an experiment. We will see in the next two appendices that this reparameterization leads to a much more efficient description of certain quantum states, which can then be leveraged to construct an efficient computational framework.

For our purposes, we will find it more useful to express quantum moments in terms of quantum cumulants. This inverse transformation can be performed by thinking of $\chi(z,z^*) = {\rm tr}\left\{ \rhou e^{i z^* \hat{b}^{\dagger}} e^{i z \hat{b}} \right\}$ as the composite of the exponential function $\mathrm{exp} (G) = e^G$ and $G (z, z^{\ast}) = \ln \mathrm{tr} \{\hat{\rho} e^{i z^{\ast} \hat{b}^{\dagger}} e^{i z \hat{b}}\}$. Again, Fa{\`a} di Bruno's formula gives
\begin{align}
    & \frac{\partial^{p+q} \chi}{\partial (i z^{\ast})^p \partial (i z)^q} (z, z^*) = \sum_{\pi} \!\! \left( \mathrm{exp}^{(| \pi |)} (G (z, z^*)) \prod_{B \in \pi} \frac{\partial^{p' + q'} G}{\partial (i z^{\ast})^{p'} \partial (i z)^{q'}} (z, z^*) \! \right)
\end{align}
where $\pi$ defines the set of possible partitions of the ordered set $\{ iz^{\ast}, \cdots, iz^{\ast}, iz, \cdots iz \}$ with $iz^{\ast}$ appearing $p$ times and $iz$ appearing $q$ times, $B$ indicates elements in this set of partitions, and $p'$ and $q'$ respectively counts the number of $iz^{\ast}$ and $iz$ in each partition element $B$. Again, noting that $G (0, 0) = \ln \mathrm{tr} (\rhou) = 0$, and then using $\mathrm{exp}^{(| \pi |)} (G(0, 0)) = (e^G)^{(| \pi |)} |_{G = 0} = 1$ we get for $n$-th order partial derivative where $n=p+q$:
\begin{align}
    \langle \hat{b}^{\dagger p} \hat{b}^q \rangle & = \left. \frac{\partial^{p+q} \chi}{\partial (i z^{\ast})^p \partial (i z)^q} (z, z
    ^{\ast}) \right|_{z=z^{\ast}=0} \nonumber \\
    & = \left. \sum_{\pi} \prod_{B \in \pi} \frac{\partial^{p' + q'} G}{\partial (i z^{\ast})^{p'} \partial (i z)^{q'}} (z, z^{\ast}) \right|_{z=z^{\ast}=0} \nonumber \\
    & = \sum_{\pi} \prod_{B \in \pi} C_{b^{\dagger p'} b^{q'}}. \label{eq:cumulant2moment}
\end{align}

Finally, it is straightforward to generalize the definition of quantum cumulants to the bosonic $N$-mode case. For a complex-valued vector $\boldsymbol{z} = (z_1, z_2, \cdots, z_N)$, we can define the generating function $G: \mathbb{C}^{N} \times \mathbb{C}^{N} \rightarrow \mathbb{R}$ as: 
\begin{align}
     G (\boldsymbol{z}, \boldsymbol{z}^{\ast}) = \ln \mathrm{tr} \left\{ \rhou e^{iz^{\ast}_1 \hat{b}_1^{\dagger}} \cdots e^{iz^{\ast}_N \hat{b}_N^{\dagger}} e^{iz_1 \hat{b}_1} \cdots e^{iz_N \hat{b}_N} \right\}. \label{eq:GenNmodecumulants}
\end{align}
The series expansion of $G(\boldsymbol{z}, \boldsymbol{z}^{\ast})$ yields all quantum cumulants
\begin{align}
    & \left. \frac{\partial^n G}{\partial (iz_1^{\ast})^{p_1} \!\cdots\! \partial (iz_N^{\ast})^{p_N} \partial (iz_1)^{q_1} \!\cdots\! \partial (iz_N)^{q_N}} (\boldsymbol{z}, \boldsymbol{z}^{\ast}) \right| _{\boldsymbol{z}=\boldsymbol{z}^{\ast}=\boldsymbol{0}} = C_{b_1^{\dagger p_1} \cdots b_N^{\dagger p_N} b_1^{q_1} \cdots b_N^{q_N}}. \label{eq:Nmodecumulants}
\end{align}
where $p_1+\cdots+p_N+q_1+\cdots+q_N = n$. The transformation between moments and cumulants in multi-mode case can be obtained by simply replacing all single-mode moments and cumulants in Eq.(\ref{eq:moment2cumulant}),~(\ref{eq:cumulant2moment}) with multi-mode moments and cumulants corresponding to normal-ordered operator $ \hat{o}_1 \hat{o}_2 \cdots \hat{o}_n = \hat{b}_1^{\dagger p_1} \cdots \hat{b}_N^{\dagger p_N} \hat{b}_1^{q_1} \cdots \hat{b}_N^{q_N}$: 
\begin{align}
    C_{o_1 o_2 \cdots o_n} & = \sum_{\pi} \left( (- 1)^{| \pi | - 1} (| \pi | - 1) ! \prod_{B \in \pi} \langle \hat{o}_{i}: i \in B \rangle \right), \label{eq:moment2cumulantN} \\
    \avg{\hat{o}_1 \hat{o}_2 \cdots \hat{o}_n} & = \sum_{\pi}\prod_{B\in \pi} C_{o_i : i \in B}, \label{eq:cumulant2momentN}
\end{align}
where $\pi$ defines the set of possible partitions of operators in the $n$-order moment, and $B$ indicates elements in this set of partitions.

\subsection{Cumulants for some simple quantum states}
\label{si:cumulantsQS}

Eqs.~(\ref{eq:moment2cumulant}),~(\ref{eq:cumulant2moment}) and their multi-mode generalization Eqs.~(\ref{eq:moment2cumulantN}),~(\ref{eq:cumulant2momentN}) show that quantum cumulants and quantum moments have a one-to-one correspondence through logarithm and exponential. On the other hand, for two complex-valued vectors $\boldsymbol{z} = (z_1, z_2, \cdots, z_N)$ and $\boldsymbol{\beta} = (\beta_1, \beta_2, \cdots, \beta_N)$, the $P$-representation of quantum state and quantum characteristic function are mutually related by Fourier transformation \cite{carmichael_statistical_2002}:
\begin{align}
    P (\boldsymbol{\beta}, \boldsymbol{\beta}^{\ast}) & = \frac{1}{\pi^2}  \int \chi (\boldsymbol{z}, \boldsymbol{z}^{\ast}) e^{- i \boldsymbol{z} \cdot \boldsymbol{\beta}^{\ast}} e^{- i \boldsymbol{z}^{\ast} \boldsymbol{\beta}}~d^2 \boldsymbol{z}, \label{eq:chi2P} \\
    \chi (\boldsymbol{z}, \boldsymbol{z}^{\ast}) & = \int P (\boldsymbol{\beta}, \boldsymbol{\beta}^{\ast}) e^{ i \boldsymbol{z} \cdot \boldsymbol{\beta}^{\ast}} e^{ i \boldsymbol{z}^{\ast} \boldsymbol{\beta}}~d^2 \boldsymbol{\beta} . \label{eq:P2chi}
\end{align}
where the inner product is conventionally $\boldsymbol{z} \cdot \boldsymbol{\beta}^{\ast} = z^{\ast}_1 \beta_1^{\ast} + \cdots + z^{\ast}_N \beta_N^{\ast}$ and $\boldsymbol{z}^{\ast} \cdot \boldsymbol{\beta} = z_1 \beta_1 + \cdots + z_N \beta_N$. This result ensures the one-to-one correspondence between a quantum state and the quantum cumulants.

Importantly, the truncation of cumulants to certain discrete orders naturally characterizes some special types of quantum states. In this section, we will show that states with only nonzero first order cumulants correspond to coherent states, while states with only nonzero first and second cumulants orders correspond to Gaussian states.

We first consider a product of coherent states $\ket{\bm{\beta}_0} \equiv \ket{\beta_1, \beta_2, \cdots, \beta_N}$. In Eq.~(\ref{eq:GenNmodecumulants}), we then set $\hat{\rho} = \ket{\boldsymbol{\beta}_0} \bra{\boldsymbol{\beta}_0}$, following which the RHS gives
\begin{align}
    & \ln \mathrm{tr} \left\{ \ket{\boldsymbol{\beta}_0} \bra{\boldsymbol{\beta}_0} e^{iz^{\ast}_1 \hat{b}_1^{\dagger}} \cdots e^{iz^{\ast}_N \hat{b}_N^{\dagger}} e^{iz_1 \hat{b}_1} \cdots e^{iz_N \hat{b}_N} \right\} \nonumber\\
    =~& \ln \left( \bra{\boldsymbol{\beta}_0} e^{iz^{\ast}_1 \hat{b}_1^{\dagger}} \cdots e^{iz^{\ast}_N \hat{b}_N^{\dagger}} e^{iz_1 \hat{b}_1} \cdots e^{iz_N \hat{b}_N} \ket{\boldsymbol{\beta}_0} \right) \nonumber\\
    =~& \ln e^{iz^{\ast}_1 \beta_1^{\ast} + \cdots + iz^{\ast}_N \beta_N^{\ast} + iz_1 \beta_1 + \cdots + iz_N \beta_N} \nonumber\\
    =~& iz^{\ast}_1 \beta_1^{\ast} + \cdots + iz^{\ast}_N \beta_N^{\ast} + iz_1 \beta_1 + \cdots + iz_N \beta_N,
\end{align}
From Eq.~(\ref{eq:Nmodecumulants}), we obtain non-trivial derivative contributions only for first-order derivatives. Thus for all $k \in \{1,2,\cdots,N\}$, we have $C_{b_k} = \beta_k$ and $C_{b_k^{\dagger}} = \beta_k^{\ast}$, so that 
\begin{subequations}
\begin{align}
    \left(C_{b_1}, C_{b_2}, \cdots, C_{b_N} \right) & = \boldsymbol{\beta}_0, \\ 
    \left(C_{b^{\dagger}_1}, C_{b^{\dagger}_2}, \cdots, C_{b^{\dagger}_N} \right) & = \boldsymbol{\beta}^{\ast}_0
\end{align}
\end{subequations}

We can also show that physical states with vanishing cumulants of order greater than one are necessarily coherent states. We begin with the normal-ordered characteristic function for such states, which is given by $\chi (\boldsymbol{z}, \boldsymbol{z}^{\ast}) = e^{G (i \boldsymbol{z}^{\ast}, i \boldsymbol{z})} = e^{i \boldsymbol{z} \cdot \boldsymbol{\beta}^{\ast}_0 + i \boldsymbol{z}^{\ast} \cdot
\boldsymbol{\beta}_0}$. The corresponding $P$-representation is simply the Fourier transform of the characteristic function, and is given by
\begin{align}
    P (\boldsymbol{\beta}, \boldsymbol{\beta}^{\ast}) & = \frac{1}{\pi^2} \int \chi(\boldsymbol{z}, \boldsymbol{z}^{\ast}) e^{- i \boldsymbol{z} \cdot \boldsymbol{\beta}^{\ast}} e^{- i \boldsymbol{z}^{\ast} \cdot \boldsymbol{\beta}}~d^2\boldsymbol{z} \nonumber\\ 
    & = \frac{1}{\pi^2}  \int e^{i \boldsymbol{z} \cdot \boldsymbol{\beta}^{\ast}_0 + i \boldsymbol{z}^{\ast} \cdot \boldsymbol{\beta}_0} e^{- i \boldsymbol{z} \cdot \boldsymbol{\beta}^{\ast}} e^{- i \boldsymbol{z}^{\ast} \cdot \boldsymbol{\beta}}~d^2 \boldsymbol{z} \nonumber\\
    & = \delta (\boldsymbol{\beta} - \boldsymbol{\beta}_0),
\end{align}
which describes a product of coherent states. As a result, the states with all cumulants or order greater than one vanishing are exactly the collection of all coherent states.

Next, we consider quantum states whose quasi-probability distribution is Gaussian. Recall that the $P$, $Q$, and Wigner phase-space representations are related by the Weierstrass transformation, which always maps one Gaussian distribution to another Gaussian distribution \cite{carmichael_statistical_2002}: 
\begin{align}
    W (\boldsymbol{\beta}, \boldsymbol{\beta}^{\ast}) & = \frac{2}{\pi}  \int d^2\boldsymbol{\widetilde{\beta}}~e^{- 2 | \boldsymbol{\beta} - \boldsymbol{\widetilde{\beta}} |^2} P (\boldsymbol{\widetilde{\beta}}, \boldsymbol{\widetilde{\beta}}^{\ast}) , \\
    Q (\boldsymbol{\beta}, \boldsymbol{\beta}^{\ast}) & = \frac{2}{\pi}  \int d^2\boldsymbol{\widetilde{\beta}}~e^{- 2 | \boldsymbol{\beta} - \boldsymbol{\widetilde{\beta}} |^2} W (\boldsymbol{\widetilde{\beta}}, \boldsymbol{\widetilde{\beta}}^{\ast}). 
\end{align}
The relation between the $P$-representation and characteristic function in Eqs.~(\ref{eq:chi2P}),~(\ref{eq:P2chi}) respectively is via Fourier transform, which also maps one Gaussian distribution to another Gaussian distribution. 

Via this series of transformations, we therefore see that a quantum state with a Gaussian Wigner distribution will also have a Gaussian characteristic function, and hence must be entirely defined by cumulants of up to second-order (by definition of a Gaussian distribution).

\newpage

\section{Truncated Equations of Motion approach applied to the single-mode Kerr \qrc{}}
\label{si:singleNodeQRC}

To benchmark the computational approach developed in this paper based on quantum cumulants, we will use a simple system as a testbed: a single coherently-drive Kerr oscillator. This system also forms a single mode of the \qrcs{} we employ. 

The resulting system schematic is shown in Fig.~\ref{fig:singleQRCClassical}. We begin by analyzing the \textit{unconditional} dynamics of the resulting system, described by the quantum master equation
\begin{align}
    d\hat{\rho} \equiv \mathcal{L}\rhou~dt = \Lqrc\hat{\rho}~dt  - i[\hat{\mathcal{H}}_d,\hat{\rho}]~dt
\end{align}
where the drive Hamiltonian in the frame rotating at the drive frequency $\omega_d$ is then
\begin{align}
\hat{\mathcal{H}}_d = \eta_b(\hat{b} + \hat{b}^{\dagger}),
\end{align}
and we have defined for convenience $\hat{b} \equiv \hat{b}_1$. The coherent drive is incident on the \qrc{} directly, via a separate input port with loss rate $\Gamma_1$. This situation describes \qrc{} processing of classical signals encoded in the generally time-dependent coherent drive amplitude and phase. The Liouvillian $\mathcal{L}_{\rm QRC}$ describing the single-mode \qrc{} takes the form
\begin{align}
    \Lqrc\hat{\rho} = -i\left[-\Delta_1\hat{b}^{\dagger}\hat{b} - \frac{\Lambda_1}{2}\hat{b}^{\dagger}\hat{b}^{\dagger}\hat{b}\hat{b},\hat{\rho} \right] + (\gamma_1+\Gamma_1) \mathcal{D}[\hat{b}]\hat{\rho}
    \label{eq:singleQRCME}
\end{align}
where $\Delta = \omega_d - \omega_1$ is the detuning of the drive frequency from the bare \qrc{} mode frequency $\omega$. For simplicity of notation, in what follows we introduce for convenience the total linear damping ${\gamma} = \gamma_1 + \Gamma_1$ of the nonlinear mode.


\begin{figure}[t]
    \centering
    \includegraphics[scale=1.0]{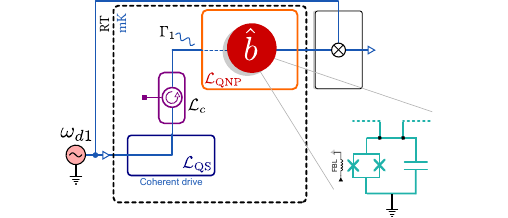}
    \caption{Schematic showing a single-mode \qrc{} as part of a simplified measurement chain: a single frequency coherent drive is incident on the \qrc{} via an input channel that is not measured, while the second port of the \qrc{} is monitored. The \qrc{} can be realized in cQED as a tunable-frequency Kerr oscillator using a capacitively shunted SQUID loop~\cite{koch_charge-insensitive_2007}.}
    \label{fig:singleQRCClassical}
\end{figure}


\subsection{Obtaining TEOMs for a single coherently-driven Kerr oscillator}
\label{si:mToC}

Our approach requires the expression of moments of a multimode quantum system in terms of their associated cumulants. Moments and cumulants are related via their generating functions, which allows one to write a general prescription for writing an arbitrary normal-ordered joint system moment of order $n$ in terms of a series of cumulants of order $1$ up to $n$. This takes the specific form given by Eq.~(\ref{eq:cumulant2momentN}),
\begin{align}
    \avg{\hat{o}_1\hat{o}_2\ldots \hat{o}_n} = \sum_{\pi}\prod_{B\in \pi} C_{o_i : i \in B},
    \label{appeq:mtoC}
\end{align}
where $\pi$ defines the set of possible partitions of operators in the $n$-order moment, and $B$ indicates elements in this set of partitions. The rather opaque form can be simplified by considering a specific case. We will use the above prescription to obtain the TEOMs for the single-mode \qrc{}.

This minimal system already includes the key nonlinear element of the \qrc{}, and therefore the most complex terms that need to be analyzed will arise already in these equations. We start by writing down the equation for $\avg{\hat{b}} = {\rm tr}\{\mathcal{L}\rhou\hat{b}\}$,
\begin{align}
    \avg{\dot{\hat{b}}} = \left(i\Delta_1 - \frac{{\gamma}}{2} \right)\avg{\hat{b}} + i\Lambda_1 \avg{\hat{b}^{\dagger}\hat{b}\hat{b}} - i\eta_b.
    \label{appeq:b}
\end{align}
To rewrite the above equation in terms of cumulants, we must express the third-order moment $\avg{\hat{b}^{\dagger}\hat{b}\hat{b}}$ in terms of cumulants. The possible partitions of the third-order moment are given by: $\pi \in \{(\hat{b}^{\dagger}\hat{b}\hat{b}),(\hat{b}^{\dagger}\hat{b},\hat{b})\times 2,(\hat{b}\hat{b},\hat{b}^{\dagger}),(\hat{b}^{\dagger},\hat{b},\hat{b}) \}$. Then, using Eq.~(\ref{appeq:mtoC}), we obtain
\begin{align}
    \avg{\hat{b}^{\dagger}\hat{b}\hat{b}} = C_{b^{\dagger}bb} + 2C_{b^{\dagger}b}\avg{\hat{b}} + C_{bb}\avg{\hat{b}^{\dagger}} + \avg{\hat{b}^{\dagger}}\avg{\hat{b}}^2.
\end{align}
Using the above, we can now rewrite Eq.~(\ref{appeq:b}) in terms of cumulants as
\begin{align}
    \avg{\dot{\hat{b}}} = \left(i\Delta_1 - \frac{{\gamma}}{2} \right)\avg{\hat{b}} + i\Lambda_1 \avg{\hat{b}^{\dagger}}\avg{\hat{b}}^2 - i\eta_b + i\Lambda_1\left(C_{bb}\avg{\hat{b}^{\dagger}} + 2C_{b^{\dagger}b}\avg{\hat{b}} + C_{b^{\dagger}bb} \right)
\end{align}

To similarly obtain dynamical equations for cumulants $C_{b^{\dagger}b}, C_{bb}$, we write equations of motion for the second-order moments $\avg{\hat{b}^{\dagger}\hat{b}}$ and $\avg{\hat{b}\hat{b}}$ respectively. Starting with the former, we find:
\begin{align}
    \frac{d}{dt} \avg{\hat{b}^{\dagger}\hat{b}} &= -{\gamma}\avg{\hat{b}^{\dagger}\hat{b}} +i\eta_b\avg{\hat{b}} - i\eta_b\avg{\hat{b}^{\dagger}}  \nonumber \\
    &= -{\gamma}C_{b^{\dagger}b} - {\gamma}\avg{\hat{b}^{\dagger}}\avg{\hat{b}} + i\eta_b\avg{\hat{b}} - i\eta_b\avg{\hat{b}^{\dagger}}
\end{align}
where in the second line we rewrite the second-order moment in terms of cumulants. From this we can easily write $dC_{b^{\dagger}b} = d\avg{\hat{b}^{\dagger}\hat{b}} - (d\avg{\hat{b}^{\dagger}})\avg{\hat{b}} - (d\avg{\hat{b}})\avg{\hat{b}^{\dagger}}$, which yields the dynamical equation for $C_{b^{\dagger}b}$:
\begin{align}
    \dot{C}_{b^{\dagger}b} = -{\gamma}C_{b^{\dagger}b} -i\Lambda_1\left(C_{bb}\avg{\hat{b}^{\dagger}}^2 - C_{bb}^*\avg{\hat{b}}^2 \right) - i\Lambda_1 \left( C_{b^{\dagger}bb}\avg{\hat{b}^{\dagger}} - C_{b^{\dagger}bb}^*\avg{\hat{b}}   \right)
\end{align}

Finally, we write the equation of motion for $\avg{\hat{b}\hat{b}}$,
\begin{align}
    \frac{d}{dt} \avg{\hat{b}\hat{b}} &= \left(2i\Delta_1-{\gamma}\right)\avg{\hat{b}\hat{b}} -2i\eta_b\avg{\hat{b}} + i\Lambda_1 \avg{\hat{b}\hat{b}} + i2\Lambda_1\avg{\hat{b}^{\dagger}\hat{b}\hat{b}\hat{b}} \nonumber \\
    &= \left(2i\Delta_1-{\gamma}+i\Lambda_1\right)(C_{bb}+\avg{\hat{b}}\avg{\hat{b}}) -2i\eta_b\avg{\hat{b}} + i2\Lambda_1\avg{\hat{b}^{\dagger}\hat{b}\hat{b}\hat{b}}
\end{align}
where we once again rewrite second-order moments in terms of cumulants in the second line. We now need to express the remaining fourth-order moment $\avg{\hat{b}^{\dagger}\hat{b}\hat{b}\hat{b}}$ in terms of cumulants. The possible partitions of this fourth-order moment are given by $\pi \in \{(\hat{b}^{\dagger}\hat{b}\hat{b}\hat{b}),(\hat{b}^{\dagger}\hat{b}\hat{b},\hat{b})\times 3,(\hat{b}\hat{b}\hat{b},\hat{b}^{\dagger}),(\hat{b}^{\dagger}\hat{b},\hat{b}\hat{b})\times 3,(\hat{b}^{\dagger}\hat{b},\hat{b},\hat{b})\times 3,(\hat{b}\hat{b},\hat{b}^{\dagger},\hat{b})\times 3,(\hat{b}^{\dagger},\hat{b},\hat{b},\hat{b}) \}$, so that
\begin{align}
    \avg{\hat{b}^{\dagger}\hat{b}\hat{b}\hat{b}} &= C_{{b}^{\dagger}{b}{b}{b}} +3\avg{\hat{b}}C_{b^{\dagger}bb} + \avg{\hat{b}^{\dagger}}C_{bbb} \nonumber \\
    &+ 3C_{b^{\dagger}b}C_{bb} + 3C_{b^{\dagger}b}\avg{\hat{b}}^2 + 3C_{bb}|\avg{\hat{b}}|^2 + \avg{\hat{b}^{\dagger}}\avg{\hat{b}}\avg{\hat{b}}\avg{\hat{b}}
\end{align}

From this we can finally write $dC_{bb} = d\avg{\hat{b}\hat{b}} - 2(d\avg{\hat{b}})\avg{\hat{b}}$, which yields the dynamical equation for $C_{bb}$:
\begin{align}
    \dot{C}_{bb} &= \left(2i\Delta_1-{\gamma}+i\Lambda_1\right)(C_{bb}+\avg{\hat{b}}^2) -2i\eta_b\avg{\hat{b}} \nonumber \\
    &~~~~+ i2\Lambda_1\left( C_{{b}^{\dagger}{b}{b}{b}} +3\avg{\hat{b}}C_{b^{\dagger}bb} + \avg{\hat{b}^{\dagger}}C_{bbb} \right) \nonumber \\
    &~~~~+ i2\Lambda_1\left( 3C_{b^{\dagger}b}C_{bb} + 3C_{b^{\dagger}b}\avg{\hat{b}}^2 + 3C_{bb}|\avg{\hat{b}}|^2 + \avg{\hat{b}^{\dagger}}\avg{\hat{b}}^3 \right) \nonumber \\
    &~~~~-2\left(i\Delta_1 - \frac{{\gamma}}{2} \right)\avg{\hat{b}}^2 - i2\Lambda_1 \avg{\hat{b}^{\dagger}}\avg{\hat{b}}^3 + 2i\eta_b\avg{\hat{b}} \nonumber \\
    &~~~~-i2\Lambda_1\left(C_{bb}|\avg{\hat{b}}|^2 + 2C_{b^{\dagger}b}\avg{\hat{b}}^2 + C_{b^{\dagger}bb}\avg{\hat{b}} \right)
\end{align}
which finally simplifies to:
\begin{align}
    &\dot{C}_{bb} = \left(2i\Delta_1-{\gamma}+i\Lambda_1\right)C_{bb} +i\Lambda_1\avg{\hat{b}}^2 +i 4\Lambda_1|\avg{\hat{b}}|^2C_{bb} + i 6\Lambda_1 C_{b^{\dagger}b}C_{bb} \nonumber \\
    &+i2\Lambda_1 C_{b^{\dagger}b}\avg{\hat{b}}^2+i2\Lambda_1 C_{b^{\dagger}bbb} + i4\Lambda_1 C_{b^{\dagger}bb}\avg{\hat{b}} +i 2\Lambda_1 C_{bbb}\avg{\hat{b}^{\dagger}}
\end{align}



Clearly, second-order cumulants couple to cumulants of third and fourth order. This is a general feature of nonlinear quantum systems: moments and thus cumulants of a certain order can couple to those of higher-order, leading to an infinite hierarchy of equations that do not form a closed set. 

However, an infinite set of normal-ordered cumulants is not necessarily required to describe all multimode quantum states. In particular, as we have seen in Sec.~\ref{si:cumulants} of the SI, a multimode quantum system in a product of coherent states is described entirely by its nonzero first-order cumulants; all cumulants of order $p+q > 1$ vanish. Multimode quantum states that are defined entirely by their first and second-order cumulants admit Gaussian phase-space representations, and are labelled Gaussian states. States with nonzero cumulants of third or higher-order are thus by definition non-Gaussian states~\cite{boutin_effect_2017, ra_non-gaussian_2020}.

Our numerical approach leverages this efficient representation of specific multimode quantum states in terms of cumulants. In particular, we consider an ansatz wherein the quantum state of the complete measurement is described entirely by cumulants up to a finite order $p+q \leq n_{\rm trunc}$; all cumulants of order $p+q > n_{\rm trunc}$ are thus set to zero, truncating the hierarchy and yielding a closed set of equations for the retained nonzero cumulants. In this work, we choose $n_{\rm trunc} = 2$ for a quantum measurement chain defined entirely by its first and second-order cumulants, although the truncation can similarly be carried out at higher order. The resulting Truncated Equations Of Motion (TEOMs), and their generalization to \textit{conditional} dynamics of multimode quantum systems, the Stochastic Truncate Equations of Motion (STEOMs), form the basis of our computational approach in this paper. We have developed a computer algebra approach that automates the above process of calculating contributions to equations of motion and expressing arbitrary moments in terms of cumulants, thus allowing the systematic truncation necessary to arrive at TEOMs (and STEOMs) introduced in the main text.

\subsection{Calculating STEOMs - stochastic measurement contributions to truncated cumulant dynamics}
\label{si:stochcumulants}

The measurement-conditioned dynamics of a quantum system under heterodyne monitoring is given by a stochastic master equation (SME). We can now calculate the contributions to equations of motion for cumulants due to stochastic contributions via an SME. To illustrate the salient features of the calculation, a single homodyne measurement superoperator for monitoring a general mode $\hat{b}_k$ suffices:
\begin{align}
    \mathcal{S}_{\rm hom}[\sqrt{\gamma_k}\hat{b}_k]\rhou = \sqrt{\frac{\gamma_k}{2}} \left( \bk{k} \rhou + \rhou \bkd{k} - \avg{\bk{k}+\bkd{k}} \right) dW_{\mathcal{I}_k}(t)
\end{align}
For convenience, we will then write a general stochastic master equation in the form:
\begin{align}
    d\rhoc = \mathcal{L}\rhoc~dt + \mathcal{S}_{\rm hom}[\sqrt{\gamma_k}\hat{b}_k]\rhoc
\end{align}
where $\mathcal{L}$ defines all contributions to the master equation governing deterministic evolution. We can thus write for the differential of conditional expectation of an arbitrary operator $\hat{o}$:
\begin{align}
    d\avgc{\hat{o}} = {\rm tr} \{\mathcal{L}\rhoc\hat{o}\}~dt + \sqrt{\frac{\gamma_k}{2}} \left( \Cc{o b_k} + \Cc{b_k^{\dagger}o} \right)dW_{\mathcal{I}_k}(t)
    \label{appeq:do}
\end{align}
Analogously, we can write for arbitrary second-order moments:
\begin{align}
    &d\avgc{\hat{o}_1\hat{o}_2} = {\rm tr} \{\mathcal{L}\rhoc\hat{o}_1\hat{o}_2\}~dt + \sqrt{\frac{\gamma_k}{2}}\left(\! \avgc{\hat{o}_1\hat{o}_2\hat{b}_k} \!+\! \avgc{\hat{b}_k^{\dagger}\hat{o}_1\hat{o}_2} \!-\! \avgc{\hat{o}_1\hat{o}_2}\avgc{\hat{b}_k} \!-\! \avgc{\hat{b}_k^{\dagger}}\avgc{\hat{o}_1\hat{o}_2} \! \right)\! dW_{\mathcal{I}_k}(t)
    \label{appeq:do1o2}
\end{align}

We will now use the above results obtain the equation of motion for an arbitrary \textit{normal-ordered} cumulant $\Cc{o_1o_2} = \avgc{\hat{o}_1\hat{o}_2} - \avgc{\hat{o}_1}\avgc{\hat{o}_2}$. The important step arises in writing the differential of this cumulant using Ito's lemma:
\begin{align}
    d\Cc{o_1o_2} &= d\avgc{\hat{o}_1\hat{o}_2} - (d\avgc{\hat{o}_1})\avgc{\hat{o}_2} - \avgc{\hat{o}_1}(d\avgc{\hat{o}_2}) \nonumber \\
    & - (d\avgc{\hat{o}_1})(d\avgc{\hat{o}_2})
    \label{appeq:dco1o2}
\end{align}
where the term on the second line arises from the modified chain rule in Ito calculus. We can now proceed to obtaining the individual contributions to the above equation.

We start with the first term in Eq.~(\ref{appeq:dco1o2}), which was calculated in Eq.~(\ref{appeq:do1o2}). To proceed, it will prove convenient to write moments higher than first-order in terms of their corresponding cumulants. We begin with the second-order moments:
\begin{align}
    &d\avgc{\hat{o}_1\hat{o}_2} = {\rm tr} \{\mathcal{L}\rhoc\hat{o}_1\hat{o}_2\}~dt + \nonumber \\
    &\sqrt{\frac{\gamma_k}{2}}\Big( \avgc{\hat{o}_1\hat{o}_2\hat{b}_k} + \avgc{\hat{b}_k^{\dagger}\hat{o}_1\hat{o}_2} - \Cc{{o}_1{o}_2}\avgc{\hat{b}_k} - \avgc{\hat{b}_k^{\dagger}}\Cc{{o}_1{o}_2} \nonumber \\
    &- \avgc{\hat{o}_2}\avgc{\hat{o}_1}\avgc{\hat{b}_k} - \avgc{\hat{o}_2}\avgc{\hat{o}_1}\avgc{\hat{b}_k^{\dagger}}  \Big)dW_{\mathcal{I}_k}(t)
\end{align}
Next, we write down the expressions relating third-order moments to their corresponding cumulants using Eq.~(\ref{appeq:mtoC}):
\begin{subequations}
\begin{align}
    \Cc{o_1o_2b_k} &= \avgc{\hat{o}_1\hat{o}_2\hat{b}_k} - \Cc{{o}_1{o}_2}\avgc{\hat{b}_k} - \Cc{{o}_1{b}_k}\avgc{\hat{o}_2} - \Cc{{o}_2{b}_k}\avgc{\hat{o}_1} \nonumber \\
    & - \avgc{\hat{o}_2}\avgc{\hat{o}_1}\avgc{\hat{b}_k} \\
    \Cc{b_k^{\dagger}o_1o_2} &= \avgc{\hat{b}_k^{\dagger}\hat{o}_1\hat{o}_2} - \Cc{{o}_1{o}_2}\avgc{\hat{b}^{\dagger}_k} - \Cc{{b}_k^{\dagger}{o}_1}\avgc{\hat{o}_2} - \Cc{{b}_k^{\dagger}{o}_2}\avgc{\hat{o}_1} \nonumber \\
    & - \avgc{\hat{o}_2}\avgc{\hat{o}_1}\avgc{\hat{b}^{\dagger}_k}
\end{align}
\end{subequations}
employing which, Eq.~(\ref{appeq:dco1o2}) becomes:
\begin{align}
    &d\avgc{\hat{o}_1\hat{o}_2} = {\rm tr} \{\mathcal{L}\rhoc\hat{o}_1\hat{o}_2\}~dt + \nonumber \\
    &\sqrt{\frac{\gamma_k}{2}}\Big( \Cc{o_1o_2b_k} + \Cc{{o}_1{b}_k}\avgc{\hat{o}_2} + \Cc{{o}_2{b}_k}\avgc{\hat{o}_1}  \Big)dW_{\mathcal{I}_k}(t) \nonumber \\
    +&\sqrt{\frac{\gamma_k}{2}}\Big( \Cc{b_k^{\dagger}o_1o_2} + \Cc{{b}_k^{\dagger}{o}_1}\avgc{\hat{o}_2} + \Cc{{b}_k^{\dagger}{o}_2}\avgc{\hat{o}_1}   \Big)dW_{\mathcal{I}_k}(t)
    \label{appeq:do1o2C}
\end{align}
where terms involving only first-order moments cancel. Note that the above includes terms of up to $O(dt)$, as $dW_{\mathcal{I}_k}(t)$ is formally $O(dt^{1/2})$.

Next, we can calculate the second and third terms in Eq.~(\ref{appeq:dco1o2}), using Eq.~(\ref{appeq:do}). These take the simple forms:
\begin{subequations}
\begin{align}
    (d\avgc{\hat{o}_1})\avgc{\hat{o}_2} &= \avgc{\hat{o}_2}{\rm tr} \{\mathcal{L}\rhoc\hat{o}_1\}~dt  + \sqrt{\frac{\gamma_k}{2}} \left( \Cc{o_1 b_k}\avgc{\hat{o}_2} + \Cc{b_k^{\dagger}o_1}\avgc{\hat{o}_2} \right)dW_{\mathcal{I}_k}(t) \label{appeq:do1xo2} \\
    (d\avgc{\hat{o}_2})\avgc{\hat{o}_1} &= \avgc{\hat{o}_1}{\rm tr} \{\mathcal{L}\rhoc\hat{o}_2\}~dt  + \sqrt{\frac{\gamma_k}{2}} \left( \Cc{o_2 b_k}\avgc{\hat{o}_1} + \Cc{b_k^{\dagger}o_2}\avgc{\hat{o}_1} \right)dW_{\mathcal{I}_k}(t) \label{appeq:do2xo1}
\end{align}
\end{subequations}
Both equations above once again contain terms up to $O(dt)$. 

Finally, we can write down the term arising from Ito's lemma. The requiring differential is that given by Eq.~(\ref{appeq:do}). However, we need only retain the lowest $O(dt)$ term here, which is given by:
\begin{align}
    (d\avgc{\hat{o}_1})(d\avgc{\hat{o}_2}) = \frac{\gamma_k}{2}\left( \Cc{o_1 b_k} + \Cc{b_k^{\dagger}o_1} \right)\left( \Cc{o_2 b_k} + \Cc{b_k^{\dagger}o_2} \right)~dt + O(dt^{3/2})
    \label{appeq:do1do2}
\end{align}
where we have used $(dW_{\mathcal{I}_k})^2 = dt$.

Finally, we can write down Eq.~(\ref{appeq:dco1o2}) by combining Eqs.~(\ref{appeq:do1o2C}),~(\ref{appeq:do1xo2}),~(\ref{appeq:do2xo1}), and (\ref{appeq:do1do2}). This finally yields:
\begin{align}
    &d\Cc{o_1o_2} = {\rm tr} \{\mathcal{L}\rhoc\hat{o}_1\hat{o}_2 \!-\! \avgc{\hat{o}_2}\mathcal{L}\rhoc\hat{o}_1  \!-\! \avgc{\hat{o}_1}\mathcal{L}\rhoc\hat{o}_2 \}dt \nonumber \\
    &~~~~~~~~~~~~- \frac{\gamma_k}{2}\left( \Cc{o_1 b_k} + \Cc{b_k^{\dagger}o_1} \right)\left( \Cc{o_2 b_k} + \Cc{b_k^{\dagger}o_2} \right)dt \nonumber \\
    &~~~~~~~~~~~~+\sqrt{\frac{\gamma_k}{2}}\left( \Cc{o_1o_2b_k} + \Cc{b_k^{\dagger}o_1o_2} \right)dW_{\mathcal{I}_k}(t)
\end{align}
The first term on the right hand side describes deterministic dynamics governed by $\mathcal{L}$. The second term is due to the measurement, but does not depend on the stochastic Wiener increment $dW_{\mathcal{I}_k}(t)$. The last term is a stochastic contribution; however note that it is related only to third-order cumulants. Within the ans\"atz used in this paper, this explicitly stochastic term vanishes.


\subsection{Numerical tests: unconditional dynamics and comparison to exact results}

We will use the single coherently-driven \qrc{} mode modeled as a Kerr oscillator to benchmark the TEOMs approach. Using the results of Sec.~\ref{si:mToC} of the SI, and following truncation with $n_{\rm trunc} = 2$, the single Kerr oscillator is defined by the first order cumulant $C_b \equiv \avg{\hat{b}}$ governed by the equation,
\begin{align}
    \avg{\dot{\hat{b}}} = \left( i\Delta_1 -\frac{{\gamma}}{2} \right)\avg{\hat{b}} + i\Lambda_1 \avg{\hat{b}^{\dagger}}\avg{\hat{b}}\avg{\hat{b}} -i \eta_b + i\Lambda_1 \left( C_{bb}\avg{\hat{b}^{\dagger}} + 2C_{b^{\dagger}b}\avg{\hat{b}} \right)
    \label{eq:b}
\end{align}
as well as the second-order cumulants $C_{bb} = \avg{\hat{b}\hat{b}}-\avg{\hat{b}}^2$, $C_{b^{\dagger}b} = \avg{\hat{b}^{\dagger}\hat{b}}-\avg{\hat{b}^{\dagger}}\avg{\hat{b}}$, defined by
\begin{subequations}
\begin{align}
    \dot{C}_{b^{\dagger}b} &= -{\gamma}C_{b^{\dagger}b} -i\Lambda_1 (C_{bb}\avg{\hat{b}^{\dagger}}^2-C_{bb}^*\avg{\hat{b}}^2) , \label{eq:CbdbFull} \\
    \dot{C}_{bb} &= \left( i2\Delta_1 - {\gamma} + i\Lambda_1(1+4|\avg{\hat{b}}|^2\!\! +6C_{b^{\dagger}b}) \right)C_{bb} + i\Lambda_1 \avg{\hat{b}}^2 (1+2C_{b^{\dagger}b}). \label{eq:CbbFull} 
\end{align}
\end{subequations}
Eqs.~(\ref{eq:b})-(\ref{eq:CbbFull}) together comprise the TEOMs for the single Kerr oscillator with $n_{\rm trunc} = 2$.

It is reasonable to assume that such a truncation must be valid only in a certain subset of the full parameter space defining the system. To identify the general constraints on such a truncation, we make extensive use of the exact quantum steady-state solution of the coherently-driven Kerr oscillator, obtained via the complex-$\mathcal{P}$ representation~\cite{drummond_quantum_1980}; details can be found in SI Sec.~\ref{app:complexP}. This provides access to exact steady-state cumulants of arbitrary order, which help us benchmark the accuracy of the TEOM approach.

One such regime should be furnished by the \textit{classical limit} of the single-mode Kerr \qrc{}, where the \qrc{} is always in a coherent state and thus described entirely by its first-order cumulants. To identify this classical limit, and thus explore deviations from it due to quantum effects, it proves convenient to work with specific scaled quantities: after introducing dimensionless time $t' = {\gamma}t$ and energy scales $(\Delta_1',\Lambda_1',\eta_b') = (\Delta_1,\Lambda_1,\eta_b)/{\gamma}$, we scale $\avg{\hat{b}}'\to \sqrt{\Lambda_1'}\avg{\hat{b}}$, and impose our truncated ansatz, following which Eq.~(\ref{eq:b}) becomes
\begin{align}
    \frac{d}{dt'}\avg{\hat{b}}' = \left( i\Delta_1' -\frac{1}{2} \right)\avg{\hat{b}}' + i \avg{\hat{b}^{\dagger}}'\avg{\hat{b}}'\avg{\hat{b}}' -i \CNL   + i\Lambda_1' \left( C_{bb}\avg{\hat{b}^{\dagger}}' + 2 C_{b^{\dagger}b}\avg{\hat{b}}' \right). 
    \label{eq:transfb}
\end{align}
The first line in Eq.~(\ref{eq:transfb}) makes no reference to second-order cumulants of the \qrc{} mode state; it thus describes dynamics when the \qrc{} is in a coherent state (for which such cumulants vanish), corresponding to the classical limit. Dynamics in this limit depend on the drive and the nonlinearity via a single dimensionless parameter~\cite{dykman_theory_1979, dykman_fluctuating_2012},
\begin{align}
    \CNL = \eta'\sqrt{\Lambda'} = \frac{\eta_b}{{\gamma}} \sqrt{\frac{ \Lambda_1}{{\gamma}}}.
    \label{eq:CNL}
\end{align}

The second line of Eq.~(\ref{eq:transfb}) does however depend on cumulants; these terms describe quantum dynamics that lead to a deviation from purely coherent state evolution. In these scaled units, the TEOMs for the second-order cumulants obtained from Eq.~(\ref{eq:CbdbFull}),~(\ref{eq:CbbFull}) take the form
\begin{subequations}
\begin{align}
    \frac{d}{dt'}{C}_{b^{\dagger}b} &= -C_{b^{\dagger}b} -i (C_{bb}(\avg{\hat{b}^{\dagger}}')^2-C_{bb}^*(\avg{\hat{b}}')^2), \label{eq:transfCbdb} \\
    \frac{d}{dt'}{C}_{bb} &= \left( i2\Delta_1' - 1 + i(\Lambda_1'+4|\avg{\hat{b}}'|^2\!\! +6\Lambda_1'C_{b^{\dagger}b}) \right)C_{bb}  + i(\avg{\hat{b}}')^2(1+2C_{b^{\dagger}b}).  \label{eq:transfCbb}
\end{align}
\end{subequations}
Consider now the transformation where $\Lambda_1' \to 0$, while $\CNL$ is held fixed (by increasing the drive strength $\eta_b'$ simultaneously). Terms in the first line of Eq.~(\ref{eq:transfb}) describing classical dynamics remain unchanged, while those in the second line describing quantum deviations become smaller, \textit{provided} second-order cumulants do not grow with $\Lambda_1'$. From Eqs.~(\ref{eq:transfCbdb}),~(\ref{eq:transfCbb}), it is clear that $C_{b^{\dagger}b}$ is independent of $\Lambda_1'$, while $C_{bb}$ has a dependence that is negligible as $\Lambda_1' \to 0$ for $|\avg{\hat{b}}'| \neq 0$. Hence by keeping $\CNL$ fixed while decreasing the nonlinearity strength, the influence of second-order cumulants on dynamics of the \qrc{} mode amplitude is suppressed, describing the classical limit of \qrc{} operation. Physically, this transformation leads to an increase in the unscaled drive $\eta'$ and \qrc{} mode amplitude $|\avg{\hat{b}}|$, and hence occupation number, decreasing the relative impact of quantum fluctuations in agreement with conventional notions of classicality.
Conversely, increasing nonlinearity strength for fixed $\CNL$ should enhance the impact of quantum fluctuations, marked by a systematic increase in second-order cumulants, and taking the Kerr \qrc{} gradually towards a quantum regime of operation.

To explore this transition, we start with the classical limit of the \qrc{} as determined by terms in the first line of Eq.~(\ref{eq:transfb}), for which the steady-state phase diagram can be easily computed (see SI~\cite{SI}). This phase diagram, shown in $(\Delta_1,\CNL)$-space in the center panel of Fig.~\ref{fig:singleQRCCumulants}(a), is the well-known result for a coherently-driven classical Kerr oscillator~\cite{dykman_theory_1979, dykman_fluctuating_2012}: the orange region depicts the classical bistability, which emerges for sufficiently negative detuning given our choice of the sign of the Kerr nonlinearity.


\begin{figure}[t]
    \centering
    \includegraphics[scale=1.0]{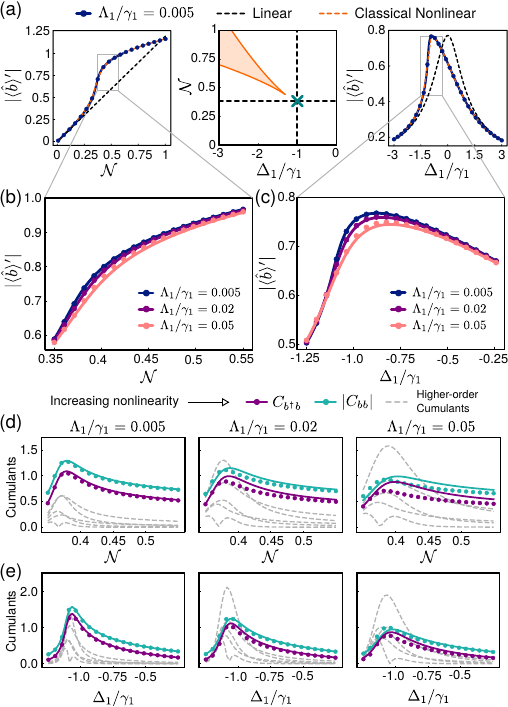}
    \caption{Classical and quantum dynamics of a single Kerr \qrc{} mode. In this plot, colored solid curves are exact complex-$\mathcal{P}$ results, colored dots are TEOMs simulations in the long-time limit, and dashed orange curves are classical results. Other curves are specified when relevant. (a) Center panel: \textit{Classical} phase diagram of the single coherently-driven Kerr \qrc{} mode in ($\Delta_1,\CNL$) space. Left panel: \qrc{} amplitude as a function of $\CNL$ for fixed detuning $\Delta_1/\gamma_1 = -1.0$ (across indicated vertical cross-section of the phase diagram). Note that the total damping rate ${\gamma} =\gamma_1+\Gamma_1$, and $\Gamma_1 = \gamma_1/2$. Right panel: \qrc{} amplitude as a function of detuning $\Delta_1$ for fixed $\CNL = 0.385$ (across indicated horizontal cross-section of the phase diagram). Dashed black lines in both plots indicate the response for a linear oscillator. (b) Steady-state scaled amplitude $|\avg{\hat{b}'}|$ for nonlinearity strengths $\Lambda_1/\gamma_1 \in [0.005,0.02,0.05]$, as a function of $\CNL$ for fixed $\Delta_1/\gamma_1 = -1.0$ (region within the gray box in (a), left panel). (c) Same as (b) but as a function of detuning $\Delta_1$ for fixed $\CNL = 0.385$ (region within the gray box in (a), right panel). (d) Steady-state second-order cumulants ($C_{b^{\dagger}b}$, purple, and $|C_{bb}|$, green) and higher (third, fourth) order cumulants (dashed gray) across the cross-section in (b); nonlinearity increases from left to right. (e) Same as (d) but now for the cross-section in (c).
    }
    \label{fig:singleQRCCumulants}
\end{figure}


This bistability is a feature specific to nonlinear systems; in its vicinity, we can expect an enhanced nonlinear response of the \qrc{} observables to \qrc{} inputs. This is explicitly observed in the scaled field amplitude $|\avg{\hat{b}}'|$, plotted in Fig.~\ref{fig:singleQRCCumulants}(b) as a function of $\CNL$ for fixed $\Delta/\gamma_1 = -1.0$ (that is, across the indicated vertical dashed line of the classical phase diagram). The solid blue curve is the exact complex-$P$ result for $\Lambda/\gamma_1 = 0.005$, while blue dots are obtained by evolving the TEOMs, Eqs.~(\ref{eq:b}),~(\ref{eq:CbdbFull}),~(\ref{eq:CbbFull}), to their steady state numerically. For small $\CNL$, the response follows that of a linear oscillator (dashed black line), but becomes nonlinear with increasing $\CNL$. Also shown in dashed orange is the response of the classical nonlinear oscillator (explicitly dropping second-order cumulants), which does not depend on the actual nonlinearity strength but only on $\CNL$. 

The influence of nonlinearity is similarly evident when plotting $|\avg{\hat{b}'}|$ as a function of $\Delta$, for fixed $\CNL=0.385$ in Fig.~\ref{fig:singleQRCCumulants}(c) (across the horizontal dashed line of the classical phase diagram). The black dashed curve indicates the standard Lorentzian response of a linear oscillator, centered around the bare frequency $\omega_b$. The response of the nonlinear \qrc{} mode is clearly that of a Lorentzian deformed towards lower frequencies due to the Kerr-induced frequency shift. 

In both results, we find excellent agreement between the exact quantum results, the TEOMs, \textit{and} classical nonlinear dynamics. This is indicative of operating parameters where not only is a truncated description using first and second-order cumulants sufficient, but the influence of second-order cumulants on first-order cumulants is negligible as well. Consequently, we expect the \qrc{} mode here to function as a classical reservoir mode, still allowing for nonlinear processing of information encoded in the drive amplitude, phase, and/or frequency.

If we now keep $\CNL$ fixed but increase the nonlinearity strength $\Lambda$, the \qrc{} can be moved to a quantum regime of operation. We plot the \qrc{} response as a function of $\CNL$ for $\Delta/\gamma_1 = -1.0$ in Fig.~\ref{fig:singleQRCCumulants}(b), and as a function of $\Delta$ for $\CNL = 0.385$ in Fig.~\ref{fig:singleQRCCumulants}(c), for three different nonlinearity strengths $\Lambda/\gamma_1 \in [0.005,0.02,0.05]$. In both cases, we see that with increasing nonlinearity, the response deviates from that of the classical nonlinear oscillator (which recall is essentially equivalent to the quantum result for weak $\Lambda/\gamma_1 = 0.005$). From Eq.~(\ref{eq:transfb}), it is clear that this deviation is due to the coupling of first-order moments to second-order cumulants; this is precisely the effect accounted for by the TEOMs, which thus agree with the full quantum result.

Also shown in Figs.~\ref{fig:singleQRCCumulants}(d) and (e) are steady-state cumulants plotted across the same cross-sections as Figs.~\ref{fig:singleQRCCumulants}(b) and (c) respectively, with panels from left to right indicating stronger nonlinearities. Second-order cumulants are plotted in green and purple (lines are exact complex-$\mathcal{P}$ results, dots are TEOMs); they clearly display non-monotonic behaviour as a function of $\CNL$ and $\Delta$, acquiring large magnitudes for specific operating parameters in the vicinity of the classical bistability. Physically, second-order cumulants are related to the maximum and minimum (dimensionless) quadrature variances of the \qrc{} modes: $\frac{1}{2} + C_{b^{\dagger}b} \pm |C_{bb}|$ respectively, and their magnitudes are thus measures of amplification and squeezing of quantum fluctuations due to the underlying Kerr nonlinearity. 

Third and fourth order cumulants are shown in the same plots in dashed gray (exact complex-$\mathcal{P}$ only).  There are two features of note. Firstly, the overall magnitude of these higher-order cumulants relative to second-order cumulants appears to increase for stronger nonlinearities. This is an indicator of the emergence of non-Gaussian features with increasing $\Lambda$, as expected, and generally leads to greater deviation between TEOMs and exact quantum results for first and second-order cumulants. Secondly, even for a fixed nonlinearity strength, the magnitude of cumulants depends on operating parameters ($\CNL$ and $\Delta$) that lead to operation in the vicinity of the classical bistability. Hence, while results obtained from TEOMs for such operating parameters can deviate from the exact quantum solution, away from these regions the agreement improves, even for strong nonlinearities.

\subsection{Numerical tests: conditional dynamics}
\label{si:condQRCdyn}

We have so far used steady-state, unconditional quantities to analyze the nonlinear processing capabilities of a single Kerr-based \qrc{} mode, and identified parameter regimes where an approach based on truncated EOMs is valid. However, to allow for processing of time-dependent information by the \qrc{}, and to analyze the output from the \qrc{} obtained as \textit{individual} measurement records, an approach is needed that is able to accurately capture the \textit{conditional, dynamical} evolution of the \qrc{} mode. This is described by the SME
\begin{align}
    d\rhoc = \mathcal{L}\rhoc~dt + \mathcal{S}[\sqrt{\gamma_1}\hat{b}]\rhoc
    \label{eq:singleQRCSME}
\end{align}
where $\mathcal{S}[\sqrt{\gamma_1}\hat{b}]$ is defined as,
\begin{align}
&\mathcal{S}[\sqrt{\gamma_1}\hat{b}_k]\rhoc = \sqrt{\frac{\gamma_1}{2}}\!\left( \hat{b}_k \rhoc + \rhoc \hat{b}_k^{\dagger} - \avgc{\hat{b}_k+\hat{b}_k^{\dagger}} \right)\! dW_{\mathcal{I}_k}\!(t)~ \nonumber \\
&~~~~~~+\sqrt{\frac{\gamma_1}{2}}\!\left( -i\hat{b}_k \rhoc + i\rhoc \hat{b}_k^{\dagger} - \avgc{-i\hat{b}_k+i\hat{b}_k^{\dagger}} \right) \!dW_{\mathcal{Q}_k}\!(t) 
\end{align}
and our current system assumes the $K=1$ case. Note the slight differences to Eq.~(\ref{eq:Smeas}) of the main text, which are only for purposes of exposition: the dissipator in Eq.~(\ref{eq:Smeas}) is absorbed into $\mathcal{L}$ here, to aid the discussion in the previous subsection. 

 Expectation values with respect to the conditional quantum state under measurement $\rhoc$ are now conditional, which we indicate by the superscript $c$. For example, the \ch{variation} $d\avgc{\hat{b}}$ \ch{in a time $dt$} is now given by ${\rm tr }\{(\mathcal{L}~dt+\mathcal{S}[\hat{b})]\rhoc\hat{b} \}$, which takes the form (as derived in Sec.~\ref{si:stochcumulants} of the SI)
\begin{align}
    &d\avgc{\hat{b}} = {\rm tr }\{(\mathcal{L}\rhoc\hat{b} \}~dt + \nonumber \\
    &\sqrt{\frac{\gamma_1}{2}} \left(\Cc{b^{\dagger}b}+\Cc{bb} \right)dW^X(t) + i\sqrt{\frac{\gamma_1}{2}}\left(\Cc{b^{\dagger}b}-\Cc{bb} \right)dW^P(t). 
    \label{eq:bstoch}
\end{align}
The first line $\propto \mathcal{L}$ simply includes terms from Eq.~(\ref{eq:b}), with expectation values replaced by their conditional counterparts. Terms in the second line are due to the measurement superoperator, and render the equation of motion for $\avgc{\hat{b}}$ \textit{stochastic}, conditioned on the trajectory-specific realizations of $dW^{X,P}(t)$. Dropping cumulants higher than second order in Eq.~(\ref{eq:bstoch}), arising in the first line, yields the \textit{stochastic} truncated EOM (STEOM) for $\avgc{\hat{b}}$. 

We can similarly obtain truncated equations of motion for the conditional cumulants $\Cc{bb}$, $\Cc{b^{\dagger}b}$ under heterodyne measurement (as derived in Sec.~\ref{si:stochcumulants} of the SI):
\begin{subequations}
\begin{align}
    d\Cc{b^{\dagger}b} &=  \left[{\rm tr }\{\mathcal{L}\rhoc\hat{b}^{\dagger}\hat{b} \}\! - \!\avgc{\hat{b}^{\dagger}}{\rm tr }\{\mathcal{L}\rhoc\hat{b} \}\! - \!\avgc{\hat{b}}{\rm tr }\{\mathcal{L}\rhoc\hat{b}^{\dagger} \}\right]\! dt \nonumber \\
    &~~~~-{\gamma_1}\left[ (\Cc{b^{\dagger}b})^2 + \Cc{bb}(\Cc{bb})^* \right]~dt, \label{eq:Cbbstoch} \\
    d\Cc{bb} &=  \left[ {\rm tr }\{\mathcal{L}\rhoc\hat{b}\hat{b} \} - 2\avgc{\hat{b}}{\rm tr }\{\mathcal{L}\rhoc\hat{b} \}\right]dt \nonumber \\
    &~~~~-2{\gamma_1}\Cc{bb}\Cc{b^{\dagger}b}~dt. \label{eq:Cbdbstoch}
\end{align}
\end{subequations}
Again, terms $\propto \mathcal{L}$ are as found in Eqs.~(\ref{eq:CbdbFull}),~(\ref{eq:CbbFull}) post-truncation, with expectation values replaced by their conditional counterparts. The second line of each equation describes the evolution due to measurement, which at first glance appears deterministic: note that Wiener increments $dW^{X,P}(t)$ make no appearance. For \textit{linear} quantum systems under continuous weak measurement, this is in fact the case: second-order cumulants form a closed set described by the above equations, and no stochastic terms arise~\cite{breslin_conditional_1997, doherty_feedback_1999, cernotik_adiabatic_2015, zhang_prediction_2017}. However, for \textit{nonlinear} quantum systems of interest here, second-order cumulants can couple to the \textit{stochastic} first-order moments (here, via terms $\propto \mathcal{L}$), rendering the conditional evolution of second-order cumulants generally stochastic as well.


\begin{figure}
    \centering
    \includegraphics[scale=1.0]{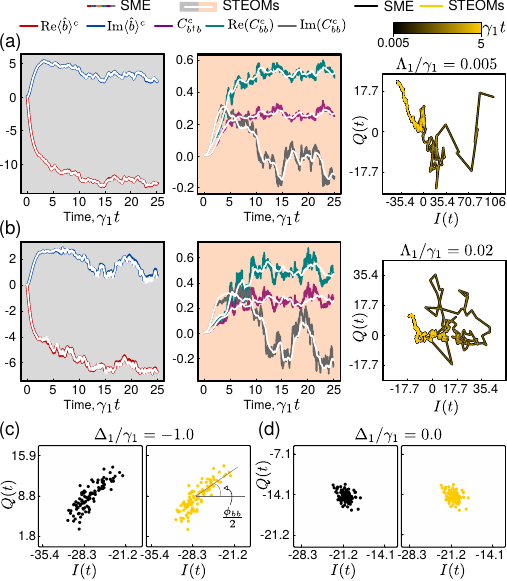}
    \caption{Benchmarking STEOMs against full SME simulations. Simulated parameters are as marked by the teal `X' in the phase diagram of Fig.~\ref{fig:singleQRCCumulants}(a): $\CNL = 0.385$, $\Delta_1/\gamma_1 =-1.0$, for (a) $\Lambda_1/\gamma_1 = 0.005$ and (b) $\Lambda_1/\gamma_1 = 0.02$. Note that the total damping rate ${\gamma} =\gamma_1+\Gamma_1$, and $\Gamma_1 = \gamma_1/2$. Left panel shows real and imaginary parts of the conditional expectation value $\avg{\hat{b}}$, and center panel shows conditional second-order cumulants $C_{bb}, C_{b^{\dagger}b}$; colored (white) curves are SME (STEOMs) simulations. Right panel shows measured quadratures $I(t)$, $Q(t)$: the SME result is plotted in solid black, while the STEOMs result is plotted in yellow, with a lighter shade indicating earlier times. (c) Measured quadratures $\{I(t),Q(t)\}$ at $\gamma t= 20.0$ obtained using SME (left panel) and STEOMs (right panel), for $\CNL = 0.385$, $\Lambda_1/\gamma_1 = 0.02$, and $\Delta_1/\gamma_1 = -1.0$. (d) Same as (c) but for $\Delta_1/\gamma_1 = 0.0$. }
    \label{fig:singleQRCSim}
\end{figure}


Eqs.~(\ref{eq:bstoch}),~(\ref{eq:Cbbstoch}),~(\ref{eq:Cbdbstoch}) define the STEOMs for the single Kerr \qrc{} mode \ch{under continuous heterodyne monitoring}. To assess their validity, we compare their simulation results against integration of the SME, Eq.~(\ref{eq:singleQRCSME}). For concreteness, we consider the operating point marked by the teal `X' in the phase diagram of Fig.~\ref{fig:singleQRCCumulants}(a): $\CNL = 0.385$, $\Delta/\gamma =-1.0$. Note that the \textit{same noise realizations} $dW^{X,P}(t)$ must always be used for both simulations to ensure the compared trajectories are conditioned on the same measurement record. Results are included in Fig.~\ref{fig:singleQRCSim}(a),~(b) for $\Lambda/\gamma \in [0.005,0.02]$ respectively, showing first-order moments (left panel) and second-order cumulants (center panel), with STEOMs results in white, and SME results obtained using QuTiP~\cite{johansson_qutip_2013} in color. Both methods agree very well, especially for first-order cumulants. In particular, note the stochastic nature of second-order cumulants, which is a direct signature of the nonlinearity of the quantum system under study, and is further enhanced at strong nonlinearities. However, for the same temporal resolution, SME simulations with a Hilbert space cutoff of 625 photons (around 3 times larger than the maximum occupation number, a typical conservative value) demand several orders of magnitude more simulation time than the STEOMs.

However, in a real experiment, such information about the \qrc{} state must be extracted from measurement records. For the case of heterodyne measurement of the single Kerr oscillator, the obtained records $\mathcal{I}(t)\equiv \mathcal{I}_1(t),\mathcal{Q}(t) \equiv \mathcal{Q}_1(t)$ are defined in Eqs.~(\ref{eq:IQkraw}) of the main text (here we take $\bar{n}_{\rm cl} = 0$, ignoring the role of classical readout noise). These records are typically processed to reduce noise, most commonly via temporal filtering; the processed records define \textit{measured quadratures} as in the main text,
\begin{align}
    I(t) = \frac{1}{\sqrt{2\mathcal{T}} }\int_{t-\mathcal{T}}^{t} d\tau~\mathcal{I}(\tau);~Q(t) = \frac{1}{\sqrt{2\mathcal{T}} }\int_{t-\mathcal{T}}^{t} d\tau~\mathcal{Q}(\tau).
    \label{eq:IXIP} 
\end{align}
which here describe processing of single-shot readout records. In this section, we take $t-\mathcal{T}=0$ for filtering over the entire collection window. Measured quadratures obtained using a single set of measurement records are shown in the measured phase space given by $(I(t),Q(t))$ in the right panel of Fig.~\ref{fig:singleQRCSim}(a),~(b) $\Lambda/\gamma \in [0.005,0.02]$ as before; SME results are shown in solid black, with the STEOMs results plotted on top, using a colorscale going from black to yellow indicating increasing time. Both show excellent agreement; this is unsurprising since ${I}(t),{Q}(t)$ yield information about the underlying conditional dynamics of $\avgc{\hat{b}}$, which also agree very well between the two methods.

Note that the measured quadratures are themselves stochastic quantities, whose statistics are correlated with the underlying measured system state. This can be seen via the measured quadrature distributions in Fig.~\ref{fig:singleQRCSim}(c), obtained by simulating several measurement records (here $100$ in total) and plotting $\{I(t),Q(t)\}$ at $\gamma t = 20.0$, for $\Delta_1/\gamma_1 = -1.0$ as before. At this operating point near the classical bistable region, the Kerr nonlinearity amplifies the magnitude of cumulants $C_{bb} \equiv |C_{bb}|e^{i\phi_{bb}}$~[see Fig.~\ref{fig:singleQRCCumulants}~(d)] leading to squeezing of the internal \qrc{} field along the axis determined by $\phi_{bb}/2$ (and amplification along the orthogonal quadrature). This internal \qrc{} squeezing manifests as squeezing of the measured quadrature distribution; the squeezing axis is unchanged since the temporal filter defining measured quadratures via Eq.~(\ref{eq:IXIP}) is quadrature-agnostic, and thus only reduces the overall noise power while preserving its relative strength amongst measured quadratures. On the other hand, for $\Delta_1/\gamma_1 = 0.0$, further away from the bistable region, $C_{bb}$ is much smaller in magnitude and the measured quadrature distribution in Fig.~\ref{fig:singleQRCSim}(d) exhibits no squeezing. Excellent agreement between SME (left) and STEOMs (right) is observed for both cases. 

\subsection{Non-Gaussian signatures captured by STEOM simulations}
\label{si:nongaussian}

The (S)TEOMs approach to nonlinear quantum systems in this work uses $n_{\rm trunc} = 2$, so that only cumulants of up to second-order are retained. Gaussian states have at most second-order cumulants as well. However, the TEOMs approach is distinct from other methods of analyzing nonlinear quantum systems where fluctuations are treated within a Gaussian approximation, such as standard linearization. There are two important indicators of this difference. First is the deviation of first-order cumulants from their classical values, observed in Fig.~\ref{fig:singleQRCCumulants}, which would not happen in standard linearization schemes. The second indicator appears in stochastic dynamics of second-order cumulants, as seen in Fig.~\ref{fig:singleQRCSim}, which would be perfectly deterministic for a quantum system whose state is defined \textit{exactly} by a Gaussian description.

These features indicate the ability of (S)TEOMs to simulate dynamics beyond those of just Gaussian states. In fact, we find that the STEOMs can exhibit non-Gaussian characteristics in measured heterodyne outputs that are also captured by SME simulations. An example is depicted in Fig.~\ref{fig:nonGaussian}, where we show measured distributions obtained using the SME and STEOMs for a Kerr oscillator operated in the vicinity of the classical Kerr bistability (for parameters, see caption). Both measured distributions clearly display `crescent'-shaped deformations typical of non-Gaussian characteristics observed in transient dynamics of Kerr-type nonlinear oscillators~\cite{boutin_effect_2017}.


\begin{figure}[t]
    \centering
    \includegraphics[scale=1.0]{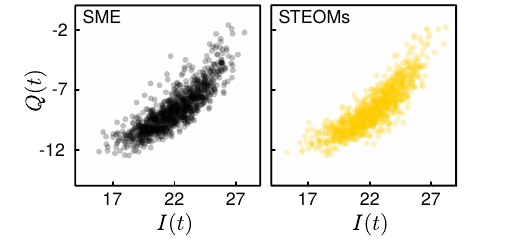}
    \caption{Non-Gaussian signatures in single Kerr oscillator heterodyne measurements. Measured quadratures $\{I(t),Q(t)\}$ at $\gamma_1 t= 60.0$ obtained using SME (left panel) and STEOMs (right panel), for $\CNL = 0.369$, $\Lambda_1/\gamma_1 = 0.0295$, and $\Delta_1/\gamma_1 = -1.0$. Note that the total damping rate ${\gamma} =\gamma_1+\Gamma_1$, and $\Gamma_1 = \gamma_1/2$.   }
    \label{fig:nonGaussian}
\end{figure}


\subsection{Validity and scalability of (S)TEOMs}

The key observations from our analysis of Kerr \qrc{} operating regimes and the generation of higher-order cumulants (summarized in Fig.~\ref{fig:singleQRCCumulants}) are twofold. Primarily, the magnitude of nonzero higher-order cumulants increases with the strength of the Kerr nonlinearity relative to the \qrc{} mode damping rate, $\Lambda/{\gamma}$ (where $\gamma$ is the total \qrc{} mode loss rate), which coincides with the \qrc{} under coherent driving transitioning from classical to quantum regimes. However, operating regimes in the vicinity of the classical bistability also lead to an enhancement in higher-order cumulants, regardless of nonlinearity strength. To be confident of the validity of the truncated cumulants approach, we therefore analyze \qrcs{} with nonlinearity strengths $\Lambda/{\gamma} \leq 0.02$, and operate \textit{near} but not within the classically-bistable region. 

This truncated cumulants approach provides a highly efficient mathematical description when applied to multimode quantum systems. For a measurement chain comprising an $M$-mode quantum system coupled to a $K$-mode \qrc{}, the composite system of $\ntot=M+K$ total quantum modes is described by $2\ntot^2 + 3\ntot$ unknowns using (S)TEOMs, scaling quadratically with $\ntot$ instead of the exponential growth in required Hilbert space size for full (S)ME simulations. This description also places no constraints on modal occupation numbers, enabling our exploration of a well-defined classical limit of the coherently-driven Kerr \qrc{}. Our (S)TEOMs approach is built to be a scalable theoretical framework via an efficient computer-algebra implementation, which allows the calculation of equations of motion of retained cumulants up to order $n_{\rm trunc}$ (here, second-order) for nonlinear quantum systems comprising arbitrary numbers of bosonic modes under continuous measurement. Supplementary benchmarking simulations for systems with more than one mode are included in Sec.~\ref{si:verify} of the SI. In the main text, we employ (S)TEOMs to simulate dynamics of measurement chains comprising measured quantum systems coupled to multimode \qrcs{}, while operating within the aforementioned constraints.

\newpage

\newpage

\section{Quantum properties of the quantum system (QS) states to be distinguished}
\label{app:qs}

In this section we analyze the quantum system (QS) defined by Eq.~(\ref{eq:hsysamp}) of the main text, whose states we wish to distinguish in classification tasks. The TEOMs that describe the dynamics of the QS are simply given by the QS sector of Eqs.~(\ref{eq:teoma}),~(\ref{eq:lyapunov}), and are \textit{exact} for $n_{\rm trunc} = 2$, owing to the QS linearity under coherent driving, and its non-reciprocal coupling to the \qrc{}. For convenience, we reproduce the TEOMs below,
\begin{subequations}
\begin{align}
    \frac{d}{dt}\avg{\hat{\bm{a}}} &= \mathbf{L}_a\avg{\hat{\bm{a}}} + \vec{\eta}, \label{appeq:QS1} \\
    \frac{d}{dt}\mathbf{C}_a &= \mathbf{L}_a\mathbf{C}_a + \mathbf{C}_a\mathbf{L}_a^T +  \mathbf{D}_a, \label{appeq:QS2}
\end{align}  
\end{subequations}
where recall that $\hat{\bm{a}} \equiv (\hat{a}_1,\hat{a}_1^{\dagger},\ldots,\hat{a}_M,\hat{a}_M^{\dagger})^T$ and $\mathbf{C}_a = \langle:\!\hat{\bm{a}}\hat{\bm{a}}^T\!\!:\rangle - \avg{ \hat{\bm{a}}}\avg{\hat{\bm{a}}^T} $. Crucially, we emphasize that the above TEOMs describe the unconditional dynamics of QS state variables \textit{irrespective} of the \qrc{}'s precise nonlinearity, due to the non-reciprocal coupling which ensures the QS drives the \qrc{}, but not vice-versa.

In what follows, we specialize to the $M=2$ mode system and its specified parameters from Sec.~\ref{subsec:taskdef}, namely $G_2 = 0$, $\phi_{1}=\frac{\pi}{2}$, and $\Gamma_m + \kappa_m \equiv \kappa~\forall~m$. Then, the drift matrix $\mathbf{L}_a$ and  diffusion matrix $\mathbf{D}_a$ defining Eqs.~(\ref{appeq:QS1}),~(\ref{appeq:QS2}) take the respective forms:
\begin{subequations}
 \begin{align}
    \mathbf{L}_a &= \!
    \begin{pmatrix}
    -\frac{\kappa}{2} & -iG_1 & 0 & -i e^{i\phi_{12}}G_{12} \\
    -iG_1 & -\frac{\kappa}{2} & i e^{-i\phi_{12}}G_{12} & 0 \\
    0 & -i e^{i\phi_{12}}G_{12} & -\frac{\kappa}{2} & 0 \\
    i e^{-i\phi_{12}}G_{12} & 0 & 0 & -\frac{\kappa}{2}
    \end{pmatrix}, \\
    \mathbf{D}_a &= \!
    \begin{pmatrix}
     -i G_1 & 0 & -i e^{i\phi_{12}}G_{12} & 0 \\
     0 & i G_1 & 0 & i e^{-i\phi_{12}}G_{12} \\
     -i e^{i\phi_{12}}G_{12} & 0 & 0 & 0 \\
     0 & i e^{-i\phi_{12}}G_{12} & 0 & 0 
    \end{pmatrix},
\end{align}   
\end{subequations}
while the coherent drive vector is $\vec{\eta} = (\eta_1,\eta_1,\eta_2,\eta_2)^T$.

The linearity of Eqs.~(\ref{appeq:QS1}),~(\ref{appeq:QS2}) renders them straightforward to solve. In particular, as we are interested in classification in the long-time limit, we simply evaluate the steady-state first-order cumulants describing the quantum system by setting $\frac{d}{dt}\avg{\hat{\bm{a}}} = 0$ in Eq.~(\ref{appeq:QS1}), obtaining
\begin{align}
    \avg{\hat{\bm{a}}} = -\mathbf{L}_a^{-1} \vec{\eta}.
    \label{appeq:qssol1}
\end{align}
Similarly, second-order cumulants in the long-time limit can be obtained by setting $\frac{d}{dt}\mathbf{C}_a = 0$ in Eq.~(\ref{appeq:QS2}); the steady state value of $\mathbf{C}_a$ then satisfies the continuous Lyapunov equation,
\begin{align}
    \mathbf{L}_a\mathbf{C}_a +\mathbf{C}_a\mathbf{L}_a^T + \mathbf{D}_a = 0.
\end{align}

In lieu of writing down the exact expressions for steady-state cumulants of the quantum system in complete generality -- which can be unwieldy -- we analyze the specific states indexed by $l$ and summarized in Fig.~\ref{fig:classifyAmpOnly} that we wish to distinguish in classification tasks. Starting with Task~1, recall that $\Gamma_2 = 0$, so that $\hat{a}_2$ is not read out, and is consequently also not coupled to the \qrc{}. Hence the value of first-order cumulants defining its quantum states does not contribution to classification. The first-order cumulants for $\hat{a}_1$ are then given by:
\begin{align}
    \avg{\hat{a}_1^{(1)}} = -\frac{2\eta_1^{(1)}}{\kappa-2G^{(1)}_1},~\avg{\hat{a}_1^{(2)}} = -\frac{2\eta_1^{(2)}}{\kappa^2-4(G^{(2)}_{12})^2}
\end{align}
By suitable choices of $\eta_1^{(1,2)}$, an example of which is provided in Fig.~\ref{fig:classifyAmpOnly}(b), we can realize states for which $\avg{\hat{a}_1^{(1)}}=\avg{\hat{a}_1^{(2)}} = \amp$. However, the realized states do differ in their second-order cumulants, which take the forms:
   \begin{align}
    \mathbf{C}^{(1)}_a = 
    \frac{1}{\kappa^2-4(G^{(1)}_1)^2}
    \begin{pmatrix}
        \kappa G^{(1)}_1 & 2(G^{(1)}_1)^2 & 0 & 0 \\
        2(G^{(1)}_1)^2 & \kappa G^{(1)}_1 & 0 & 0 \\
        0 & 0 & 0 & 0 \\
        0 & 0 & 0 & 0 
    \end{pmatrix},~~\mathbf{C}^{(2)}_a = 
    \frac{1}{\kappa^2-4(G^{(2)}_{12})^2}
    \begin{pmatrix}
        0 & 2(G^{(2)}_{12})^2 & -i\kappa G^{(2)}_{12} & 0 \\
        2(G^{(2)}_{12})^2 & 0 & 0 & i\kappa G^{(2)}_{12} \\
        -i\kappa G^{(2)}_{12} & 0 & 0 & 2(G^{(2)}_{12})^2 \\
        0 & i\kappa G^{(2)}_{12} & 2(G^{(2)}_{12})^2 & 0 
    \end{pmatrix}
\end{align} 
It is clear that the structure of covariance matrices is very distinct, in particular as seen in the top-left sector corresponding to mode $\hat{a}_1$, which is the only mode read out in Task~1. 

For Task 2, we set $\Gamma_2 = \Gamma_1 = \kappa/2$ so that both system modes $\hat{a}_1$ and $\hat{a}_2$ are read out. In this case, the first-order cumulants are given by
\begin{subequations}
\begin{align}
    \avg{\hat{a}^{(3,4)}_1} &= -\frac{2(\kappa \eta^{(3,4)}_1 - i 2G^{(l)}_{12}\eta^{(3,4)}_2 e^{i\phi^{(l)}_{12}}) }{\kappa^2-4(G^{(l)}_{12}))^2}  \\
    \avg{\hat{a}^{(3,4)}_2} &= -\frac{2(\kappa \eta^{(3,4)}_2 - i 2G^{(l)}_{12}\eta^{(3,4)}_1 e^{i\phi^{(l)}_{12}}) }{\kappa^2-4(G^{(l)}_{12}))^2}
\end{align}  
\end{subequations}

Again, by appropriate choice of $\eta_1^{(3,4)}$, an example of which is provided in Fig.~\ref{fig:classifyAmpOnly}(c), we can realize states for which $\avg{\hat{a}_{1,2}^{(3)}}=\avg{\hat{a}_{1,2}^{(4)}} = \amp$. The second-order cumulants are now given by
\begin{align}
    \mathbf{C}^{(3,4)}_a = 
    \frac{1}{\kappa^2-4(G^{(3,4)}_{12})^2}
    \begin{pmatrix}
        0 & 2(G^{(3,4)}_{12})^2 & -i\kappa G^{(3,4)}_{12}e^{i\phi_{12}^{(3,4)}} & 0 \\
        2(G^{(3,4)}_{12})^2 & 0 & 0 & i\kappa G^{(3,4)}_{12}e^{-i\phi_{12}^{(3,4)}} \\
        -i\kappa G^{(3,4)}_{12}e^{i\phi_{12}^{(3,4)}} & 0 & 0 & 2(G^{(3,4)}_{12})^2 \\
        0 & i\kappa G^{(3,4)}_{12}e^{-i\phi_{12}^{(3,4)}} & 2(G^{(3,4)}_{12})^2 & 0 
    \end{pmatrix}
\end{align} 
Note that states $l=3,4$ differ only in their two-mode interaction phase $\phi_{12}$. Furthermore, this appears only in the off-diagonal sector of the covariance matrix, and therefore is a property of the joint cumulants of the quantum system. 

The above completes our characterization of the QS states we wish to classify in Tasks~I and II considered in the main text. In a measurement chain where the signals from the QS are processed by a downstream \qrc{}, the QS state determines the \qrc{} state; this relationship is quantified in SI Sec.~\ref{app:lin}.

\newpage

\section{Nonlinear van Kampen expansion of the Fokker-Planck equation describing the complete measurement chain}
\label{app:lin}

The (S)TEOMs, an example of which was derived in SI Sec.~\ref{si:singleNodeQRC}, provide a useful approximate numerical scheme to simulate the dynamics of measurement chains with \qrcs{}. However the close connection between quantum cumulants that underpin the (S)TEOMs framework and the positive-$P$ representation (discussed in SI Sec.~\ref{app:cumulants}) allows us to further develop an approximate \textit{analytic} approach to describing the dynamics of \qrcs{} under quantum input signals. Arriving at this analytic description is the subject of the present section. 

\subsection{Fokker-Planck equation in the Positive-$P$ representation}




Our approach will begin by considering an alternative representation of the general quantum state $\rhou$ of the complete measurement chain: the positive-$P$ phase space representation~\cite{carmichael_statistical_2002}. The positive-$P$ representation is obtained via a projection of $\rhou$ onto a non-diagonal basis of coherent state projectors:
\begin{align}
    \rhou(t) = \int d\vec{Z}~\mathcal{P}(\vec{Z},t) \bigotimes_i \frac{ \ket{\alpha_i}\langle\alpha_i^{\dagger *}|}{e^{\alpha_i\alpha_i^{\dagger} }} \bigotimes_j \frac{ \ket{\beta_j}\langle\beta_j^{\dagger *}| }{e^{\beta_j\beta_j^{\dagger} }}
    \label{appeq:prepdef}
\end{align}
The positive-$P$ representation is defined in a phase space spanned by $\mathbb{C}$-number variables $\vec{Z} = (\alpha_1,\alpha_1^{\dagger},\ldots,\alpha_M,\alpha_M^{\dagger},\beta_1,\beta_1^{\dagger},\ldots,\beta_K,\beta_K^{\dagger})^T$, which are associated with the respective quantum operators $\hat{\bm{z}} = (\hat{a}_1,\hat{a}_1^{\dagger},\ldots,\hat{a}_M,\hat{a}_M^{\dagger},\hat{b}_1,\hat{b}_1^{\dagger},\ldots,\hat{b}_K,\hat{b}_K^{\dagger})^T$. The computational importance of this correspondence is that it enables \textit{normal-ordered} expectation values with respect to the quantum state to be computed by instead evaluating averages of phase space variables with respect to the positive-$P$ distribution:
\begin{align}
   {\rm tr}\{ (:\hat{z}_i\cdots\hat{z}_k:) \rhou(t) \} = \avg{:\hat{z}_i\cdots\hat{z}_k:} = \avg{Z_i\cdots Z_k} = \int d\vec{\zeta}~(Z_i\cdots Z_k)\mathcal{P}(\vec{Z},t)
   \label{appeq:fpeexpdef}
\end{align}
Evaluating expectation values via averages computed against the positive-$P$ representation can in many cases be extremely efficient, circumventing the Hilbert space restrictions that constrain evolution of $\rhou$.

However, using Eq.~(\ref{appeq:fpeexpdef}) requires knowledge of the positive-$P$ representation $\mathcal{P}(\vec{Z},t)$, which in turn requires a governing equation for $\mathcal{P}(\vec{Z},t)$ that could formally be solved for its time evolution and steady-state, the analog to the master equation for the quantum state $\rhou$. Using Eq.~(\ref{appeq:prepdef}), the master equation can be mapped to this desired equation, which for a large class of bosonic nonlinear systems takes the form of a Fokker-Planck equation (FPE). The general form of an FPE for the positive-$P$ representation $\mathcal{P}$ is given by: 
\begin{align}
    d_{\tau} \mathcal{P}(\vec{Z},\tau) = \left( -\frac{\partial}{\partial Z_i} {A}_i + \frac{1}{2}\frac{\partial^2}{\partial Z_i \partial Z_j}\mathbf{D}_{ij} \right) \! \mathcal{P}(\vec{Z},\tau)
    \label{appeq:fpefull}
\end{align}
where we have for convenience also introduced dimensionless time and nonlinearity, 
\begin{align}
    \tau = \gamma t,~\tbar{\Lambda}_k = \Lambda_k/\gamma
\end{align}
where $\gamma$ is the common monitored channel loss rate of all \qrc{} modes, $\gamma_k = \gamma~\forall~k$. Then the dynamics of the FPE are entirely governed by the dimensionless drift vector $\vec{A}$ with elements ${A}_i$, and the dimensionless diffusion matrix $\mathbf{D}$ with elements $\mathbf{D}_{ij}$; repeated indices are summed over. Both $\vec{A}$ and $\mathbf{D}$ are completely defined by the underlying master equation governing $\rhou$, from the transformation of which the FPE is obtained. 

For a measurement chain of comprising multiple, possibly nonlinear bosonic modes as described by Eq.~(\ref{eq:sme}), we can further write general forms for the drift vector and diffusion matrix:
\begin{subequations}
\begin{align}
    \vec{A}(\vec{Z}) &= \mathbf{A}^L \vec{Z} + \tbar{\bm{\Lambda}}\vec{A}^N\!(\vec{Z}) + \vec{f}, \label{appeq:Af} \\
    \mathbf{D}(\vec{Z}) &= \mathbf{D}^L + \tbar{\bm{\Lambda}}\mathbf{D}^N\!(\vec{Z}). \label{appeq:Df}
\end{align}
\end{subequations}
Here, the matrices $\mathbf{A}^L$ and $\mathbf{D}^L$ define the respective and sole contributions to the drift and diffusion terms of Eq.~(\ref{appeq:fpefull}) if all bosonic modes in the measurement chain were linear. The inhomogeneous terms $\vec{f}$ further includes the entire effect of coherent driving. Then, $\tbar{\bm{\Lambda}}$ is a matrix describing the nonlinearity; here it is a diagonal (due to the local nature of the nonlinearities) matrix of the \qrc{} mode Kerr nonlinearity strengths, and hence only vanishes if $\tbar{\Lambda}_k = 0~\forall~k$. The vector $\vec{A}^N$ and matrix $\mathbf{D}^N$ then respectively describe the nonlinear contributions to the drift and diffusion terms respectively; these have a more complex dependence on the phase space variables $\vec{Z}$ than their corresponding linear counterparts do, and will be the focus of our analysis here. 

In what follows, we will assume $\tbar{\Lambda}_k \equiv \tbar{\Lambda}~\forall~k$, the case considered in the main text. This is not necessary for the remainder of the calculation, or for \qrcs{} in general, but enables us to simplify our notation by foregoing a matrix representation of the nonlinearity. We thus let $\tbar{\bm{\Lambda}} \to \tbar{\Lambda}$ from here on.

We next introduce a displacement of the phase space variables $\vec{Z}$, around a quantum state of the measurement chain where the phase space variables take the expectation values $\avg{\vec{\tbar{Z}}(t)}$. Formally, we write the original phase space variables in the form
\begin{align}
    \vec{Z} = {\tbar{\Lambda}}^{-\frac{1}{2}}( \avg{\vec{\tbar{Z}}(\tau)} + {\tbar{\Lambda}}^q \vec{\zeta} )
    \label{appeq:expf}
\end{align}
where the variables $\vec{\zeta}$ describe the quantum statistics of fluctuations around the quantum state we are expanding about. We can therefore define a positive-$P$ distribution $\tbar{\mathcal{P}}$ in the phase space of fluctuations. which is related to the original positive-$P$ distribution via,
\begin{align}
    \tbar{\mathcal{P}}(\vec{\zeta},\tau) \equiv {\tbar{\Lambda}}^{q-\frac{1}{2}} \cdot \mathcal{P}( {\tbar{\Lambda}}^{-\frac{1}{2}} \avg{\vec{\tbar{Z}}(\tau)} + {\tbar{\Lambda}}^{q-\frac{1}{2}} \vec{\zeta},\tau)
    \label{appeq:prepf}
\end{align}
where the prefactor ensures the correct normalization of the positive-$P$ distribution of fluctuations.



Our aim now is to convert Eq.~(\ref{appeq:fpefull}) to an equivalent FPE for in the phase space of \textit{fluctuations} alone. To this end, we first make use of the chain rule as implied by Eq.~(\ref{appeq:expf}),
\begin{subequations}
\begin{align}
    \frac{\partial}{\partial \avg{\tbar{Z}_j}} &= \frac{\partial \zeta_j}{\partial \avg{\tbar{Z_j}} }\frac{\partial}{\partial \zeta_j}  = {\tbar{\Lambda}}^{-q}\frac{\partial}{\partial \zeta_j}, \\
    \frac{\partial}{\partial Z_j} &= \frac{\partial \zeta_j}{\partial Z_j}\frac{\partial}{\partial \zeta_j}  = {\tbar{\Lambda}}^{-(q-\frac{1}{2})}\frac{\partial}{\partial \zeta_j}.
    \label{appeq:chain}
\end{align}   
\end{subequations}
Then, to determine the left hand side of Eq.~(\ref{appeq:fpefull}), we first compute the time derivative of Eq.~(\ref{appeq:prepf}),
\begin{align}
    d_{\tau} \tbar{\mathcal{P}} &= {\tbar{\Lambda}}^{q-\frac{1}{2}} \cdot \left( \frac{\partial \mathcal{P}}{\partial \avg{\tbar{Z}_i} }\frac{\partial \avg{\tbar{Z}_i}  }{\partial \tau} + \frac{\partial \mathcal{P}}{\partial \tau} \right) \nonumber \\
    &=\left( {\tbar{\Lambda}}^{-q} \frac{\partial \avg{\tbar{Z}_i} }{\partial \tau} \right)\frac{\partial \tbar{\mathcal{P}}}{\partial \zeta_i}  +  \frac{\partial \tbar{\mathcal{P}}}{\partial \tau} 
    \label{appeq:fpefluc}
\end{align}
Now, the final term on the right hand side defines the partial derivative of $\tbar{\mathcal{P}}$ with respect to time; using Eq.~(\ref{appeq:expf}), this can directly be related to the FPE for the original positive-$P$ distribution, Eq.~(\ref{appeq:fpefull} under the expansion of Eq.~(\ref{appeq:expf}). We first replace phase space derivatives in Eq.~(\ref{appeq:fpefull} via the chain rule, Eq.~(\ref{appeq:chain}), and then simplify to arrive at:
\begin{align}
    d_{\tau} \tbar{\mathcal{P}}(\vec{Z},\tau) = \left( {\tbar{\Lambda}}^{-q} \frac{\partial \avg{\tbar{Z}_i} }{\partial \tau}\frac{\partial}{\partial \zeta_i} \! -{\tbar{\Lambda}}^{-(q-\frac{1}{2})}\frac{\partial}{\partial \zeta_i} {A}_i + \frac{1}{2}{\tbar{\Lambda}}^{-2(q-\frac{1}{2})}\frac{\partial^2}{\partial \zeta_i \partial \zeta_j}\mathbf{D}_{ij} \right) \! \tbar{\mathcal{P}}(\vec{Z},\tau)
    \label{appeq:fpeexp}
\end{align}
The above takes the form of an FPE, now for the positive-$P$ distribution of fluctuations, $\tbar{\mathcal{P}}(\vec{Z},\tau)$. However, an important complexity is hidden in the notation. Note that both $A_i$ and $\mathbf{D}_{ij}$ depend on phase space variables, and therefore must also be evaluated under the expansion of Eq.~(\ref{appeq:expf}) to arrive at the final desired FPE.

\subsection{Expanding the drift term}

We first write the drift term following the expansion of Eq.~(\ref{appeq:expf}). For simplicity, we introduce the right-bar notation,
\begin{align}
    \vbar{F} \equiv \left( F(\vec{Z})\Big|_{\vec{Z}\to{\tbar{\Lambda}}^{-\frac{1}{2}}\avg{\vec{\tbar{Z}}}} \right)
    \label{appeq:rightbar}
\end{align}
to describe the evaluation of any quantity $F$ (including vectors and matrices) of phase space variables $\vec{Z}$ around the expansion point defined by Eq.~(\ref{appeq:expf}).

In particular we will consider a Taylor expansion of the drift vector,
\begin{align}
    \tbar{\Lambda}^{\frac{1}{2}}{A}_i &= \tbar{\Lambda}^{\frac{1}{2}} \vbar{A_i} + \sum_j \tbar{\Lambda}^{\frac{1}{2}}\!\vvbar{\frac{\partial A_i}{\partial Z_j}}~ {\tbar{\Lambda}}^{q-\frac{1}{2}}\zeta_j  + \frac{1}{2}\sum_{jk}\tbar{\Lambda}^{\frac{1}{2}}\!\vvbar{\frac{\partial^2 A_i}{\partial Z_j\partial Z_k}}~{\tbar{\Lambda}}^{2(q-\frac{1}{2})}\zeta_j\zeta_k + \ldots
\end{align}
Using Eq.~(\ref{appeq:Af}), the above can be written to explicitly extract the dependence on nonlinearity:
\begin{align}
    \tbar{\Lambda}^{\frac{1}{2}}A_i &= \left(\sum_j\tbar{\Lambda}^{\frac{1}{2}}\mathbf{A}^L_{ij}\tbar{\Lambda}^{-\frac{1}{2}}\avg{\tbar{Z}_j}\right) +  \tbar{\Lambda}^{\frac{3}{2}}\vbar{A_i^N}  + \tbar{\Lambda}^{\frac{1}{2}}f_i \nonumber \\
    &+ \tbar{\Lambda}^q \Bigg\{ \sum_j\tbar{\Lambda}^{\frac{1}{2}}\mathbf{A}^L_{ij}\tbar{\Lambda}^{-\frac{1}{2}}{\zeta}_j + \sum_j \tbar{\Lambda}^{\frac{3}{2}} \vvbar{ \frac{\partial A^N_i}{\partial Z_j} }~  \tbar{\Lambda}^{-\frac{1}{2}}\zeta_j \Bigg\} \nonumber \\
    &+ \tbar{\Lambda}^{2q}\Bigg\{ \frac{1}{2}\sum_{jk} \tbar{\Lambda}^{\frac{3}{2}} \vvbar{ \frac{\partial^2 A^N_i}{\partial Z_j\partial Z_k} }~\tbar{\Lambda}^{-1}\zeta_j\zeta_k \Bigg\} + \ldots
    \label{appeq:driftexp1}
\end{align}

The evaluation (~$\vbar{}$~) of nonlinear terms $A_i^N$ and their derivatives at the expansion point will return additional factors of $\Lambda$, subject to how the $A_i^N$ depend on phase space variables $\vec{Z}$. This is ultimately determined by the order of nonlinearity under consideration. For example, as we will see for the Kerr model, $A_i^N$ are \textit{cubic} in phase space variables $Z_j$; as a result, evaluating the nonlinear drift terms yields factors of $(\tbar{\Lambda}^{-\frac{1}{2}})^3$,
\begin{align}
    \vbar{A_i^N(\vec{Z})} = \tbar{\Lambda}^{-\frac{3}{2}}A_i^N (\avg{\vec{\tbar{Z}}}) \equiv \tbar{\Lambda}^{-\frac{3}{2}}\tbar{A}_i^N
\end{align}
so that we have now defined $\tbar{A}_i^N$ as the nonlinear drift term only in terms of $\avg{\vec{\tbar{Z}}}$. Analogous relations can be obtained if the nature of nonlinearity is different from the Kerr model considered here. 

We can straightforwardly compute derivatives of the left hand side with respect to phase space variables, as required in Eq.~(\ref{appeq:driftexp1}). First, we note that since $A_i^N$ depends at most cubically on the phase space variables, derivatives of order $n > 3$ with respect to these variables must vanish. Derivatives of lower order can be evaluated with the help of the chain rule,
\begin{align}
    \vvbar{ \frac{\partial^{(n)} A_i^N }{\prod_j^n \partial Z_j} } = 
    \begin{cases}
    \tbar{\Lambda}^{-\frac{3}{2}} \prod_j^n \frac{\partial \avg{\tbar{Z}_j} }{\partial Z_j }  \frac{\partial^{(n)}\tbar{A}_i^N }{\prod_j^n \partial \avg{\tbar{Z}_j}},~0 \leq n \leq 3  \\
    0,~{\rm otherwise}
    \end{cases}
\end{align}
Here, we note from Eq.~(\ref{appeq:expf}) that $\prod_j^n\frac{\partial \avg{\tbar{Z}_j} }{\partial Z_j } = \tbar{\Lambda}^{\frac{n}{2}}$. Simplifying and rearranging, we can write derivatives of order $n$ entirely as derivatives of the scaled nonlinear drift terms $\tbar{A}_i^N$,
\begin{align}
    \tbar{\Lambda}^{\frac{3}{2}}\cdot \vvbar{\frac{\partial^{(n)}\! A^N_i}{\prod_j^n \partial Z_j}}   =
    \begin{cases}
    \tbar{\Lambda}^{\frac{n}{2}} \frac{\partial^{(n)}\! \tbar{A}^N_i}{\prod_j^n \partial \avg{\tbar{Z}_j} } ,~0 \leq n \leq 3  \\
    0,~{\rm otherwise}
    \end{cases}
    \label{appeq:driftder}
\end{align}

Substituting Eq.~(\ref{appeq:driftder}) into Eq.~(\ref{appeq:driftexp1}), we finally obtain the Taylor expansion of the drift vector elements, written explicitly up to order $\Lambda^{2q}$,
\begin{align}
    \tbar{\Lambda}^{\frac{1}{2}}\!{A}_i = \left[\sum_j\mathbf{A}^L_{ij}\avg{\tbar{Z}_i} +  \tbar{A}^N_i + \tbar{f}_i\right]  + \tbar{\Lambda}^q  \sum_j \! \left(\mathbf{A}^L_{ij} + \frac{\partial \tbar{A}^N_i}{\partial \avg{\tbar{Z}_j} } \right) \! \zeta_j  + \tbar{\Lambda}^{2q} \Bigg( \frac{1}{2}\sum_{jk} \frac{\partial^2 \tbar{A}^N_i}{\partial \avg{\tbar{Z}_j}\partial \avg{\tbar{Z}_k} } \Bigg) \zeta_j\zeta_k + O(\Lambda^{3q})
    \label{appeq:driftexpf}
\end{align}
where we have also introduced the scaled inhomogeneous terms $\tbar{f}_i \equiv \tbar{\Lambda}^{\frac{1}{2}}f_i$.


\subsection{Expanding the diffusion term}

Analogously, we consider the Taylor expansion of the diffusion matrix terms around the expansion point,
\begin{align}
    \mathbf{D}_{ij} &= \vbar{\boldsymbol{{\rm D}}_{ij}} + \sum_k \vvbar{ \frac{\partial \boldsymbol{{\rm D}}_{ij} }{\partial Z_k} }~\tbar{\Lambda}^{q-\frac{1}{2}}\zeta_k + \ldots 
\end{align}
Using Eq.~(\ref{appeq:Df}), the above can be written explicitly as well:
\begin{align}
    \mathbf{D}_{ij} &=\mathbf{D}^L_{ij} + \tbar{\Lambda}\vbar{\boldsymbol{{\rm D}}^N_{ij}} + \tbar{\Lambda}^q\Bigg\{ \sum_k \tbar{\Lambda} \vvbar{ \frac{\partial \boldsymbol{{\rm D}}^N_{ij} }{\partial Z_k} }~ \tbar{\Lambda}^{-\frac{1}{2}}\zeta_k  \Bigg\} + \ldots
    \label{appeq:diffexp1}
\end{align}
Once more, the nonlinear diffusion terms can be written by taking into account their dependence on the phase space variables. For the Kerr model, the diffusion matrix depends \textit{quadratically} on the phase space variables; as a result, evaluation at the expansion point yields factors of $(\Lambda^{-\frac{1}{2}})^2$,
\begin{align}
    \vbar{\boldsymbol{{\rm D}}^N_{ij}(\vec{Z})} = \tbar{\Lambda}^{-1} \mathbf{D}^N_{ij}(\avg{\vec{\tbar{Z}}}) \equiv \Lambda^{-1} \tbar{\mathbf{D}}^N_{ij}
\end{align}
and so we have again defined $\tbar{\mathbf{D}}^N_{ij}$ as the nonlinear diffusion matrix elements in terms of $\avg{\vec{\tbar{Z}}}$ alone. 

The evaluation of derivatives proceeds as before: derivatives of order $n>2$ vanish since the diffusion matrix elements depend at most quadratically on phase space variables, while derivatives of lower order can be written as derivatives with respect to $\avg{\tbar{Z}_j}$ variables using the chain rule,
\begin{align}
    \vvbar{ \frac{\partial^{(n)} \boldsymbol{{\rm D}}^N_{ij} }{\prod_k^n \partial Z_k} } = 
    \begin{cases}
        \tbar{\Lambda}^{-1} \prod_k^n \frac{\partial \avg{\tbar{Z}_k} }{\partial Z_k }   \frac{\partial^{(n)}\tbar{\mathbf{D}}_{ij}^N }{\prod_k^n \partial \avg{\tbar{Z}_k}},~0 \leq n \leq 2  \\
    0,~{\rm otherwise}
    \end{cases}
\end{align}
Rearranging and simplifying, we arrive at the general form of derivatives of order $n$ of the diffusion matrix elements,
\begin{align}
    \tbar{\Lambda}\cdot \vvbar{\frac{\partial^{(n)} \boldsymbol{{\rm D}}^N_{ij} }{\prod_k^n \partial Z_k} }   =
    \begin{cases}
    \tbar{\Lambda}^{\frac{n}{2}} \frac{\partial^{(n)} \tbar{\mathbf{D}}^N_{ij}}{\prod_k^n \partial \avg{\tbar{Z}_k} } ,~0 \leq n \leq 2  \\
    0,~{\rm otherwise}
    \end{cases}
    \label{appeq:diffder}
\end{align}

Finally, substituting Eq.~(\ref{appeq:diffder}) into Eq.~(\ref{appeq:diffexp1}) yields the Taylor expansion of the diffusion matrix elements, which we write explicitly up to order $\Lambda^q$,
\begin{align}
    \mathbf{D}_{ij} &= \left[\mathbf{D}^L_{ij} + \tbar{\mathbf{D}}^N_{ij}\right] +\tbar{\Lambda}^q\Bigg\{ \sum_k \frac{\partial \tbar{\mathbf{D}}^N_{ij}}{\partial \avg{\tbar{Z}_k} } \zeta_k  \Bigg\} + O(\tbar{\Lambda}^{2q})
    \label{appeq:diffexpf}
\end{align}

\subsection{Fokker-Planck equation of fluctuations to lowest nontrivial order in \qrc{} nonlinearity}

We can now substitute the final expressions for the Taylor expansions of the drift and diffusion matrices, Eqs.~(\ref{appeq:driftexpf}),~(\ref{appeq:diffexpf}) respectively, into Eq.~(\ref{appeq:fpeexp}). This finally yields the FPE for fluctuations in increasing powers of $\tbar{\Lambda}^q$, 
\begin{align}
    d_{\tau} \tbar{\mathcal{P}}(\vec{\zeta},\tau) &= \tbar{\Lambda}^{-q}\left( \frac{\partial\avg{\tbar{Z}_i}}{\partial \tau} - \left[\sum_j\mathbf{A}^L_{ij}\avg{\tbar{Z}_j} + \tbar{A}^N_i + \tbar{f}_i\right] \right)\frac{\partial \tbar{\mathcal{P}}}{\partial \zeta_i}  \nonumber \\
    &~~~~-\frac{\partial}{\partial \zeta_i} \! \left[ \tbar{\Lambda}^0\sum_j\Bigg( \mathbf{A}^L_{ij} + \frac{\partial \tbar{A}^N_i}{\partial \avg{\tbar{Z}_j} }   \Bigg)\zeta_j + \tbar{\Lambda}^{q}\Bigg( \frac{1}{2}\sum_{jk} \frac{\partial^2 \tbar{A}^N_i}{\partial \avg{\tbar{Z}_j}\partial \avg{\tbar{Z}_k} } \zeta_j\zeta_k \Bigg) + O(\tbar{\Lambda}^{2q}) \right]\!\tbar{\mathcal{P}}(\vec{\zeta},\tau) \nonumber \\
    &~~~~+ \frac{1}{2}\frac{\partial^2}{\partial \zeta_i \partial\zeta_j}\!\left[ \tbar{\Lambda}^{-2(q-\frac{1}{2})} \Big( \mathbf{D}^L_{ij} + \tbar{\mathbf{D}}^N_{ij} \Big)+\tbar{\Lambda}^{1-q} \left( \sum_k \frac{\partial \tbar{\mathbf{D}}^N_{ij}}{\partial \avg{\tbar{Z}_k} } \zeta_k  \right) +  O(\tbar{\Lambda}) \right]\!\tbar{\mathcal{P}}(\vec{\zeta},\tau)
    \label{appeq:fpefluc2}
\end{align}    
The leading term of order $\tbar{\Lambda}^{-q}$ will diverge as $\tbar{\Lambda} \to 0$ and therefore its coefficient must be set to zero; this yields a set of TEOMs for first-order cumulants defining the expansion point around which we are considering fluctuations, 
\begin{align}
    \frac{\partial}{\partial \tau} \avg{\vec{\tbar{Z}}} = \mathbf{A}^L\avg{\vec{\tbar{Z}}} + \vec{\tbar{A}}^N\!(\avg{\vec{\tbar{Z}}}) + \vec{\tbar{f}}
\end{align}
Recalling that $\vec{\tbar{A}}^N$ are nonlinear functions of $\avg{\vec{\tbar{Z}}}$, the above system of equations is clearly nonlinear.

Satisfying Eq.~(\ref{appeq:scaled1}) ensures that the $\tbar{\Lambda} \to 0$ limit does not diverge. However, to obtain a well-defined distribution function in this limit, with both finite drift and diffusion contributions, we require the leading order drift term, $O(\tbar{\Lambda}^0)$, and the leading diffusion term, $O(\tbar{\Lambda}^{-2(q-\frac{1}{2})})$, to be of the same order. This naturally enforces $q = \frac{1}{2}$ as the choice of the expansion parameter. 

Our analysis thus far is exact, since all terms in the expansion are retained. We can now make an approximation in the strength of the nonlinearity by dropping terms in Eq.~(\ref{appeq:fpefluc2}). However, it is clear that under the choice $q = \frac{1}{2}$, the second order drift and diffusion terms both appear at the same order of nonlinearity. 

The resulting approximate FPE of fluctuations then takes the compact form,
\begin{align}
    d_{\tau} \tbar{\mathcal{P}}(\vec{\zeta},\tau) \simeq  \! \Bigg(\! -\frac{\partial}{\partial \zeta_i} \Big[ \tbar{\Lambda}^0\sum_j\tbar{\mathbf{J}}_{ij}{\zeta}_j + \tbar{\Lambda}^{\frac{1}{2}}\sum_{jk} \tbar{\mathbf{H}}_{ijk} \zeta_j\zeta_k \Big] +  \frac{1}{2}\frac{\partial^2}{\partial \zeta_i \partial\zeta_j} \Big[\tbar{\Lambda}^0\tbar{\mathbf{B}}_{ij} + \sum_k O(\tbar{\Lambda}^{\frac{1}{2}})\zeta_k\Big] \Bigg)\tbar{\mathcal{P}}(\vec{\zeta},\tau)
    \label{appeq:fpefluc3}
\end{align}    
where we have defined $\tbar{\mathbf{J}}$ as the familiar Jacobian describing the measurement chain,
\begin{align}
    \tbar{\mathbf{J}}_{ij} = 
    \mathbf{A}^L_{ij} + \frac{\partial \tbar{A}^N_i}{\partial \avg{\tbar{Z}_j} }
    \label{appeq:jacobian}
\end{align}
which clearly includes contributions from the first derivative of the nonlinearity vector $\vec{\tbar{A}}^N$. Secondly, we have also introduced the linearized diffusion matrix $\tbar{\mathbf{B}}$ evaluated at the expansion point,
\begin{align}
    \tbar{\mathbf{B}}_{ij} = \mathbf{D}^L_{ij}+ \tbar{\mathbf{D}}^N_{ij}
    \label{appeq:lindiff}
\end{align}

The most interesting term is the final drift term of $O(\tbar{\Lambda}^{\frac{1}{2}})$, where we have introduced $\tbar{\mathbf{H}}_{ijk}$ as the \textit{Hessian tensor} of the nonlinearity vector $\vec{\tbar{A}}^N$; its elements are given by
\begin{align}
    \mathbf{H}_{ijk} \equiv \frac{1}{2}\frac{\partial^2 \tbar{A}^N_i}{\partial \avg{\tbar{Z}_j}\partial \avg{\tbar{Z}_k} }
    \label{appeq:hessian}
\end{align}
Here, it plays the role of mapping second-order cumulants to first-order cumulants.

\subsection{TEOMs to lowest order in the \qrc{} nonlinearity using the NVK approximation}

Under the ansatz of Eq.~(\ref{appeq:expf}), the desired first-order cumulants in the original basis can be related to cumulants of the fluctuation variables:
\begin{align}
    \avg{Z_i} = \tbar{\Lambda}^{-\frac{1}{2}}\avg{\tbar{Z}_i} + \avg{\zeta_i} 
    \label{appeq:ciexp}
\end{align}
where we have used $q=\frac{1}{2}$. Similarly, second-order cumulants in the original basis can be related to cumulants of the fluctuation variables:
\begin{align}
    \mathbf{C}_{ij} &= \avg{Z_iZ_j} - \avg{Z_i}\avg{Z_j} \nonumber \\
    &= \tbar{\Lambda}^{-1}\avg{\tbar{Z}_i}\avg{\tbar{Z}_j} + \tbar{\Lambda}^{-\frac{1}{2}}\avg{\tbar{Z}_i}\avg{\zeta_j}  + \tbar{\Lambda}^{-\frac{1}{2}}\avg{\tbar{Z}_j}\avg{\zeta_i} + \avg{\zeta_i\zeta_j}  - \left( \tbar{\Lambda}^{-1}\avg{\tbar{Z}_i}\avg{\tbar{Z}_j} + \tbar{\Lambda}^{-\frac{1}{2}}\avg{\tbar{Z}_i}\avg{\zeta_j}  + \tbar{\Lambda}^{-\frac{1}{2}}\avg{\tbar{Z}_j}\avg{\zeta_i} + \avg{\zeta_i}\avg{\zeta_j} \right) \nonumber \\
    &= \avg{\zeta_i\zeta_j} - \avg{\zeta_i}\avg{\zeta_j}
    \label{appeq:cijexp}
\end{align}
Therefore, if we can obtain TEOMS for the fluctuation variables, these can be solved to obtain cumulants in the original basis. Our work in deriving the FPE for fluctuation variables, Eq.~(\ref{appeq:fpefluc3}), provides us exactly the path to deriving such simplified TEOMs within the NVK approximation. In particular, we note that the positive-$P$ distribution that governs fluctuations can be used to compute expectation values of any function $O$ of phase space fluctuation variables $\zeta_i$ via the definition $\avg{O} = \int d\vec{\zeta}~O\tbar{\mathcal{P}}(\vec{\zeta},\tau)$. Computing time derivatives of this definition, we find:
\begin{align}
    \frac{d}{d\tau}\avg{O} = \int d\vec{\zeta}~O~ d_{\tau}\tbar{\mathcal{P}}(\vec{\zeta},\tau)
\end{align}
The time derivative in the integrand is simply the definition of the FPE governing fluctuations, Eq.~(\ref{appeq:fpefluc3}). Making use of the FPE, the time derivative in the integrand can be eliminated in favour of derivatives of $\tbar{\mathcal{P}}$ with respect to phase space variables. Standard applications of integration by parts can then be used until the integrand is defined only in terms of $\tbar{\mathcal{P}}$ and not its derivatives; for full details we refer the reader to standard texts such as Ref.~\cite{carmichael_statistical_2002}.

This approach can be used to obtain equations of motion for first-order cumulants of fluctuations, namely $O \to \zeta_i$:
\begin{align}
    \frac{\partial}{\partial \tau} \avg{\zeta_i} &\simeq \sum_j \tbar{\mathbf{J}}_{ij} \avg{\zeta_j} + \tbar{\Lambda}^{\frac{1}{2}}\sum_{jk} \tbar{\mathbf{H}}_{ijk} \avg{\zeta_j\zeta_k}.
\end{align}
Similarly, we can obtain equations of motion for the second-order quantities by setting $O \to \zeta_i\zeta_j$. To lowest nontrivial order in the nonlinearity, we obtain:
\begin{align}
    \frac{d}{d\tau} \avg{\zeta_i\zeta_j} &=  \sum_k \tbar{\mathbf{J}}_{ik} \avg{\zeta_k\zeta_j} + \sum_k \tbar{\mathbf{J}}_{jk} \avg{\zeta_k\zeta_i} + \tbar{\mathbf{B}}_{ij}  + O(\tbar{\Lambda}^{\frac{1}{2}}).
\end{align}

From the above equations, we are able to make some useful deductions. First, we see that $\avg{\zeta_i\zeta_j}$ will in general be at least $O(\tbar{\Lambda}^0)$, provided $\tbar{\mathbf{B}}$ is nonvanishing. Secondly, we see that $\avg{\zeta_i}$ is at least $O(\tbar{\Lambda}^{\frac{1}{2}})$; dropping the term $\propto \tbar{\Lambda}^{\frac{1}{2}}$ yields $\avg{\zeta_i} \to 0$. Consequently, $\mathbf{C}_{jk} = \avg{\zeta_j\zeta_k} + O(\tbar{\Lambda}^{1})$. Therefore, when retaining terms up to order $O(\tbar{\Lambda}^{\frac{1}{2}})$, we can with reasonable confidence approximate $\mathbf{C}_{jk} \simeq \avg{\zeta_j\zeta_k}$. Under this approximation, we can therefore write for the equations of motion of first-order cumulants of fluctuations:
\begin{align}
    \frac{\partial}{\partial \tau} \avg{\zeta_i} &\simeq \sum_j \tbar{\mathbf{J}}_{ij} \avg{\zeta_j} + \tbar{\Lambda}^{\frac{1}{2}}\sum_{jk} \tbar{\mathbf{H}}_{ijk} \mathbf{C}_{jk} + O(\tbar{\Lambda}^{\frac{3}{2}})
    \label{appeq:scaledf}
\end{align}
and similarly for the second-order cumulants:
\begin{align}
    \frac{d}{d\tau} \avg{\zeta_i\zeta_j} \simeq \frac{d}{d\tau} \mathbf{C}_{ij} &=  \sum_k \tbar{\mathbf{J}}_{ik} \mathbf{C}_{kj} + \sum_k \tbar{\mathbf{J}}_{jk} \mathbf{C}_{ki} +  \tbar{\mathbf{B}}_{ij}  + O(\tbar{\Lambda}^{\frac{1}{2}})
\end{align}
The first set of equations can be compactly expressed in matrix form as:
\begin{align}
    \frac{\partial}{\partial \tau} \avg{\vec{\zeta}} \simeq \tbar{\mathbf{J}} \avg{\vec{\zeta}} + \tbar{\Lambda}^{\frac{1}{2}} \tbar{\mathbf{H}} : \mathbf{C},
    \label{appeq:scaledf2}
\end{align}
where $(:)$ represents the tensor double contraction, while the second set takes the form of a Lyapunov system,
\begin{align}
    \frac{d}{d\tau}\mathbf{C} = \tbar{\mathbf{J}}\mathbf{C} + \mathbf{C}\tbar{\mathbf{J}}^T + \tbar{\mathbf{B}}.
    \label{appeq:scaled2}
\end{align}
Solving the above sets of equations requires solving for the expectation value $\avg{\vec{\tbar{Z}}}$, governed by the system of equations, 
\begin{align}
    \frac{\partial}{\partial \tau} \avg{\vec{\tbar{Z}}} = \mathbf{A}^L\avg{\vec{\tbar{Z}}} + \vec{\tbar{A}}^N\!(\avg{\vec{\tbar{Z}}}) + \vec{\tbar{f}}.
    \label{appeq:scaled1}
\end{align}
Eqs.~(\ref{appeq:scaledf2}),~(\ref{appeq:scaled2}), and (\ref{appeq:scaled1}) together allow us to solve for the first and second-order cumulants of fluctuations within the NVK approximation. Finally, using Eq.~(\ref{appeq:ciexp}),~(\ref{appeq:cijexp}), these solutions allow us to determine the first and second-order cumulants in the original basis to lowest order in the nonlinearity strength. In SI Sec.~\ref{app:FI}, we use the solution of these TEOMs for the calculation of quantities that are relevant to quantum state classification considered in the main text.

Finally, we will also require the unequal-time correlations of fluctuation variables, $\mathbf{C}_{ij}(\tau+\theta,\tau) = \avg{Z_i(\tau+\theta)Z_j(\tau)} - \avg{Z_i(\tau+\theta)}\avg{Z_j(\tau)}$, for the calculation of measured covariances (see SI Sec.~\ref{si:io2}). For linear(ized) systems governed by $\tbar{\mathbf{J}}$, the required correlation matrix can be written entirely in terms of its equal-time counterpart~\cite{carmichael_statistical_2002}, again to lowest nontrivial order in the nonlinearity,
\begin{subequations}
    \begin{align}
    \mathbf{C}(\tau+\theta,\tau) &= e^{\tbar{\mathbf{J}}\theta}\mathbf{C}(\tau)  \label{appeq:corrTocovP0} \\
    \mathbf{C}(\tau-\theta,\tau) &= \mathbf{C}(\tau-\theta) e^{\tbar{\mathbf{J}}^T\theta}  \label{appeq:corrTocovM0}
\end{align}
\end{subequations}
where the second equation is obtained from the first by computing the transpose, replacing $\tau \to \tau - \theta$, and recalling that the normal-ordered covariance matrix is a symmetric matrix.

\subsection{Comparison of full (S)TEOMs and the NVK approximation}

The NVK approximation we have developed solves for the first and second-order cumulants of the complete quantum state of the measurement chain to lowest order in the \qrc{} nonlinearity. This approximation drops certain terms from the full (S)TEOMs even at the same truncation order $n_{\rm ord}$. Such terms are clearly seen even in the simple example of TEOMs of a single Kerr oscillator in SI Sec.~\ref{si:singleNodeQRC}: here, TEOMs for second-order cumulants, e.g. Eqs.~(\ref{eq:transfCbdb}),~(\ref{eq:transfCbb}), include terms nonlinear in the cumulants that are absent from Eqs.~(\ref{appeq:scaled2}). Therefore, we find that the NVK approximation and full (S)TEOMs agree for weaker nonlinearity strengths, while for increasing nonlinearity the (S)TEOMs provide a numerical scheme that is valid beyond the NVK approximation. A clear example of the comparison between the NVK approximation and full (S)TEOMs is seen in the exploration of \qrc{} performance against nonlinearity strength in Fig.~\ref{fig:ampClassifyDrive} of the main text.

\newpage

\section{Quantum statistics of measured quadratures under continuous heterodyne measurement}
\label{app:io}

The STEOMs framework developed in SI Sec.~\ref{si:condQRCdyn} provides an efficient numerical scheme to sample measurement records obtained from measurement chains deploying \qrcs{} as processors of quantum signals from a QS. Consequently, it provides an approach to determining the quantum statistics of these measurement records by computing empirical averages and variances, analogous to how such statistics may be obtained in any real experiment. The quantum statistics of these measurement records must be related to the quantum state of the \qrc{}, and ultimately to the upstream QS whose signals are being processed by the \qrc{}. In this section, we clarify the first of these connections. In doing so, we also build towards a semi-analytic scheme that allows determining the quantum statistics of measurement records by solving the deterministic TEOMs, as opposed to the more computationally expensive stochastic integration of the STEOMs.

We will begin by identifying precisely the information accessible to an experimentalist performing heterodyne measurements on quantum modes in a measurement chain. Then, we show how these quantities can be related to expectation values of appropriate dynamical variables computed with respect to the quantum state $\rhou$ of the measurement chain. From a theoretical standpoint, this is simply an application of quantum input-output theory, presented here in the language of quantum trajectories in the Schr\"odinger picture, as opposed to the perhaps more common description in the Heisenberg representation. 


\subsection{Quadrature expectation values}

Recall that the heterodyne measured quadratures extracted from the measurement chain are as defined in Eq.~(\ref{eq:IQkraw}) of the main text. We are interested in the mean and variance of these measured quadratures. It is sufficient to consider the measured quadratures of just one mode,
\begin{align}
    \bm{\mu}_k(t) \equiv \mathbb{E}
    \!\left[
    \begin{pmatrix}
        I_k(t) \\
        Q_k(t)
    \end{pmatrix}
    \right] = \frac{1}{\sqrt{2\mathcal{T}}} \int_{t-\mathcal{T}}^t d\tau~\mathbb{E}
    \!\left[
        \begin{pmatrix}
        \mathcal{I}_k(\tau)  \\
        \mathcal{Q}_k(\tau) 
    \end{pmatrix}
    \right]. 
    \label{appeq:muk}
\end{align}
The required expectation value of measured quadratures can be computed using Eqs.~(\ref{eq:IQkraw}) of the main text,
\begin{align}
    \mathbb{E}\!\left[
    \begin{pmatrix}
        \mathcal{I}_k(\tau) \\
        \mathcal{Q}_k(\tau)
    \end{pmatrix}\right] \!\! &= \! \mathbb{E}\!\left[\!
    \begin{pmatrix}
        \xi_{\mathcal{I}_k}(\tau) + \!\sqrt{\gammahet}\left[ \avg{\hat{X}_k(\tau)} + \xi^{\rm qm}_{\mathcal{I}_k}(\tau) \right]\! + \! \sqrt{\bar{n}_{\rm cl}}~\xi^{\rm cl}_{\mathcal{I}_k}(\tau)  \\
        \xi_{\mathcal{Q}_k}(\tau) + \!\sqrt{\gammahet}\left[ \avg{\hat{P}_k(\tau)} + \xi^{\rm qm}_{\mathcal{Q}_k}(\tau) \right]\! + \! \sqrt{\bar{n}_{\rm cl}}~\xi^{\rm cl}_{\mathcal{Q}_k}(\tau)
    \end{pmatrix} 
    \!\right] \! \nonumber \\
    &= 
    \sqrt{\gammahet}\!
    \begin{pmatrix}
        \avg{\hat{X}_k(\tau)} \\
        \avg{\hat{P}_k(\tau)}
    \end{pmatrix} 
\end{align}
where we have used the fact that the expectation values of all the noise terms vanish; for the quantum noise contributions, this follows since the expectation value of conditional quantum trajectories is equal to their unconditional expectation value, $\mathbb{E}[\avgc{\hat{O}(t)}]=\mathbb{E}[{\rm tr}\{\hat{O}\rhoc(t)\}] = {\rm tr}\{\hat{O}\mathbb{E}[\rhoc(t)]\} = {\rm tr}\{\hat{O}\rhou(t)\} = \avg{\hat{O}(t)}$.

The calculation thus far is general; we now specialize to the requirement of steady-state measured quadratures for the quantum state classification task under consideration. For long enough $t$, the right hand side above reaches steady state. Substituting the result into Eq.~(\ref{appeq:muk}), the time integral can be trivially computed, yielding
\begin{align}
    \bm{\mu}_k = \mathbb{E}\left[
    \begin{pmatrix}
        I_k \\
        Q_k
    \end{pmatrix}
    \right]
    = \sqrt{\frac{\mathcal{T}}{2}}\cdot
    \sqrt{\gammahet}
    \begin{pmatrix}
        \avg{\hat{X}_k} \\
        \avg{\hat{P}_k}
    \end{pmatrix} 
\end{align}
where for compactness we suppress time labels to denote steady state (time-independent) quantities.

We see that expectation values of measured quadratures directly probe internal quadratures of the modes being monitored, which are themselves related to first-order cumulants. However, note that the modes that are monitored are often a subset of the complete set of modes comprising a quantum measurement chain. It therefore proves useful to relate the measured expectation values to the complete set of first-order cumulants describing the quantum measurement chain. To do so, we introduce the vector defining operators for the $\ntot=M+K$ modes of the complete measurement chain, $\hat{\bm{z}} = (\hat{a}_1,\hat{a}_1^{\dagger},\ldots,\hat{a}_M,\hat{a}_M^{\dagger},\hat{b}_1,\hat{b}_1^{\dagger},\ldots,\hat{b}_K,\hat{b}_K^{\dagger})^T$. Then the vector of first-order cumulants is simply $\avg{\hat{\bm{z}}}$. Then, the mode quadratures can be written in terms of first-order cumulants via
\begin{align}
    \sqrt{\gammahet}
    \begin{pmatrix}
        \avg{\hat{X}_k} \\
        \avg{\hat{P}_k}
    \end{pmatrix}
    =
    \mathbf{M}_k\avg{\hat{\bm{z}}}
\end{align}
where we have introduced the \textit{heterodyne measurement matrix} $\mathbf{M}_k$, which simply serves to extract the appropriate subset of mode quadratures in the canonical basis. It is a 2-by-2$\ntot$ matrix with nonzero entries only in columns corresponding to the $k$th mode in $\hat{\bm{z}}$:
\begin{align}
    \mathbf{M}_k = \sqrt{\gammahet}
    \begin{pmatrix}
        \mathbf{0} & \cdots & \underbrace{\mathbf{U}}_{k{\rm th~mode}} & \cdots & \mathbf{0}
    \end{pmatrix},
    \label{appeq:measmat}
\end{align}
where $\mathbf{U}$ is the standard unitary matrix defining a change-of-basis from canonical quadratures to normal-ordered first-order cumulants,
\begin{align}
    \mathbf{U} = 
    \frac{1}{\sqrt{2}}
    \begin{pmatrix}
        1 & 1 \\
        -i & i
    \end{pmatrix}.
\end{align}
We can hence finally write the expectation values of measured quadratures in terms of first-order cumulants obtained via TEOMs,
\begin{align}
    \bm{\mu}_k = \mathbb{E}\left[
    \begin{pmatrix}
        I_k \\
        Q_k
    \end{pmatrix}
    \right]
    =  
    \sqrt{\frac{\mathcal{T}}{2}}\cdot
    \mathbf{M}_k
    \avg{\hat{\bm{z}}}
    \label{appeq:mukgeneral}
\end{align}

\subsection{Quadrature noise and correlations}
\label{si:io2}

We are also interested in the noise of measured quadratures, including correlations between any pair of distinct quadratures. This information can be accessed by computing quadrature variances. To this end, we introduce the measured quadrature covariance matrix $\mathbf{\Sigma}_{jk}$ which describes the correlations between the measured quadratures of any pair $(j,k)$ of modes:
\begin{align}
    \mathbf{\Sigma}_{jk}(t) \equiv 
    \mathbb{E}\!\left[
    \begin{pmatrix}
        I_j(t) \\
        Q_j(t)
    \end{pmatrix}
    \!\!
    \begin{pmatrix}
        I_k(t) \\
        Q_k(t)
    \end{pmatrix}^{\!\!\! T}
    \right]
    \!
    -
    \mathbb{E}\!\left[
    \begin{pmatrix}
        I_j(t) \\
        Q_j(t)
    \end{pmatrix}
    \right]
    \!
    \mathbb{E}\!\left[
    \begin{pmatrix}
        I_j(t) \\
        Q_j(t)
    \end{pmatrix}
    \right]^{T}
\end{align}
Using the definition of measured quadratures, we can write the above as:
\begin{align}
    \mathbf{\Sigma}_{jk}(t) &= \frac{\bar{n}_{\rm cl}}{2}\mathbf{I}_2\delta_{jk} + 
    \frac{1}{2\mathcal{T}}\int_{t-\mathcal{T}}^t \int_{t-\mathcal{T}}^t d\tau d\tau' \times \Bigg\{\!\mathbb{E}\!\left[
        \begin{pmatrix}
        \mathcal{I}_j(\tau') \\
        \mathcal{Q}_j(\tau')
    \end{pmatrix}\!\!
    \begin{pmatrix}
        \mathcal{I}_k(\tau) \\
        \mathcal{Q}_k(\tau)
    \end{pmatrix}^{\!\!\! T}
    \right] 
    \!\!
    -\mathbb{E}\!
    \left[
        \begin{pmatrix}
        \mathcal{I}_j(\tau') \\
        \mathcal{Q}_j(\tau')
    \end{pmatrix}\right]\!
    \mathbb{E}\!
    \left[
    \begin{pmatrix}
        \mathcal{I}_k(\tau) \\
        \mathcal{Q}_k(\tau)
    \end{pmatrix}
    \right]^T\!\!
    \Bigg\} \nonumber \\
    &\equiv  \frac{\bar{n}_{\rm cl}}{2}\mathbf{I}_2\delta_{jk} + \frac{1}{2\mathcal{T}}\int_{t-\mathcal{T}}^t \int_{t-\mathcal{T}}^t d\tau d\tau'~\bm{\mathcal{S}}_{jk}(\tau',\tau)
    \label{appeq:VV}
\end{align}
where we have introduced $\bm{\mathcal{S}}_{jk}(t',t)$ as the correlation matrix of measured heterodyne records prior to any post-processing. A standard calculation~\cite{wiseman_quantum_2009} can be used to relate this measured correlation matrix to correlations of mode operators:
\begin{align}
    \bm{\mathcal{S}}_{jk}(\tau',\tau) = \delta_{jk}\delta(\tau-\tau')\mathbf{I}_2 + \mathbf{M}_j\mathbf{C}(\tau',\tau)\mathbf{M}_k^T
    \label{appeq:VC}
\end{align}
Here, the first term arises simply due to the $\delta$-correlation of measurement noise, which only contributes when autocorrelations are being calculated. The second term is the important contribution that depends on the state of the quantum modes being probed, via $\mathbf{C}(\tau',\tau)$ which is the unconditional normal-ordered correlation matrix of the complete measurement chain. The elements of the correlation matrix are defined explicitly as
\begin{align}
    [\mathbf{C}(\tau',\tau)]_{ij} = \avg{:\!\hat{z}_i(\tau')\hat{z}_j(\tau)\!:} - \avg{\hat{z}_i(\tau')}\avg{\hat{z}_j(\tau)}
\end{align}
where $\avg{:\!\!\hat{f}\!\!:}$ implies an expectation value where the arbitrary operator $\hat{f}$ is written in normal-ordered form. The measurement matrices in Eq.~(\ref{appeq:VC}) then once again extract the sectors of the correlation matrix pertinent to the correlations of mode pair $(j,k)$.



Then, substituting Eq.~(\ref{appeq:VC}) into Eq.~(\ref{appeq:VV}), the measured covariance matrix takes the  form:
\begin{align}
    \mathbf{\Sigma}_{jk}(t) = \sigma_{\rm vac}^2\mathbf{I}_2(\bar{n}_{\rm cl}+1)\delta_{jk} +\frac{1}{2\mathcal{T}}\mathbf{M}_j\!\left[\int_{t-\mathcal{T}}^t \int_{t-\mathcal{T}}^t d\tau d\tau'~\mathbf{C}(\tau',\tau)\right]\!\mathbf{M}_k^T
    \label{appeq:measV}
\end{align}
where $\sigma_{\rm vac}^2 = \frac{1}{2}$ as introduced in the main text. The first term then describes the vacuum noise contribution that is always present. The second term is the filtered contribution from monitoring the quantum system. In particular, it depends non-trivially on the filter window $\mathcal{T}$.

It is clear that for the completely general filter window $\mathcal{T}$, we will also require an expression for the correlation matrix $\mathbf{C}(\tau',\tau)$ for \textit{unequal} times $\tau \neq \tau'$. We first consider a special case of the measured covariances for short filter windows, where Eq.~(\ref{appeq:measV}) simplifies considerably. We then consider approximations of measurement chains under which simple analytic expressions for the correlation matrix can be found.

\subsection{Steady-state measured covariances in the short-filter limit}
\label{si:outputcovlimits}

A simple case of heterodyne monitoring involves a short filter window $\mathcal{T}\to 0$. We can then make the assumption that the correlation matrix $\mathbf{C}(\tau',\tau)$ is effectively unchanged over the course of the short window $[t-\mathcal{T},t]$. We can then approximate $\tau \simeq \tau' \to t$ over the integration window, thereby replacing the correlation matrix by its equal-time counterpart, the \textit{covariance} matrix $\mathbf{C}(t)$. Extracting the covariance matrix outside of the integral in Eq.~(\ref{appeq:measV}) (in analogy with standard Markov-like approximations) yields:
\begin{align}
    \mathbf{\Sigma}_{jk}(t) \stackrel{\mathcal{T}\to 0}{\simeq} \sigma_{\rm vac}^2\mathbf{I}_2(\bar{n}_{\rm cl}+1)\delta_{jk} +\frac{1}{2\mathcal{T}}\mathbf{M}_j\mathbf{C}(t)\!\left[\int_{t-\mathcal{T}}^t \int_{t-\mathcal{T}}^t d\tau d\tau'\right]\!\mathbf{M}_k^T,
\end{align}
which immediately simplifies to:
\begin{align}
    \mathbf{\Sigma}_{jk}(t) \stackrel{\mathcal{T}\to 0}{\simeq} \sigma_{\rm vac}^2\mathbf{I}_2(\bar{n}_{\rm cl}+1)\delta_{jk} +\frac{\mathcal{T}}{2}\cdot\mathbf{M}_j\mathbf{C}(t)\mathbf{M}_k^T
    \label{appeq:measVT0a}
\end{align}
Therefore, for short enough filter windows, the measured covariances can be related entirely to the measured mode covariance matrix $\mathbf{C}(t)$. We emphasize that Eq.~(\ref{appeq:measVT0}) holds generally for \textit{any} quantum measurement chain under heterodyne monitoring. In the next subsection, we will make use of the properties of linear(ized) systems to obtain expressions for the measured covariances beyond the assumption of short filter window. As we will show, for long enough filter windows, even equal-time covariances of measured quadratures will include a temporal dependence on correlations of the measured modes at different times. 

\subsection{Steady-state measured covariances for linear(ized) systems under the NVK approximation}
\label{si:outputcov}

In SI Sec.~\ref{app:lin}, we have shown that for linear(ized) systems under an appropriate assumption of weak nonlinearity - made precise within the NVK approximation - the correlation matrix $\mathbf{C}(\tau',\tau)$ is governed by the linearized Jacobian matrix $\tbar{\mathbf{J}}$ and the steady-state equal-time covariance matrix $\mathbf{C}(\tau)$ by Eqs.~(\ref{appeq:corrTocovP0}),~(\ref{appeq:corrTocovM0}). For convenience, we reproduce these equations below:
\begin{subequations}
    \begin{align}
    \mathbf{C}(\tau+\theta,\tau) &= e^{\tbar{\mathbf{J}}\theta}\mathbf{C}(\tau)  \label{appeq:corrTocovP} \\
    \mathbf{C}(\tau-\theta,\tau) &= \mathbf{C}(\tau-\theta) e^{\tbar{\mathbf{J}}^T\theta}  \label{appeq:corrTocovM}
\end{align}
\end{subequations}

This fact allows us to simplify the expression for measured covariances, Eq.~(\ref{appeq:measV}), for such linear(ized) systems. In particular, we can first rewrite the integral in Eq.~(\ref{appeq:measV}) as two terms:
\begin{align}
    \int_{t-\mathcal{T}}^t \int_{t-\mathcal{T}}^t d\tau d\tau'~\mathbf{C}(\tau',\tau)   =\int_{t-\mathcal{T}}^t \int_{t-\mathcal{T}}^{\tau} d\tau d\tau'~\mathbf{C}(\tau',\tau) + \int_{t-\mathcal{T}}^t \int_{\tau}^{t} d\tau d\tau'~\mathbf{C}(\tau',\tau) 
\end{align}
where in the first term $\tau' < \tau$, while in the second term $\tau' > \tau$. Next, we introduce an alternate, positive integration variable $\theta$ to denote the temporal separation of $\tau$ and $\tau'$: precisely, we make the substitution $\tau-\tau' = \theta$ in the first term and $\tau'-\tau = \theta$ in the second, to obtain
\begin{align}
    \int_{t-\mathcal{T}}^t \int_{t-\mathcal{T}}^t d\tau d\tau'~\mathbf{C}(\tau',\tau)  =\int_{t-\mathcal{T}}^t \int_{0}^{\tau-(t-\mathcal{T})} \!\!\!\!\!\!\!\! d\tau d\theta~\mathbf{C}(\tau-\theta,\tau) + ~~~\int_{t-\mathcal{T}}^t \int_{0}^{t-\tau} d\tau d\theta~\mathbf{C}(\tau+\theta,\tau)
    \label{appeq:corrInt}
\end{align}

Substituting Eqs.~(\ref{appeq:corrTocovP}),~(\ref{appeq:corrTocovM}) into Eq.~(\ref{appeq:corrInt}) and simplifying thus allows us to write the correlation matrix contribution entirely in terms of covariance matrices,
\begin{align}
    \int_{t-\mathcal{T}}^t \int_{t-\mathcal{T}}^t d\tau d\tau'~\mathbf{C}(\tau',\tau) =\left(\int_{t-\mathcal{T}}^t \int_{0}^{\tau-(t-\mathcal{T})} \!\!\!\!\!\!\!\! d\tau d\theta~\mathbf{C}(\tau-\theta)e^{\tbar{\mathbf{J}}^T\theta}\right) +  \left(\int_{t-\mathcal{T}}^t \int_{0}^{t-\tau} d\tau d\theta~e^{\tbar{\mathbf{J}}\theta}\mathbf{C}(\tau)\right)
\end{align}
We can now specialize to the case of steady-state covariances. In particular, the minimal value of integration variables $\tau, \theta$ is $t-\mathcal{T}$. Provided $t-\mathcal{T}$ is long enough in comparison to the relaxation rates of the measurement chain (by choosing large enough $t$), the time dependence of the covariance matrix $\mathbf{C}(\tau)$ itself can be neglected. In this case, we can replace $\mathbf{C}(\tau) \to \mathbf{C}$, and simplify the filtered correlation matrix,
\begin{align}
    \int_{t-\mathcal{T}}^t \int_{t-\mathcal{T}}^t d\tau d\tau'~\mathbf{C}(\tau',\tau)  = \mathbf{C}\left(\int_{t-\mathcal{T}}^t \int_{0}^{\tau-(t-\mathcal{T})} \!\!\!\!\!\!\!\! d\tau d\theta~e^{\tbar{\mathbf{J}}^T\theta}\right) + \left(\int_{t-\mathcal{T}}^t \int_{0}^{t-\tau} d\tau d\theta~e^{\tbar{\mathbf{J}}\theta}\right)\mathbf{C}
\end{align} 
Hence we are only required to compute integrals over the $e^{\tbar{\mathbf{J}}\theta}$ and its transpose. To this end, the diagonal representation of the matrix $\tbar{\mathbf{J}} = \mathbf{P}\mathbf{N}\mathbf{P}^{-1}$ proves very useful, as $e^{\mathbf{A}\theta} = \mathbf{P}e^{\mathbf{N}\theta}\mathbf{P}^{-1}$ where $\mathbf{N}$ is a diagonal matrix of the eigenvalues of $\tbar{\mathbf{J}}$. With this result, we can simplify the second of the two integrals above,
\begin{align}
    \int_{t-\mathcal{T}}^t \int_{0}^{t-\tau} d\tau d\theta~e^{\tbar{\mathbf{J}}\theta} &= \mathbf{P}\int_{t-\mathcal{T}}^t d\tau \left[\int_{0}^{t-\tau} d\theta ~e^{\mathbf{N}\theta}\right]\mathbf{P}^{-1} \nonumber \\
    &= \mathbf{P}\mathbf{N}^{-1}\int_{t-\mathcal{T}}^t d\tau\left[e^{\mathbf{N}(t-\tau)}-\mathbf{I}\right]\mathbf{P}^{-1} \nonumber \\
    &= \mathbf{P}\mathbf{N}^{-2}\left[e^{\mathbf{N}\mathcal{T}}-\mathbf{I}\right]\mathbf{P}^{-1}-\mathcal{T}\tbar{\mathbf{J}}^{-1} \nonumber \\
    &= \tbar{\mathbf{J}}^{-2}\left[ e^{\tbar{\mathbf{J}}\mathcal{T}}-\mathbf{I}\right] - \mathcal{T}\tbar{\mathbf{J}}^{-1}
\end{align}
where we have used $\tbar{\mathbf{J}}^n = \mathbf{P}\mathbf{N}^n\mathbf{P}^{-1}$ for all $n \in \mathbb{Z}$. The other required integral can be analogously computed, using $(\tbar{\mathbf{J}}^T)^n = (\mathbf{P}^T)^{-1}\mathbf{N}^n\mathbf{P}^{T}$ for all $n \in \mathbb{Z}$,
\begin{align}
    \int_{t-\mathcal{T}}^t \int_{0}^{\tau-(t-\mathcal{T})} \!\!\!\!\!\!\!\! d\tau d\theta~e^{\tbar{\mathbf{J}}^T\theta} &= (\mathbf{P}^T)^{-1}\!\!\int_{t-\mathcal{T}}^t \!\! d\tau \left[\int_{0}^{\tau-(t-\mathcal{T})}\!\!\!\!\! d\theta ~e^{\mathbf{N}\theta}\right]\mathbf{P}^{T} \nonumber \\
    &= (\mathbf{P}^T)^{-1}\mathbf{N}^{-1}\!\!\int_{t-\mathcal{T}}^t d\tau\left[e^{\mathbf{N}(\tau-(t-\mathcal{T}))}-\mathbf{I}\right]\mathbf{P}^{T} \nonumber \\
    &= (\mathbf{P}^T)^{-1}\mathbf{N}^{-2}\left[e^{\mathbf{N}\mathcal{T}}-\mathbf{I}\right]\mathbf{P}^{T}-\mathcal{T}(\tbar{\mathbf{J}}^T)^{-1} \nonumber \\
    &= (\tbar{\mathbf{J}}^T)^{-2}\left[ e^{\tbar{\mathbf{J}}^T\mathcal{T}}-\mathbf{I}\right] - \mathcal{T}(\tbar{\mathbf{J}}^T)^{-1}
\end{align}
The computed integrals thus allow us to determine the integral of the general correlation matrix as defined in Eq.~(\ref{appeq:corrInt}). This finally allows us to write down, after some rearrangement, the measured covariance matrix for mode pair $(j,k)$ for linear(ized) systems via Eq.~(\ref{appeq:measV}):
\begin{align}
    \mathbf{\Sigma}_{jk}(t) = \sigma_{\rm vac}^2(\bar{n}_{\rm cl}+1)\mathbf{I}_2\delta_{jk} -\frac{1}{2}\mathbf{M}_j\!\left[\mathbf{C}(\tbar{\mathbf{J}}^T)^{-1} + \tbar{\mathbf{J}}^{-1} \mathbf{C}\right]\!\mathbf{M}_k^T  + \frac{1}{2\mathcal{T}}\mathbf{M}_j\!\left[\mathbf{C}(\tbar{\mathbf{J}}^T)^{-2}\big(e^{\tbar{\mathbf{J}}^T\mathcal{T}}-\mathbf{I}\big) + \tbar{\mathbf{J}}^{-2}\left( e^{\tbar{\mathbf{J}}\mathcal{T}}-\mathbf{I}\right)\mathbf{C}\right]\!\mathbf{M}_k^T
    \label{appeq:Vfiltered}
\end{align}


\subsection{Short and long filter limits of measured covariances under the NVK approximation}
\label{si:outputcov2}

Eq.~(\ref{appeq:Vfiltered}) is somewhat unwieldy; we therefore evaluate some useful limits depending on the filter window length $\mathcal{T}$. First we consider $\mathcal{T}\to 0$, which corresponds to a very short filter window. In this case, the exponential terms in Eq.~(\ref{appeq:Vfiltered}) can be expanded to lowest order,
\begin{align}
    &\mathbf{\Sigma}_{jk}(t) \stackrel{\mathcal{T}\to 0}{\simeq} \sigma_{\rm vac}^2(\bar{n}_{\rm cl}+1)\mathbf{I}_2\delta_{jk} -\frac{1}{2}\mathbf{M}_j\!\left[\mathbf{C}(\tbar{\mathbf{J}}^T)^{-1} + \tbar{\mathbf{J}}^{-1} \mathbf{C}\right]\!\mathbf{M}_k^T  \nonumber \\
    &~~~~~+\frac{1}{2\mathcal{T}}\mathbf{M}_j\!\Big[\mathbf{C}(\tbar{\mathbf{J}}^T)^{-2}\left(\mathbf{I}+\tbar{\mathbf{J}}^T\mathcal{T}+\frac{1}{2}(\tbar{\mathbf{J}}^T)^2\mathcal{T}^2 + \ldots -\mathbf{I}\right) + \tbar{\mathbf{J}}^{-2}\left(\mathbf{I}+\tbar{\mathbf{J}}\mathcal{T}+\frac{1}{2}\tbar{\mathbf{J}}^2\mathcal{T}^2 + \ldots -\mathbf{I}\right)\!\mathbf{C}\Big]\mathbf{M}_k^T 
\end{align}
Expanding the second line, it becomes immediately clear that contributions of $O(\mathcal{T}^0)$ cancel contributions of the same order in the first line. As a result, the lowest order system-dependent contribution to measured covariances is $O(\mathcal{T})$,
\begin{align}
    \mathbf{\Sigma}_{jk}(t) \stackrel{\mathcal{T}\to 0}{\simeq} \sigma_{\rm vac}^2(\bar{n}_{\rm cl}+1)\mathbf{I}_2\delta_{jk} + \frac{\mathcal{T}}{2}\cdot\mathbf{M}_j \mathbf{C}\mathbf{M}_k^T.
    \label{appeq:measVT0}
\end{align}
which is exactly the steady-state version of the result obtained in Eq.~(\ref{appeq:measVT0a}). In the short filter limit, the covariances of measured quadratures are therefore only sensitive to covariances of the measured modes, and not to their temporal correlations.

We now consider the opposite limit of $\gamma \mathcal{T} \gg 1$. The requirement of stability of the quantum measurement chain requires $\tbar{\mathbf{J}}$ to have negative-definite eigenvalues $\mathbf{N}_{jj}~\forall~j$. Hence, provided $\mathcal{T} \gg {\rm max}_j\{\mathbf{N}^{-1}_{jj}\}$, $e^{\tbar{\mathbf{J}}\mathcal{T}}\to 0$. For simplicity, we simply use the shorthand $\gamma \mathcal{T} \gg 1$ to refer to this regime, although formally ${\rm max}_j\{\mathbf{N}^{-1}_{jj}\}$ could be smaller than $\gamma$. Regardless, we find,
\begin{align}
    &\mathbf{\Sigma}_{jk}(t) \stackrel{\gamma \mathcal{T} \gg 1}{\simeq} \sigma_{\rm vac}^2(\bar{n}_{\rm cl}+1)\mathbf{I}_2\delta_{jk}  -\frac{1}{2}\mathbf{M}_j\!\left[\mathbf{C}(\tbar{\mathbf{J}}^T)^{-1}\! +\! \tbar{\mathbf{J}}^{-1} \mathbf{C} +\frac{1}{\mathcal{T}}\!\left(\mathbf{C}(\tbar{\mathbf{J}}^T)^{-2}\! +\! \tbar{\mathbf{J}}^{-2}\mathbf{C} \right)  \right]\!\mathbf{M}_k^T,  
    \label{appeq:measVTInf}
\end{align}
which is now effectively an expansion in $\frac{1}{\mathcal{T}}$. For such filters, the measured covariances are no longer given only by mode covariances $\mathbf{C}$ as in Eq.~(\ref{appeq:measVT0}), but also by their correlation times via the propagator $\tbar{\mathbf{J}}^{-1}$. Furthermore, we find that that for the filter windows considered in this work, the contributions of $O(\frac{1}{\mathcal{T}})$ are non-negligible and must be taken into account.


The final expression can be further simplified using the Lyapunov equation that defines the steady-state $\mathbf{C}$, namely by setting the time-derivative in Eq.~(\ref{appeq:scaled2}) to zero. This yields: 
\begin{align}
    \mathbf{0} = \tbar{\mathbf{J}}\mathbf{C} + \mathbf{C}\tbar{\mathbf{J}}^T + \tbar{\mathbf{B}}
    \label{sieq:sscov}
\end{align}
Multiplying the above through on the left by $\tbar{\mathbf{J}}^{-1}$ and on the right by $(\tbar{\mathbf{J}}^T)^{-1}$ and rearranging, we arrive at:
\begin{align}
    \mathbf{C}(\tbar{\mathbf{J}}^T)^{-1} + \tbar{\mathbf{J}}^{-1}\mathbf{C} = - \tbar{\mathbf{J}}^{-1}\tbar{\mathbf{B}}(\tbar{\mathbf{J}}^T)^{-1}
    \label{sieq:sscovi}
\end{align}
If we consider the measured covariances Eq.~(\ref{appeq:measVTInf}) to lowest order in $O(\frac{1}{\mathcal{T}})$, Eq.~(\ref{sieq:sscovi}) can be used to obtain a simplified form:
\begin{align}
    \mathbf{\Sigma}_{jk}(t) \stackrel{\gamma \mathcal{T} \gg 1}{\simeq} &\sigma_{\rm vac}^2(\bar{n}_{\rm cl}+1)\mathbf{I}_2\delta_{jk}  -\frac{1}{2}\mathbf{M}_j\!\left[\mathbf{C}(\tbar{\mathbf{J}}^T)^{-1}\! +\! \tbar{\mathbf{J}}^{-1} \mathbf{C}  \right]\!\mathbf{M}_k^T  \nonumber \\
    =~&\sigma_{\rm vac}^2(\bar{n}_{\rm cl}+1)\mathbf{I}_2\delta_{jk}  + \frac{1}{2}\mathbf{M}_j\!\left[\tbar{\mathbf{J}}^{-1}\tbar{\mathbf{B}}(\tbar{\mathbf{J}}^T)^{-1} \right]\!\mathbf{M}_k^T
    \label{appeq:measVTInf2}
\end{align}
so that the measured covariances in the $\gamma \mathcal{T} \gg 1$ can be related directly to the linear(ized) drift (i.e. Jacobian) matrix $\tbar{\mathbf{J}}$ and linear(ized) diffusion matrix $\tbar{\mathbf{B}}$ alone.

\newpage

\section{Quantum state discrimination metrics under the NVK approximation}
\label{app:FI}

In this section, we will derive analytic expressions to quantify the fidelity of classification of QS states using the \qrc{}. Our results here will make use of the work carried out in all three prior sections. We first require the properties of the general QS states, determined in SI Sec.~\ref{app:qs}). Then, we use the results of SI Sec.~\ref{app:lin}) to determine the \qrc{} dynamics under input signals from the QS in the same measurement chain to lowest nontrivial order in nonlinearity. Finally, we must calculate the quantum statistics of measured outputs from this measurement chain (SI Sec.~\ref{app:io}) to determine classification fidelity metrics, more precisely Fisher's discriminant.

\subsection{QS and \qrc{} TEOMs under the NVK approximation}

We begin by explicitly specifying the FPE of the measurement chain we analyze in the main text, Eq.~(\ref{eq:sme}). As given by Eq.~(\ref{appeq:fpefull}), this is parameterized by the drift vector and diffusion matrix introduced in Eqs.~(\ref{appeq:Af}),~(\ref{appeq:Df}). For convenience, we will separate these matrices into QS and \qrc{} sectors, as will be relevant for the final expressions presented in the main text. Starting with the linear and nonlinear drift contributions respectively, these take the form:
\begin{align}
    \mathbf{A}^L &= 
    \begin{pmatrix}
        \mathbf{L}_a & \mathbf{0} \\ 
        -\bm{\Gamma} & \mathbf{L}_b 
    \end{pmatrix}\!,~\vec{f} = 
    \begin{pmatrix}
        \vec{\eta} \\
        \vec{0}
    \end{pmatrix},~
    \nonumber \\
    \vec{A}^N &=
    \begin{pmatrix}
        \vec{0} \\
        \vec{N}_b
    \end{pmatrix}
    \!,~\vec{N}_b = 
    \begin{pmatrix}
        +i{\beta}_1^{\dagger}(\beta_1)^2 \\
        -i({\beta}_1^{\dagger})^2\beta_1 \\
        \vdots \\
        +i{\beta}_K^{\dagger}(\beta_K)^2 \\
        -i({\beta}_K^{\dagger})^2\beta_K 
    \end{pmatrix}
    \label{appeq:driftexplicit}
\end{align}
Here $\mathbf{L}_a$ describes the linear dynamics of the QS, as introduced in SI Sec.~\ref{app:qs}. Then, $\mathbf{L}_b$ describes the corresponding linear dynamics of the \qrc{}, where $\mathbf{L}_b = \frac{1}{\gamma}{\rm diag}(\chi_1^{-1},\chi_1^{*-1},\ldots,\chi_K^{-1},\chi_K^{*-1})$, and $\chi_k^{-1} = i\Delta_k - \frac{\gamma_k+\Gamma_k}{2}$ defines the linear susceptibility of the \qrc{} modes. Then, the nonlinear terms are entirely localized to the \qrc{} modes via the vector $\vec{N}_b$, and depends cubically on phase space variables due to the Kerr nonlinearity. Finally, $\bm{\Gamma}$ defines the non-reciprocal coupling between the QS and \qrc{}, introduced in the main text via Eq.~(\ref{eq:couplingCirc}).

Similarly, we can define the constant and nonlinear contributions to the diffusion matrix respectively,
\begin{align}
    \mathbf{D}^L &= 
    \begin{pmatrix}
        \mathbf{D}_a & \mathbf{0} \\
        \mathbf{0} & \mathbf{0} 
    \end{pmatrix}\!, \nonumber \\
    \mathbf{D}^N &= 
    \begin{pmatrix}
        \mathbf{0} & \mathbf{0} \\
        \mathbf{0} & \mathbf{D}_b 
    \end{pmatrix}\!,~
    \mathbf{D}_b =
    \left(\!\!
    \begin{array}{ccccc}
        i\beta_1^2 &  & & &  \\
         & -i\beta_1^{\dagger 2} &  &  & \hspace{-5mm}\bigzero \\
         &  & \ddots &  & \\ 
         & \bigzero &  & i\beta_K^2 &  \\
         &  &  &  & -i\beta_K^{\dagger 2}
    \end{array}
    \!\!\right)
    \label{appeq:diffexplicit}
\end{align}
Here, $\mathbf{D}_a$ defines the linear part of the diffusion matrix due to the QS, and is the same as that defined in Eq.~(\ref{appeq:Df}). In contrast, $\mathbf{D}_b$ is the nonlinear part of the diffusion matrix that arises due to the Kerr nonlinearity of the \qrc{} modes, and as mentioned earlier depends quadratically on phase space variables.

Using the above, we can now use the results of SI Sec.~\ref{app:lin} to simply write down the TEOMs for quantum cumulants. First, we transition from the general notation used in SI Sec.~\ref{app:lin}, valid for general bosonic nonlinear systems, to the measurement chains we consider, with a QS and \qrc{}. In particular, we have for the expectation values at the expansion point, $\avg{\vec{\tbar{Z}}} = \begin{psmallmatrix}\avg{\hat{\tbar{\bm{a}}}} \\ \avg{\hat{\tbar{\bm{b}}}} \end{psmallmatrix}$. The expectation values of fluctuation variables are equivalently represented as $\avg{\vec{\zeta}} = \begin{psmallmatrix} \avg{\bm{\delta}\hat{\bm{a}}} \\ \avg{\bm{\delta}\hat{\bm{b}}} \end{psmallmatrix}$.

We can immediately write down the TEOMs for the first-order cumulants defining the expansion point, Eq.~(\ref{appeq:scaled1}), 
\begin{align}
    \frac{d}{dt} \!\!   
    \begin{pmatrix}
        \avg{\hat{\tbar{\bm{a}}}} \\
        \avg{\hat{\tbar{\bm{b}}}} 
    \end{pmatrix} =
    \begin{pmatrix}
        \mathbf{L}_a & \mathbf{0} \\ 
        -\bm{\Gamma} & \mathbf{L}_b 
    \end{pmatrix}
    \!\!
        \begin{pmatrix}
        \avg{\hat{\tbar{\bm{a}}}} \\
        \avg{\hat{\tbar{\bm{b}}}} 
    \end{pmatrix}
    + 
    \begin{pmatrix}
        \vec{0} \\
        \vec{\tbar{N}}_b \\
    \end{pmatrix}
    + 
    \begin{pmatrix}
    \vec{\tbar{\eta}} \\
    \vec{0} \\
    \end{pmatrix}
    \label{appeq:scaled1explicit}
\end{align}
Explicitly writing out the above immediately yields Eqs.~(\ref{eq:teoma}),~(\ref{eq:teomb}) of the main text. The above also determines the values of $\avg{\hat{\tbar{\bm{b}}}} $, which are necessary to determine all the subsequent quantities in the linearized analysis, as we will see shortly.



We next define the TEOMs for the second-order cumulants of the measurement chain. To do so we must obtain the Jacobian of the measurement chain introduced in Eq.~(\ref{appeq:jacobian}); using Eqs.~(\ref{appeq:driftexplicit}),~(\ref{appeq:diffexplicit}), this is straightforwardly calculated:
\begin{align}
    \tbar{\mathbf{J}} &=
    \begin{pmatrix}
        \mathbf{L}_a & \mathbf{0} \\ 
        -\bm{\Gamma} & \tbar{\mathbf{J}}_b 
    \end{pmatrix}\!,\nonumber \\
    \tbar{\mathbf{J}}_b &\equiv
    \mathbf{L}_b + \!
    \left(\!\!
    \begin{array}{ccccc}
        i2|\avg{\hat{\tbar{b}}_1}|^2 & i\avg{\hat{\tbar{b}}_1}^2  & & &  \\
         -i\avg{\hat{\tbar{b}}_1^{\dagger}}^2 & -i2|\avg{\hat{\tbar{b}}_1}|^2 &  &  & \hspace{-20mm}\bigzero \\
         &  & \ddots &  & \\ 
         & \hspace{-10mm}\bigzero &  & i2|\avg{\hat{\tbar{b}}_K}|^2 & i\avg{\hat{\tbar{b}}_K}^2 \\
         &  &  &  -i\avg{\hat{\tbar{b}}_K^{\dagger}}^2 & -i2|\avg{\hat{\tbar{b}}_K}|^2
    \end{array}
    \!\!\right)
    \label{appeq:jacobianexplicit}
\end{align}
We can also write down the linearized diffusion matrix $\tbar{\mathbf{B}}$ from Eq.~(\ref{appeq:lindiff}) as:
\begin{align}
    \tbar{\mathbf{B}} = 
    \begin{pmatrix}
        \mathbf{D}_a & \mathbf{0} \\
        \mathbf{0} & \tbar{\mathbf{D}}_b
    \end{pmatrix}\!,~
    \tbar{\mathbf{D}}_b = \!
    \left(\!\!
    \begin{array}{ccccc}
        i\avg{\hat{\tbar{b}}_1}^2\!\! &  & & &  \\
         & \!\!\!\!-i\avg{\hat{\tbar{b}}_1^{\dagger}}^2\!\! &  &  & \hspace{-5mm}\bigzero \\
         &  & \!\!\ddots\! &  & \\ 
         & \!\!\bigzero &  & \!\!\!i\avg{\hat{\tbar{b}}_K}^2\!\! &  \\
         &  &  &  & \!\!\!\!-i\avg{\hat{\tbar{b}}_K^{\dagger}}^2\!\!
    \end{array}
    \!\!\right)
    \label{appeq:difflinexplicit}
\end{align}

With these expressions, we can now write down Eq.~(\ref{appeq:scaled2}), the TEOMs for second-order cumulants to lowest order in the nonlinearity, as
\begin{align}
    \frac{d}{dt}\mathbf{C} = 
    \frac{d}{dt} 
    \begin{pmatrix}
    \mathbf{C}_a & \mathbf{C}_{ab} \\
    \mathbf{C}^T_{ab} & \mathbf{C}_b
    \end{pmatrix} 
    =
    \begin{pmatrix}
        \mathbf{L}_a & \mathbf{0} \\ 
        -\bm{\Gamma} & \tbar{\mathbf{J}}_b 
    \end{pmatrix}
    \begin{pmatrix}
    \mathbf{C}_a & \mathbf{C}_{ab} \\
    \mathbf{C}^T_{ab} & \mathbf{C}_b
    \end{pmatrix} 
    + 
    \begin{pmatrix}
    \mathbf{C}_a & \mathbf{C}_{ab} \\
    \mathbf{C}^T_{ab} & \mathbf{C}_b
    \end{pmatrix} 
    \begin{pmatrix}
        \mathbf{L}_a^T & -\bm{\Gamma}^T  \\ 
        \mathbf{0} &  \tbar{\mathbf{J}}_b^T
    \end{pmatrix}
    +
    \begin{pmatrix}
        \mathbf{D}_a & \mathbf{0} \\
        \mathbf{0} & \tbar{\mathbf{D}}_b
    \end{pmatrix}
    \label{appeq:scaled2explicit}
\end{align} 
which is simply Eq.~(\ref{eq:lyapunov}) of the main text.



Finally, once the steady-state first and second-order cumulants of the measurement chain are known to lowest order in nonlinearity, we can calculate the first-order cumulants of fluctuations by solving Eq.~(\ref{appeq:scaledf2}). First, this system can be compactly written in matrix form as
\begin{align}
    \frac{d}{dt}
    \begin{pmatrix}
    \avg{\bm{\delta}\hat{\bm{a}}} \\
    \avg{\bm{\delta}\hat{\bm{b}}}
    \end{pmatrix}
    =
    \begin{pmatrix}
        \mathbf{L}_a  & \mathbf{0} \\ 
        -\bm{\Gamma} & \tbar{\mathbf{J}}_b
    \end{pmatrix}
    \begin{pmatrix}
    \avg{\bm{\delta}\hat{\bm{a}}} \\
    \avg{\bm{\delta}\hat{\bm{b}}}
    \end{pmatrix}
    +\tbar{\Lambda}^{\frac{1}{2}} \!
    \begin{pmatrix}
        \vec{0} \\
        \vec{\tbar{h}}_b
    \end{pmatrix}.
\end{align}
First we immediately note that the steady-state expression for first-order cumulants of QS fluctuations simply vanishes, $\avg{\bm{\delta}\hat{\bm{a}}} = \vec{0}$, due to the absence of an inhomogeneous term on the right hand side. \qrc{} first-order cumulants, on the other hand, have such a term, which is $\propto \tbar{\Lambda}^{\frac{1}{2}}$ and arises from the Hessian tensor (see Eq.~(\ref{appeq:hessian})) contracted with the covariance matrix. This term is localized to the \qrc{} sector via the vector $\vec{h}_b$ as expected, since this is where the nonlinearity is present. Then, the steady-state for fluctuations in the first-order cumulants of the \qrc{} is given by 
\begin{align}
    \avg{\bm{\delta}\hat{\bm{b}}} = -\tbar{\Lambda}^{\frac{1}{2}}\tbar{\mathbf{J}}_b^{-1} \vec{\tbar{h}}_b.
    \label{appeq:QRCfSol1}
\end{align}
Furthermore, for the Kerr model, the vector $\vec{\tbar{h}}_b$ takes the simple form,
\begin{align}
    \vec{\tbar{h}}_b =  \tbar{\mathbf{H}}_b : \mathbf{C}_b =    
    \begin{pmatrix}
        \tbar{\mathbf{H}}_{b_1} : \mathbf{C}_{b_1} \\
        \tbar{\mathbf{H}}_{b_1^{\dagger}} : \mathbf{C}_{b_1} \\
        \vdots \\
        \tbar{\mathbf{H}}_{b_K} : \mathbf{C}_{b_K} \\
        \tbar{\mathbf{H}}_{b_K^{\dagger}} : \mathbf{C}_{b_K} 
    \end{pmatrix}
\end{align}
where $(:)$ denotes the standard tensor double contraction. Each element in $\vec{h}_b$ is hence obtained via the double contraction of a pair of 2-by-2 matrices, $\tbar{\mathbf{H}}_{b_k}$ or $\tbar{\mathbf{H}}_{b_k^{\dagger}}$,
\begin{align}
    \tbar{\mathbf{H}}_{b_k} = 
    \begin{pmatrix}
        +i\avg{\hat{\tbar{b}}^{\dagger}_k} &  +i\avg{\hat{\tbar{b}}_k}  \\
        +i\avg{\hat{\tbar{b}}_k} & 0
    \end{pmatrix}\!,~
\tbar{\mathbf{H}}_{b_k^{\dagger}} = 
    \begin{pmatrix}
        0 &  -i\avg{\hat{\tbar{b}}^{\dagger}_k}  \\
        -i\avg{\hat{\tbar{b}}^{\dagger}_k} & -i\avg{\hat{\tbar{b}}_k}
    \end{pmatrix},   
\end{align}
and $\mathbf{C}_{b_k}$ simply defines the local second-order cumulants for the $k$th \qrc{} mode:
\begin{align}
   \mathbf{C}_{b_k} = 
    \begin{pmatrix}
        C_{b_kb_k} & C_{b_k^{\dagger}b_k} \\
        C_{b_k^{\dagger}b_k} &  C_{b_k^{\dagger}b_k^{\dagger}}
    \end{pmatrix}.
\end{align}

This form indicates that the dynamics of the $k$th \qrc{} mode is ``driven'' by the covariances of the $k$th mode. Cross-cumulants such as $C_{b_kb_j}$ for $j\neq k$ do not appear in $\vec{h}_b$ since the nonlinear drift vector $\vec{\tbar{A}}_b^N$ that determines the Hessian tensor involves no nonlinear coupling terms between distinct \qrc{} modes. 

\subsection{Semi-analytic expressions for measured quadrature means, covariances, and Fisher's discriminant}

Having obtained the TEOMs to lowest nontrivial order in the nonlinearity for cumulants defining the internal dynamics of the measurement chain, we can finally compute the mean and covariances of measured output quadratures for steady-state quantum state classification. The steady-state mean value of measured \qrc{} quadratures is simply given by Eq.~(\ref{appeq:mukgeneral}):
\begin{align}
    \bm{\mu} &\equiv 
    \begin{pmatrix}
    \bm{\mu}_1 \\
    \vdots \\
    \bm{\mu}_K 
    \end{pmatrix}
    =
    \mathbb{E}\left[
    \begin{pmatrix}
        I_1 \\
        Q_1 \\
        \vdots \\
        I_K \\
        Q_K 
    \end{pmatrix}
    \right] =\sqrt{\frac{\mathcal{T}}{2}}\!
    \begin{pmatrix}
        \mathbf{M}_1 \\
        \vdots \\
        \mathbf{M}_K
    \end{pmatrix}
    \!\!
    \avg{\vec{Z}} \nonumber \\
    \implies \bm{\mu} &=
    \sqrt{\frac{\gammahet\mathcal{T}}{2}} \! \left( \tbar{\Lambda}^{-\frac{1}{2}}\avg{\hat{\tbar{\bm{b}}}} + \avg{\bm{\delta}\hat{\bm{b}}} \right)
\end{align}
where the second line is obtained since we are only reading out quadratures of \qrc{} modes using heterodyne readout. The mean separation that is required to calculate Fisher's discriminant is then immediately given by:
\begin{align}
    \delta\bm{\mu} = \bm{\mu}^{(l)}-\bm{\mu}^{(p)} = \sqrt{\frac{\gammahet\mathcal{T}}{2}} \! \left( \avg{\bm{\delta}\hat{\bm{b}}^{(l)}} - \avg{\bm{\delta}\hat{\bm{b}}^{(p)}} \right)
\end{align}
where we make use of the fact that $\avg{\hat{\tbar{\bm{b}}}^{(l)}}=\avg{\hat{\tbar{\bm{b}}}^{(p)}}$ for the binary state classification tasks we have considered. Hence Fisher discriminant is nonzero only due to the mean values of fluctuation variables. Using Eq.~(\ref{appeq:QRCfSol1}), we can finally write for the mean separation,
\begin{align}
    \delta\bm{\mu} = -\sqrt{\frac{\gammahet\mathcal{T}}{2}}\cdot \sqrt{\frac{\Lambda}{\gamma}}~\tbar{\mathbf{J}}_b^{-1} \tbar{\mathbf{H}}_b : [ \mathbf{C}^{(l)}_b - \mathbf{C}^{(p)}_b ]  = -\sqrt{\frac{\gammahet\mathcal{T}}{2}}\cdot \sqrt{\frac{\Lambda}{\gamma}}~\tbar{\mathbf{J}}_b^{-1}    
    \begin{pmatrix}
        \mathbf{H}_{b_1} : [ \mathbf{C}_{b_1}^{(l)}-\mathbf{C}_{b_1}^{(p)} ] \\
        \mathbf{H}_{b_1^{\dagger}} : [ \mathbf{C}_{b_1}^{(l)}-\mathbf{C}_{b_1}^{(p)} ] \\
        \vdots \\
        \mathbf{H}_{b_K} : [ \mathbf{C}_{b_K}^{(l)}-\mathbf{C}_{b_K}^{(p)} ] \\
        \mathbf{H}_{b_K^{\dagger}} : [ \mathbf{C}_{b_K}^{(l)}-\mathbf{C}_{b_K}^{(p)} ] 
    \end{pmatrix}
    \label{appeq:deltamuexp}
\end{align}

Of course Fisher's discriminant also requires us to obtain the noise properties of the measured quadratures in the steady-state. For simplicity, we consider the long-filter limit $\gamma \mathcal{T}\gg 1$ of measured covariances between \qrc{} modes $\hat{b}_k$ and $\hat{b}_j$ to lowest order in $\frac{1}{\mathcal{T}}$ via Eq.~(\ref{appeq:measVTInf2}),
\begin{align}
    \mathbf{\Sigma}_{jk}\!\! \stackrel{\gamma \mathcal{T}\gg 1}{\simeq} \!\! \sigma_{\rm vac}^2(\bar{n}_{\rm cl}+1)\mathbf{I}_2\delta_{jk} + \frac{1}{2}\mathbf{M}_j\!\left[\tbar{\mathbf{J}}^{-1}\tbar{\mathbf{B}}(\tbar{\mathbf{J}}^T)^{-1} \right]\!\mathbf{M}_k^T
    \label{appeq:covinter1}
\end{align}
Note that we are ultimately interested in the measured covariance matrix over all $K$ modes of the \qrc{} corresponding to the QS state $l$; this can be constructed from Eq.~(\ref{appeq:covinter1}) as
\begin{align}
    \mathbf{\Sigma}^{(l)} = 
    \begin{pmatrix}
        \mathbf{\Sigma}^{(l)}_{11} & \ldots & \mathbf{\Sigma}^{(l)}_{1K} \\
        \vdots & \ddots & \vdots \\
        \mathbf{\Sigma}^{(l)}_{K1} & \ldots & \mathbf{\Sigma}^{(l)}_{KK} 
    \end{pmatrix},
    \label{appeq:vtotexp1}
\end{align}
Recalling that the measurement matrices serve to extract the specific modes being observed, $\mathbf{\Sigma}^{(l)}$ will depend only on the \qrc{} sector of the matrix $\left[\tbar{\mathbf{J}}^{-1}\tbar{\mathbf{B}}(\tbar{\mathbf{J}}^T)^{-1} \right]$. We now proceed to simplify this term in particular, using the block form of the Jacobian matrix, Eq.~(\ref{appeq:jacobianexplicit}). More precisely, the inverse of the Jacobian matrix can be written explicitly in block form as:
\begin{align}
    \tbar{\mathbf{J}}^{-1} = 
    \begin{pmatrix}
        \mathbf{L}_a^{-1} & \mathbf{0}  \\
         \tbar{\mathbf{J}}_b^{-1}\bm{\Gamma}\mathbf{L}_a^{-1} & \tbar{\mathbf{J}}_b^{-1}
    \end{pmatrix}
\end{align}
Making further use of the block form of $\tbar{\mathbf{B}}$, Eq.~(\ref{appeq:difflinexplicit}), we find:
\begin{align}
    \tbar{\mathbf{J}}^{-1}\tbar{\mathbf{B}}(\tbar{\mathbf{J}}^T)^{-1} &=     
    \begin{pmatrix}
        \mathbf{L}_a^{-1} & \mathbf{0}  \\
         \tbar{\mathbf{J}}_b^{-1}\bm{\Gamma}\mathbf{L}_a^{-1} & \tbar{\mathbf{J}}_b^{-1}
    \end{pmatrix}
    \begin{pmatrix}
        \mathbf{D}_a & \mathbf{0} \\
        \mathbf{0} & \tbar{\mathbf{D}}_b 
    \end{pmatrix}
\begin{pmatrix}
        (\mathbf{L}^T_a)^{-1} & (\mathbf{L}^T_a)^{-1}\bm{\Gamma}^T(\tbar{\mathbf{J}}^T_b)^{-1}  \\
         \mathbf{0} & (\tbar{\mathbf{J}}^T_b)^{-1}
    \end{pmatrix} \nonumber \\
    &=
    \begin{pmatrix}
    \mathbf{L}_a^{-1}\mathbf{D}_a & \mathbf{0} \\
     \tbar{\mathbf{J}}_b^{-1}\bm{\Gamma}\mathbf{L}_a^{-1}\mathbf{D}_a & \tbar{\mathbf{J}}_b^{-1}\tbar{\mathbf{D}}_b
    \end{pmatrix}
        \begin{pmatrix}
        (\mathbf{L}^T_a)^{-1} & (\mathbf{L}^T_a)^{-1}\bm{\Gamma}^T(\tbar{\mathbf{J}}^T_b)^{-1}  \\
         \mathbf{0} & (\tbar{\mathbf{J}}^T_b)^{-1}
    \end{pmatrix} \nonumber \\
    &= 
    \begin{pmatrix}
     \mathbf{L}_a^{-1}\mathbf{D}_a(\mathbf{L}_a^T)^{-1} & & \mathbf{L}_a^{-1}\mathbf{D}_a(\mathbf{L}_a^T)^{-1}\bm{\Gamma}^T(\tbar{\mathbf{J}}_b^T)^{-1} \\
    \tbar{\mathbf{J}}_b^{-1}\bm{\Gamma}\mathbf{L}_a^{-1}\mathbf{D}_a(\mathbf{L}_a^T)^{-1} & & \tbar{\mathbf{J}}_b^{-1}\tbar{\mathbf{D}}_b(\tbar{\mathbf{J}}_b^T)^{-1} + \tbar{\mathbf{J}}_b^{-1}\bm{\Gamma}\mathbf{L}_a^{-1}\mathbf{D}_a(\mathbf{L}_a^T)^{-1}\bm{\Gamma}^T(\tbar{\mathbf{J}}_b^T)^{-1}
    \end{pmatrix}
\end{align}
The measurement matrices - encoding the fact that we are only monitoring \qrc{} modes - will serve to extract only the lower-right block in the above matrix. We can therefore write Eq.~(\ref{appeq:vtotexp1}) explicitly as:
\begin{align}
    \mathbf{\Sigma} \!\! \stackrel{\gamma \mathcal{T}\gg 1}{\simeq} \!\! \sigma_{\rm vac}^2(\bar{n}_{\rm cl}+1)\mathbf{I}_K + \sigma_{\rm vac}^2\frac{\gammahet}{\gamma}\!\mathbf{U}_K\left[ \tbar{\mathbf{J}}_b^{-1}\tbar{\mathbf{D}}_b(\tbar{\mathbf{J}}_b^T)^{-1} + \tbar{\mathbf{J}}_b^{-1}\bm{\Gamma}\mathbf{L}_a^{-1}\mathbf{D}_a(\mathbf{L}_a^T)^{-1}\bm{\Gamma}^T(\tbar{\mathbf{J}}_b^T)^{-1} \right]\mathbf{U}_K^T
    \label{appeq:vtotexp}
\end{align}
where the factor $\frac{\gammahet}{\gamma}$ arises due to the measurement matrices, and we have rewritten $\sigma_{\rm vac}^2 = \frac{1}{2}$. The above can be written to the form in Eq.~(\ref{eq:nvksigma}) of the main text. This finally allows us to construct the averaged covariance matrix $\mathbf{V} = \frac{1}{2}\left( \mathbf{\Sigma}^{(l)} + \mathbf{\Sigma}^{(p)}  \right)$. 

Finally, substituting Eqs.~(\ref{appeq:deltamuexp}),~(\ref{appeq:vtotexp}) into Eq.~(\ref{eq:fda}) of the main text, reproduced below for convenience:
\begin{align}
    \FD = \delta \bm{\mu}^T \cdot \mathbf{V}^{-1} \cdot \delta\bm{\mu},
\end{align}
yields the semi-analytic expression for the Fisher's discriminant for quantum state classification under heterodyne measurement of \qrc{} modes, to lowest nontrivial order in the nonlinearity $\Lambda$. This expression is used to calculate the approximate results labelled `NVK' presented throughout the main text.

\newpage

\section{Characterizing noise in a measurement chain using a phase-preserving parametric amplifier}
\label{app:noise}

In Sec.~\ref{sec:comp} of the main text, we analyze how nonlinear \qrcs{} incorporated in quantum measurement chains enable quantum state classification schemes that are more robust to added classical noise power than comparable approaches using linear amplifiers and nonlinear post-processing. This analysis requires a calibration of the added classical noise power, which we present in this section using a standard quantum measurement chain with phase-preserving amplifiers.


For concreteness, we consider the measurement chain shown in Fig.~\ref{fig:noiseChar}(a), which consists of a single-port phase-preserving (i.e. non-degenerate) parametric amplifier, one mode of which is monitored via homodyne measurement. The measurement chain is described by the SME
\begin{align}
d\rhoc = -i[\hat{H}_d,\rhoc]dt + \sum_n\gamma_{dn}\mathcal{D}[\hat{d}_n]\rhoc dt + \sqrt{\gamma_{d1}}\mathcal{S}_{\rm hom}[\hat{d}_1]\rhoc
\label{appeq:noisecharsme}
\end{align}
where $\hat{H}_d$ is the Hamiltonian describing the phase-preserving amplifier,
\begin{align}
    \hat{H}_d &= -\sum_n \Delta_{dn} \hat{d}_n^{\dagger}\hat{d}_n  + \eta_1(-i\hat{d}_1+i\hat{d}_1^{\dagger}) +G_{12}(-i\hat{d}_1\hat{d}_2 + h.c.),
 \label{eq:lnoise}
\end{align}
while the stochastic measurement superoperator for homodyne measurement is given by:
\begin{align}
    \mathcal{S}_{\rm hom}[\hat{d}_1]\rhoc = \left( \hat{d}_1 \rhoc + \rhoc \hat{d}_1^{\dagger} - \avgc{\hat{d}_1+\hat{d}_1^{\dagger}} \right)\! dW_{\mathcal{I}_d}(t)
    \label{appeq:shom}
\end{align}
For simplicity, we consider the case where $\Delta_{d1} = \Delta_{d2} = 0$, and $\gamma_{d1} = \gamma_{d2} \equiv \gamma_d$.

Note that being a linear system, the Hamiltonian in Eq.~(\ref{eq:lnoise}) that governs the phase-preserving amplifier is simply a special case of the general linear QS we have analyzed in SI Sec.~\ref{app:qs}. The correspondence simply requires setting $\kappa \to \gamma_d$, $\phi_{12} \to \frac{\pi}{2}$, and $\eta_2, G_1, G_2 \to 0$. This allows us to directly use solutions for the steady-state first and second-order cumulants derived earlier. Defining the corresponding vector of mode operators $\hat{\bm{d}} = (\hat{d}_1,\hat{d}_1^{\dagger},\hat{d}_2,\hat{d}_2^{\dagger})^T$, the solution for steady-state first-order cumulants is given by:
\begin{align}
    \avg{\hat{\bm{d}}} &= -\mathbf{L}_d^{-1}\vec{\eta} \nonumber \\
    &= 
    \frac{1}{\gamma_d^2-4G_{12}^2}
    \begin{pmatrix}
        2\gamma_d & 0 & 0 & 4G_{12} \\
        0 & 2\gamma_d & 4G_{12} & 0 \\
        0 & 4G_{12} & 2\gamma_d & 0 \\
        4G_{12} & 0 & 0 & 2\gamma_d 
    \end{pmatrix}
    \!\!
    \begin{pmatrix}
        \eta_1 \\
        \eta_1 \\
        0 \\
        0
    \end{pmatrix}
    \label{appeq:noisesol1}
\end{align}
Similarly, the matrix of second-order cumulants is given by:
\begin{align}
    \mathbf{C}_d = 
    \frac{1}{\gamma_d^2-4G_{12}^2}
    \begin{pmatrix}
        0 & 2G_{12}^2 & \gamma_d G_{12} & 0 \\
        2G_{12}^2 & 0 & 0 & \gamma_d G_{12} \\
        \gamma_d G_{12} & 0 & 0 & 2G_{12}^2 \\
        0 & \gamma_d G_{12} & 2G_{12}^2 & 0 
    \end{pmatrix}
    \label{appeq:noisesol2}
\end{align}

To characterize the noise in measured quadratures, we must compute the quantum statistics of the obtained measurements as described in SI Sec.~\ref{app:io}. However, the use of homodyne and not heterodyne detection means that expressions presented in that section for quantum statistics of measurements are modified, albeit in a simple way which we now explain.

\subsection{Measured homodyne records and amplifier gain}

For homodyne measurements described by the stochastic measurement superoperator in Eq.~(\ref{appeq:shom}), and using a single port used for both inputs and outputs, the obtained homodyne measurement records are given by:
\begin{align}
    \mathcal{I}_{d}(t) &= \xi_{\mathcal{I}_d} + \sqrt{\gamma_{d}}\left[ -\mathcal{I}_{d}^{\rm in}  + \sqrt{2}(\avg{\hat{X}_{d_1}} + \xi^{\rm qm}_{\mathcal{I}_d}) \right] + \sqrt{\bar{n}_{\rm cl}}\xi_{\mathcal{I}_d}^{\rm cl},
    \label{eq:Jdx}
\end{align}
The white noise terms $\xi$ are defined as in the main text (e.g. $\xi_{\mathcal{I}_d} = \frac{dW_{\mathcal{I}_d}}{dt}$), and $\mathcal{I}_{d}^{\rm in} = 2\eta_1/\gamma_d$ defines the reflected part of the coherent input drive that also appears as part of the output since we are now considering monitoring in reflection. The multiplicative factor of $\sqrt{2}$ enhances the signal in comparison to heterodyne monitoring; this simply emphasizes that homodyne readout is able to provide a higher signal-to-noise ratio than heterodyne readout, at the expense of monitoring only a single quadrature. Measured quadratures defined by filtering homodyne measurement records are defined as before, $I_d(t) = \frac{1}{\sqrt{2\mathcal{T}}}\int_{t_0}^{t_0+\mathcal{T}} d\tau~\mathcal{I}_{d}(\tau)$. 

In comparison to the analysis in SI Sec.~\ref{app:io}, the difference in homodyne monitoring is simply that only a single quadrature of a given mode is monitored. This difference is easily accounted by defining a new \textit{homodyne} measurement matrix for mode $\hat{d}_1$,
\begin{align}
    \mathbf{M}_{d_1} = \sqrt{\gamma_{d}}
    \begin{pmatrix}
     1 & 1 & 0 & 0
    \end{pmatrix}
\end{align}
Clearly $\mathbf{M}_{d_1}$ is a 1-by-2$\ntot$ matrix for a system of $\ntot$ modes (here, $\ntot=2$), describing the extraction of a single measured quadrature from the quantum system.

The expectation value of measurement records is then given by an expression analogous to Eq.~(\ref{appeq:mukgeneral}), except using the homodyne measurement matrix,
\begin{align}
    \mu_d = \mathbb{E}[I_d] = \sqrt{\frac{\mathcal{T}}{2}}\left( -\sqrt{\gamma_d}\cdot\mathcal{I}_d^{\rm in} + \mathbf{M}_{d_1}\avg{\hat{\bm{d}}} \right)
\end{align}
Using Eq.~(\ref{appeq:noisesol1}), we therefore obtain the steady-state reflection gain $\mathcal{G}$ of the phase-preserving amplifier,
\begin{align}
    \sqrt{\mathcal{G}} \equiv \frac{\mu_d}{\sqrt{\frac{\mathcal{T}}{2}}\mathcal{I}_{d}^{\rm in}} =  \frac{\gamma_d^2+4G_{12}^2}{\gamma_d^2-4G_{12}^2}
    \label{appeq:noisegain}
\end{align}


\begin{figure}[t]
    \centering
    \includegraphics[scale=1.0]{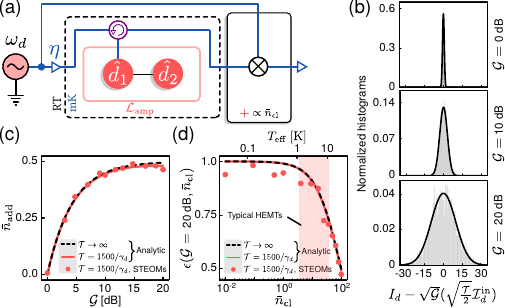}
    \caption{(a) Measurement chain for homodyne measurement of a phase-preserving amplifier. (b) Histograms of steady-state measured homodyne quadratures for different values of phase-preserving reflection gain $\mathcal{G}$, indicating increase in noise with gain. Here $\mathcal{T} = 1500/\gamma_d$. (c) Added noise power referred to the input, $\bar{n}_{\rm add}$, as a function of phase-preserving reflection gain $\mathcal{G}$, computed using both variances estimated using STEOM simulations (dots), as well as analytic calculations (solid and dashed lines) for different filter lengths. Here excess classical noise power $\bar{n}_{\rm cl} = 0$. (d) Quantum efficiency of the measurement chain as a function of excess classical noise power $\bar{n}_{\rm cl}$ and effective noise temperature $T_{\rm eff}$. Ranges of these two quantities for typical measurement chains in cQED are illustrated by the red shaded region.}
    \label{fig:noiseChar}
\end{figure}


\subsection{Covariances of measured quadratures and added quantum noise}

We can now calculate the variance of the measured quadrature, now for homodyne measurement, in which case the covariance ``matrix'' is now just a single number $ \Sigma_{d_1}$, and is defined via the homodyne measurement matrix. For simplicity we consider the expression analogous to Eq.~(\ref{appeq:measVTInf}), namely the $\mathcal{T}\to\infty$ limit often considered in the noise analysis of quantum amplifiers (referred to as calculating the noise at zero frequency). Then, to lowest order in $\frac{1}{\mathcal{T}}$,
\begin{align}
    \Sigma_{d_1} \!\! &= \mathbb{E}[I_{d_1}^2] - (\mathbb{E}[I_{d_1}])^2 \nonumber \\
    &\!\!\!\!\stackrel{\mathcal{T}\to\infty}{=} \sigma_{\rm vac}^2(\bar{n}_{\rm cl} + 1) -\frac{1}{2}\mathbf{M}_{d_1}\left[\mathbf{C}_d(\mathbf{L}_d^T)^{-1}\! +\! \mathbf{L}_d^{-1} \mathbf{C}_d\right] \mathbf{M}_{d_1}^T \nonumber \\
    &= \sigma_{\rm vac}^2(\bar{n}_{\rm cl} + 1) +\gamma_d\!\left(\!\frac{4\gamma_d C_{d_1^{\dagger}d_1}+4G_{12}(C_{d_1d_2}+C_{d_1^{\dagger}d_2^{\dagger}})  }{\gamma_d^2-4G_{12}^2}\! \right)
\end{align}
where we have used Eqs.~(\ref{appeq:noisesol1}),~(\ref{appeq:noisesol2}) in obtaining the second line. We immediately note that the measured variance does not only depend on covariances of mode $\hat{d}_1$; it also contains contributions due to correlations between modes $\hat{d}_1$ and $\hat{d}_2$, which are nonzero when the modes are coupled via a non-degenerate squeezing interaction, necessary for phase-preserving amplification. This contribution is the source of added noise during phase-preserving amplification.

Using Eqs.~(\ref{appeq:noisesol2}), together with the expression for $\mathcal{G}$ in Eq.~(\ref{appeq:noisegain}), we can simplify the expression for the measured variance further to:
\begin{align}
    \Sigma_{d_1} &= \sigma_{\rm vac}^2(\bar{n}_{\rm cl} + 1) + 2\sigma_{\rm vac}^2(\mathcal{G}-1) \nonumber \\
    &= \sigma_{\rm vac}^2 \bar{n}_{\rm cl} + \mathcal{G}\sigma_{\rm vac}^2 + (\mathcal{G}-1)\sigma_{\rm vac}^2
    \label{appeq:noiseexp}
\end{align}
where we have rearranged the expression into a more standard form in the second line. When $\bar{n}_{\rm cl}=0$ and for unit gain, the measured quadrature variance simply reduces to the vacuum value, $\sigma_{\rm vac}^2$. Contributions to the quadrature variance beyond this value arise from phase-preserving gain, as well as the classical contribution $\propto\bar{n}_{\rm cl}$. In Fig.~\ref{fig:noiseChar}(b), we plot histograms of steady-state measured quadratures $I_d$ with a filtering window $\mathcal{T}=1500/\gamma_d$, for a selection of increasing values of gain $\mathcal{G}$. The histograms are centred by subtracting off the mean, which from Eq.~(\ref{appeq:noisegain}) is simply equal to the amplified coherent input signal, $\sqrt{\mathcal{G}}(\sqrt{\frac{\mathcal{T}}{2}}\mathcal{I}_d^{\rm in})$. Moving from the top panel to the bottom panel, the increase in gain clearly leads to an increase in the added noise power, growing the variance of the measured amplified quadratures (indicated by the width of the histograms).

A standard metric to parameterize this added noise is by ``referring it to the amplifier's input'', namely normalizing the total noise by the phase-preserving gain and subtracting off the vacuum contribution to compute the total excess. The resulting added noise referred to the input $\bar{n}_{\rm add}$ is then given by:
\begin{align}
    \bar{n}_{\rm add}(\mathcal{G},\bar{n}_{\rm cl}) \equiv  \frac{\Sigma_{d_1}}{\mathcal{G}} - \sigma_{\rm vac}^2 =  \sigma_{\rm vac}^2\left[\frac{\bar{n}_{\rm cl}}{\mathcal{G}} \!+\! \left(1-\frac{1}{\mathcal{G}}\right)\right]
\end{align}
In the absence of excess classical noise ($\bar{n}_{\rm cl}=0$) the above expression for $\bar{n}_{\rm add}$ is plotted as a function of gain $\mathcal{G}$ in Fig.~\ref{fig:noiseChar}(c), together with its numerical calculation from measured quadratures obtained by integrating STEOMs of the phase-preserving amplifier. In the large $\mathcal{G}$ limit, $\bar{n}_{\rm add}$ approaches $\sigma_{\rm vac}^2 = \frac{1}{2}$, often referred to as the standard quantum limit of added noise for phase-preserving amplification: ``half a photon of noise referred to the input''. 

\subsection{Quantifying excess classical noise power}

The general form of measured noise from a measurement chain with a phase-preserving amplifier enables us to characterize the excess classical noise power $\bar{n}_{\rm cl}$ in a number of different ways. First, we note from Eq.~(\ref{appeq:noiseexp}) that $\bar{n}_{\rm cl}$ can be interpreted as excess noise power in units of vacuum state variance $\sigma_{\rm vac}^2$. Hence $\bar{n}_{\rm cl}$ units of classical noise added increases the variance of measured quadratures by $\bar{n}_{\rm cl}$ times the vacuum noise power.  

The units of excess classical noise are also commonly interpreted as a ``photon'' number, and converted to an effective temperature $T_{\rm eff}$ via the Bose-Einstein distribution at a representative system frequency $\Omega$, namely $\bar{n}_{\rm cl} \simeq \frac{k_b T_{\rm eff}}{\hbar\Omega}$. Typical HEMT amplifiers have noise temperatures $T_{\rm eff}$ of around 1-10~${\rm K}$, which corresponds to an effective classical noise power $\bar{n}_{\rm cl}$ of around 10-30 photons at microwave (GHz) frequencies.

The efficiency $\epsilon(\mathcal{G},\bar{n}_{\rm cl})$ of the measurement chain when incorporating a phase-preserving amplifier with gain $\mathcal{G}$ is defined as
\begin{align}
    \epsilon(\mathcal{G},\bar{n}_{\rm cl}) = \frac{\bar{n}_{\rm add}(\mathcal{G},\bar{n}_{\rm cl}=0)}{\bar{n}_{\rm add}(\mathcal{G},\bar{n}_{\rm cl})}
\end{align}
Unit efficiency is achieved when the amount of noise added is the minimum required by quantum mechanics, $\bar{n}_{\rm add}(\mathcal{G},\bar{n}_{\rm cl}=0) \to \frac{1}{2}$ in the large gain limit. The quantum efficiency is plotted as a function of the classical noise power in Fig.~\ref{fig:noiseChar}(d) for a typical phase-preserving gain of $\mathcal{G} =$~20~dB. The dashed vertical line marks $\bar{n}_{\rm cl} = 30$, which corresponds to a typical quantum efficiency of $\epsilon \simeq 0.77$, and to an added noise referred to the input $\bar{n}_{\rm add} \simeq 0.65 $. These are typical numbers for a modern measurement chain in cQED, and is what we use to label the typical excess classical noise powers $\bar{n}_{\rm cl}$ and effective temperatures $T_{\rm eff}$ in the red shaded region of Fig.~\ref{fig:ampComp} in the main text.

\newpage

\section{Supplementary benchmarking master equation simulations: Multimode measurement chain}
\label{si:verify}

In SI Sec.~\ref{si:singleNodeQRC}, we provided results benchmarking the performance of (S)TEOMs against full (S)ME simulations for a single coherently-driven \qrc{} mode undergoing heterodyne measurement. In this section, we present results for a more complex measurement chain that includes the key features common to the classification tasks considered in the main text: the inclusion of a measured QS and its coupling to the \qrc{}.

The specific measurement chain we consider is a simplified version of the full measurement chain employed for quantum state classification tasks in Fig.~\ref{fig:classifyAmpStates}(a) of the main text, and is depicted in Fig.~\ref{fig:appverification}(a). More precisely, the QS is as defined in Eq.~(\ref{eq:hsysamp}) of the main text but with $G_{12} = 0$ and $\Gamma_2 = 0$, thereby reducing the QS to a single mode $\hat{a}_1$. The rest of the measurement chain is unchanged: the mode $\hat{a}_1$ is coupled to the single \qrc{} mode via a non-reciprocal hopping interaction $\Gamma_1$, and the \qrc{} mode undergoes continuous heterodyne measurement. We wish to use this system to again classify two distinct states with equal means and distinct covariance, which we call Task~IV; the states to be distinguished are indexed $l=5$ and $l=6$, and are defined by the QS parameters in Table~\ref{apptab:task4}.


\begin{table}[h]
    \caption{QS parameters ($M=1$ only) defining the fourth classification task considered, in units of QS mode total loss rate $\kappa$.}
      \centering
    \begin{tabular}{m{0.035\columnwidth}<{\centering}|m{0.06\columnwidth}<{\centering}|m{0.055\columnwidth}<{\centering}|m{0.095\columnwidth}<{\centering}}
    \multicolumn{4}{c}{Task~IV} \\
    $l$ & $G_{1}^{(l)}$ & $\phi_{1}^{(l)}$ & $\eta_1^{(l)}$  \\
    \hline
    \hline
    {\color{st7}$7$} & {\color{st7}$0.3$} & {\color{st7}$+\frac{\pi}{2}$} & {\color{st7}$0.20$}$\amp$   \\
    {\color{st8}$8$} & {\color{st8}$0.3$} & {\color{st8}$-\frac{\pi}{2}$} & {\color{st8}$0.80$}$\amp$  \\
    \end{tabular}
    \label{apptab:task4}
\end{table}


\subsection{Individual quantum trajectories and measurement records}


\begin{figure}[t]
    \centering
    \includegraphics[scale=1.25]{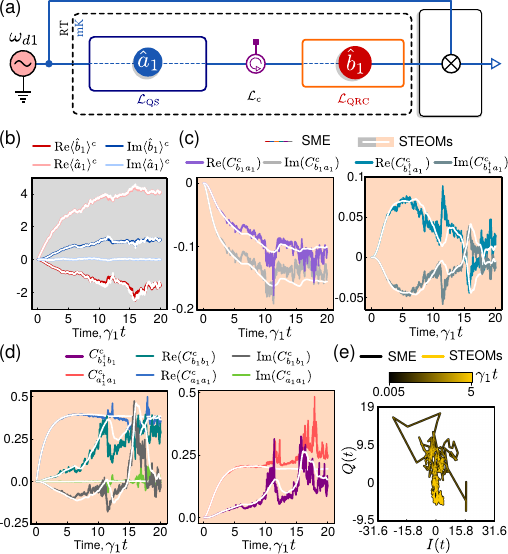}
    \caption{Benchmarking STEOMs against full SME simulations for the two-mode measurement chain depicted in (a). The single-mode measured quantum system is initialized in state $l=5$ as defined in Table~\ref{apptab:task4}, and with $\amp \simeq 4.5$. We choose $\kappa_1 = \Gamma_1 = \frac{1}{2}\gammahet$, and $\gammahet = \frac{2}{3}\gamma$ (hence $\kappa=\kappa_1+\Gamma_1=\frac{2}{3}\gamma$).  The \qrc{} parameters are given by ${\Delta}_1/{\gamma} = -\frac{2}{3}$, and ${\Lambda}_1/{\gamma} = 0.067$ such that $\CNLE = 0.385$ (see Eq.~\ref{eq:cnle}). Simulation results for a single measurement trajectory showing (b) first-order cumulants, (c) second-order cross-cumulants, (d) second-order self-cumulants, and (e) \qrc{} measured quadratures $\{I(t),Q(t)\}$.}
    \label{fig:appverification}
\end{figure}


We begin by simulating an individual quantum trajectory for the modes of the measurement chain for the state $l=5$ as defined in Table~\ref{apptab:task4}. For fair benchmarking comparisons, we choose parameters where the \qrc{} enables classification of the QS states; from results in the main text, we choose ${\Delta}_1/{\gammahet} = -\frac{2}{3}$ and $\CNLE = 0.385$ (see Eq.~\ref{eq:cnle}).  Unfortunately, increasing the number of modes immediately makes the SME simulations much more computationally taxing. To lower the modal occupation numbers, we keep $\CNLE$ fixed while using a relatively strong nonlinearity strength of ${\Lambda}/{\gamma} = 0.067$, which requires choosing a low $\amp \simeq 4.5$, corresponding to the drive strength $\eta_1/{\gamma} \simeq 0.6$. We note that this nonlinearity strength is about five times stronger than the largest nonlinearity used in Sec.~\ref{si:singleNodeQRC} of the SI. We simulate trajectories for a total time $\gamma_1 t = 20.0$. All other parameters of the measurement chain are summarized in the caption of Fig.~\ref{fig:appverification}.

The complete measurement chain of $\ntot=K+M=2$ quantum modes is determined by $2\ntot^2+3\ntot=14$ independent degrees of freedom: $\{\avgc{\hat{b}_1},\avgc{\hat{a}_1},\Cc{b_1b_1},\Cc{a_1a_1},\Cc{b_1a_1},\Cc{b_1^{\dagger}a_1} \}$ and their complex conjugates, as well as the real-valued cumulants $\{\Cc{b_1^{\dagger}b_1},\Cc{a_1^{\dagger}a_1}\}$.  We plot these quantities in Fig.~\ref{fig:appverification}(b)-(e) for a single measurement trajectory obtained using the STEOMs (white) and full SME simulations (color). We find that even though the nonlinearity is much stronger than those considered in the main text, there is in general good agreement between the methods. We also emphasize that for the same time step (here, $\gamma_1\Delta t = 5e^{-5}$), SME simulations with a Hilbert space cutoff of $40$ photons per quantum mode take about 37 hours to complete~\cite{johansson_qutip_2013} for each quantum trajectory; the STEOMs require 4 minutes.


\begin{figure}[t]
    \centering
    \includegraphics[scale=1.0]{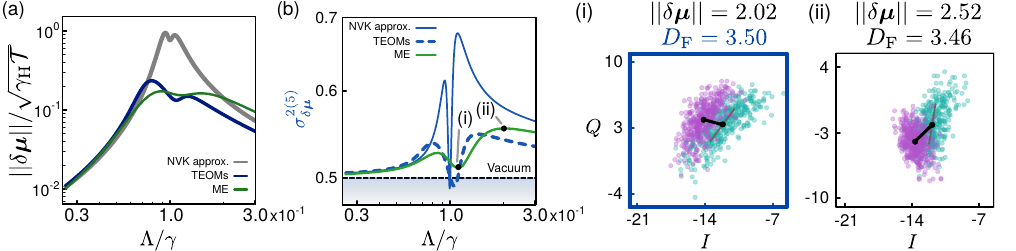}
    \caption{(a) Mean displacement $||\delta\bm{\mu}||$ and (b) Projected noise $\sigma^{2(7)}_{\delta\bm{\mu}}$ in the short-filter limit for Task~IV, calculated using TEOMs, ME integration, as well as the NVK approximation as a function of nonlinearity ${\Lambda}/\gammahet$ for fixed detuning ${\Delta}_1/{\gammahet} = -\frac{2}{3}$. (c) Measured distributions obtained using SME simulations for nonlinearity values marked (i) and (ii) in (b). }
    \label{fig:noiseManipVerify}
\end{figure}


\subsection{Mean displacement and projected noise for quantum state classification}

We now consider the full classification task, Task~IV, defined in Table~\ref{apptab:task4}. In particular, we wish to evaluate the quantities that define Fisher's discriminant for classification using both the TEOMs approach and its analytic NVK approximation, as done in Fig.~\ref{fig:ampClassifyLambda} of the main text; furthermore, we now wish to compare these results against exact ME and SME integration. Evaluating the required quantities of the mean separation and the projected noise as a function of \qrc{} parameters such as nonlinearity strength requires several independent simulations, and hence places an even larger computational overhead. We therefore consider an even lower $\amp = 3.5$ than in the previous subsection, to minimize modal occupation numbers. For (S)ME integration, we also consider a Hilbert space of $42$ photons for $\hat{a}_1$ and $17$ photons for mode $\hat{b}_1$. 

The steady-state mean separation $||\delta\bm{\mu}||$ only requires single operator, equal-time expectation values and therefore an efficient exact evaluation scheme can integrate the ME directly. In Fig.~\ref{fig:noiseManipVerify}(a), we plot $||\delta\bm{\mu}||$ using the NVK approximation, the TOEMs, as well as the full ME simulation as a function of nonlinearity strength over a large range. We note the existence of peaks in the mean separation where the \qrc{} response increases. The TEOMs exhibit the same qualitative behaviour, but with lower and broader peaks, due to the saturating effect of the nonlinearity, as also observed in Fig.~\ref{fig:ampClassifyDrive}. The ME simulations agree qualitatively with the TEOMs, although presenting slightly broader features. However, the most important feature of exhibiting a non-zero $||\delta\bm{\mu}||$ is robustly observed in the exact ME simulations, validating the result of the NVK approximation even at strong nonlinearities. \\

The calculation of projected noise $\sigma^{2(l)}_{\delta\bm{\mu}}$ is slightly more involved: depending on the length of the filter window $\mathcal{T}$ in comparison to the timescale of dynamics $\sim 1/\gamma$, the measured covariance matrix $\mathbf{\Sigma}^{(l)}$ will have contributions from correlations at different times. However, in the short filter limit, $\mathbf{\Sigma}^{(l)}$ and hence $\sigma^{2(l)}_{\delta\bm{\mu}}$ is defined entirely by equal time covariances (see Sec.~\ref{si:outputcovlimits} of the SI), and can again be efficiently calculated using ME simulations. We plot $\sigma^{2(l)}_{\delta\bm{\mu}}$ for $l=5$ in Fig.~\ref{fig:noiseManipVerify}(b) using the NVK approximation, TEOMs, and ME simulations, assuming $\gamma\mathcal{T} = 0.1 \ll 1$. We again see qualitatively similar behaviour across all three methods: the projected noise achieves a minimum for a specific nonlinearity value while also exhibiting nonzero $||\delta\bm{\mu}||$. The projected noise can be manipulated by adjusting the \qrc{} nonlinearity alone, again validating a key principle uncovered by the NVK approximation in this paper. We again observe the broadening of features via the TEOMs and ME approaches in comparison to the NVK approximation.

Finally, we would also like to evaluate the long-filter limit of readout, as $||\delta\bm{\mu}||$ scales directly with $\sqrt{\gamma\mathcal{T}}$. However, long filters probe the steady-state temporal correlation functions of the monitored modes (see Sec.~\ref{si:outputcov} of the SI). To evaluate via ME simulations, this requires the use of the quantum regression theorem and places a high demand on computational memory. An alternative approach is to integrate the SME to obtain measurement records, which provide the statistics of the filtered measured distributions directly as in the STEOMs approach. We plot measured distributions for the two marked nonlinearity values (i) and (ii) using $400$ simulated measurement records per state, and show the results in Fig.~\ref{fig:noiseManipVerify}(c). We note that here we use $t - \mathcal{T} = 0$ to maximize $||\delta\bm{\mu}||$ without having to simulate the SME for even longer, due to computational constraints. This leads to non-Gaussian signatures in transient dynamics appearing in the observed output, similar to Fig.~\ref{fig:nonGaussian} in SI Sec.~\ref{si:singleNodeQRC}.

The value (i) is a local minimum of projected noise, according to the ME results in Fig.~\ref{fig:noiseManipVerify}(b); we find that the position of the projected noise minimum is generally not changed by the filter window length. Here, we see that the two distributions obtained via SME simulations are almost parallel, so that the noise along the mean separation direction is close to minimal, consistent with the ME results. On the other hand, the value (ii) has a larger projected noise. We note that the measured distributions are no longer parallel. Crucially, we note that in (ii), the obtained $||\delta\bm{\mu}||$ is larger than in (i), by a factor of $1.25$. If the projected noise was unchanged, this would lead to a larger value of Fisher's discriminant, by a factor of $(1.25)^2$. However, the obtained value of Fisher's discriminant $\FD$ is in fact \textit{smaller} than in (i). This is because the mean separation is not entirely along the minimum noise eigenvector. The SME simulations again validate the QRC's ability to coherently manipulate quantum fluctuations.

\newpage

\section{Complex-$\mathcal{P}$ representation of single coherently-driven Kerr oscillator}
\label{app:complexP}

Recall that a single mode of our \qrc{} model - a single coherently-driven Kerr oscillator coupled to a zero temperature bath - is defined by the master equation
\begin{align}
     \Ltot\hat{\rho} = &-i\left[-\Delta_1\hat{b}^{\dagger}\hat{b} - \frac{\Lambda_1}{2}\hat{b}^{\dagger}\hat{b}^{\dagger}\hat{b}\hat{b} + \eta_b(e^{-i\varphi_{\eta}}\hat{b} + e^{i\varphi_{\eta}}\hat{b}^{\dagger}),\hat{\rho} \right] + \gamma \mathcal{D}[\hat{b}]\hat{\rho},
\end{align}
where $\gamma = \gamma_1+\Gamma_1$. For this system, it can be shown that the steady-state complex-$\mathcal{P}$ distribution $\mathcal{P}_{\rm ss}(\beta,\beta^{\dagger})$, in the phase space of variables $\beta,\beta^{\dagger}$ associated with operators $\hat{b},\hat{b}^{\dagger}$ respectively, can be found exactly by the method of potentials~\cite{drummond_quantum_1980, gardiner_stochastic_2009}. To arrive at the desired result, we note first that the drive phase $\varphi_{\eta}$ can simply be absorbed into the definition of the operators $\hat{b} \to \hat{b}e^{i\varphi_{\eta}},\hat{b}^{\dagger} \to \hat{b}^{\dagger}e^{-i\varphi_{\eta}}$, while leaving their commutator unchanged. The drive amplitude can thus be chosen to be completely real. Following this transformation, the steady-state complex-$\mathcal{P}$ distribution takes the form
\begin{align}
    &\mathcal{P}_{\rm ss}(\beta,\beta^{\dagger}) = (\beta)^{(c-2)}(\beta^{\dagger})^{\left(c^*-2\right)}\exp\left\{\frac{2\eta_b}{\Lambda_1}\!\left(\frac{1}{\beta}+\frac{1}{\beta^{\dagger}} \right) + 2\beta^{\dagger}\beta \right\}
\end{align}
where 
\begin{align}
    c = \frac{-i\Delta_1 + \frac{\gamma}{2}}{-i\Lambda_1}.
\end{align}
Knowledge of the exact steady-state complex $\mathcal{P}$-distribution allows one to calculate arbitrary moments of the quantum steady-state of the driven nonlinear mode by integrating over complex phase space. We find for arbitrary normal-ordered steady-state moments~\cite{drummond_quantum_1980}
\begin{align}
    &\avg{(\hat{b}^{\dagger})^j(\hat{b})^i} =  e^{-i\varphi_{\eta}(i-j)}\left|\frac{2\eta_b}{\Lambda_1}\right|^2\!\!\!\! \frac{\Gamma(c)\Gamma(c^*)}{\Gamma(c+i)\Gamma(c^*+j)}\frac{h(c+i,c^*+j,8|\eta_b/\Lambda_1|^2)}{h(c,c^*,8|\eta_b/\Lambda_1|^2)},
    \label{eq:prepmoments}
\end{align}
where we have reintroduced the drive phase by simply undoing the earlier transformation on operators $\hat{b},\hat{b}^{\dagger}$. Here the function $h(x,y,z)$ is given by the hypergeometric series
\begin{align}
    h(x,y,z) = \sum_{n=0}^{\infty} \frac{z^n}{n!}\frac{\Gamma(x)\Gamma(y)}{\Gamma(x+n)\Gamma(y+n)}
\end{align}
and $\Gamma(x)$ is the Gamma function.

Eq.~(\ref{eq:prepmoments}) is used to calculate exact steady-state first-order moments for the coherently-driven single-mode Kerr \qrc{} in SI Sec.~\ref{si:singleNodeQRC}.

\begin{center}
    \rule{10cm}{1pt}
\end{center}

\end{widetext}

\end{document}